\documentclass[12pt]{article}
\usepackage{booktabs}
\usepackage{tabulary}
\usepackage{multirow}
\usepackage{rotating}
\usepackage{epsfig}
\usepackage{psfrag}
\usepackage{latexsym}
\usepackage{url}
\usepackage{indentfirst}
\usepackage{fancyhdr}
\usepackage{dsfont}
\usepackage{adjustbox}
\usepackage{amsmath}
\usepackage{amssymb}
\usepackage{amsfonts}
\usepackage{mathrsfs}
\usepackage{amsthm}
\usepackage{pifont}
\usepackage{dsfont}
\usepackage{multirow}
\usepackage{array}
\usepackage{chngpage}
\usepackage{longtable}
\usepackage{cite}
\usepackage{bbold}
\usepackage{color}
\usepackage{braket}
\usepackage{colordvi}
\usepackage{fancybox}
\usepackage[footnotesize]{caption2}
\usepackage{graphicx}
\usepackage[center,footnotesize,hang]{subfigure}
\usepackage{bm}
\usepackage{bbm}
\usepackage{url}
\usepackage{gensymb}
\usepackage[table]{xcolor}
\usepackage[colorlinks, linkcolor=red, anchorcolor=black, citecolor=green]{hyperref}
\usepackage{multirow}
\usepackage{harpoon}
\usepackage{textcomp}  
\usepackage{enumitem}
\usepackage{mathrsfs}  
\usepackage[normalem]{ulem}

\usepackage{tikz}
\usepackage{diagbox}
\usepackage{enumitem}

\newcommand{\PreserveBackslash}[1]{\let\E^c=\\#1\let\\=\E^c}
\newcolumntype{C}[1]{>{\PreserveBackslash\centering}p{#1}}
\newcolumntype{R}[1]{>{\PreserveBackslash\raggedleft}p{#1}}
\newcolumntype{L}[1]{>{\PreserveBackslash\raggedright}p{#1}}
\allowdisplaybreaks  

\newcommand{\bq}{\begin{eqnarray}}
\newcommand{\nq}{\end{eqnarray}}

\newcommand{\ignore}[1]{}

\numberwithin{equation}{section}

\begin{document}

\title{
{\Large \bf
Discrete flavour and CP symmetries in light of JUNO and neutrino global fit\\[2mm]}
\date{}
\author{
Gui-Jun~Ding$^{a,b}$\footnote{E-mail: {\tt
dinggj@ustc.edu.cn}},  \
Cai-Chang Li$^{c,d,e}$\footnote{E-mail: {\tt ccli@nwu.edu.cn}},  \
Jun-Nan Lu$^{a}$\footnote{E-mail: {\tt
junnanlu@ustc.edu.cn}},  \
S.~T.~Petcov$^{\,f,g}$\footnote{Also at Institute of Nuclear Research and Nuclear Energy, Bulgarian Academy of Sciences, 1784 Sofia, Bulgaria.},
\\*[20pt]
\centerline{
\begin{minipage}{\linewidth}
\begin{center}
$^a${\it \small Department of Modern Physics and Anhui Center for fundamental sciences in theoretical physics,\\
University of Science and Technology of China, Hefei, Anhui 230026, China}\\[2mm]
$^b${\it \small College of Physics, Guizhou University, Guiyang 550025, China}\\ [2mm]
$^c${\it\small School of Physics, Northwest University, Xi'an 710127, China}\\[2mm]
$^d${\it\small Shaanxi Key Laboratory for Theoretical Physics Frontiers, Xi'an 710127, China}\\[2mm]
$^e${\it\small NSFC-SPTP Peng Huanwu Center for Fundamental Theory, Xi'an 710127, China}\\[2mm]
$^{f}$\,{\it \small INFN/SISSA, Via Bonomea 265, 34136 Trieste, Italy} \\
\vspace{2mm}
$^{g}$\,{\it\small Kavli IPMU (WPI), University of Tokyo,
5-1-5 Kashiwanoha, 277-8583 Kashiwa, Japan} \\
\vspace{2mm}
\end{center}
\end{minipage}}
\\[10mm]}}

\maketitle
\thispagestyle{empty}

\begin{abstract}

Working within the reference three-neutrino mixing framework, we confront the lepton mixing predictions derived using non-Abelian discrete flavour and CP symmetries with the first JUNO data on the solar neutrino mixing parameters $\sin^2\theta_{12}$
and with the results of the latest global neutrino data analysis. We focus on symmetry breaking patterns for which the lepton PMNS mixing matrix depends only on one or two free real parameters. Performing a comprehensive statistical analysis in each of the considered cases, we report the best fit values, the $3\sigma$ C.L. allowed ranges and the $\chi^2$-distributions of the lepton mixing observables - the three mixing angles and the three CP-violation phases. We find that the JUNO measurements can disfavour or rule out a number of the mixing patterns associated with specific types of breaking of the discrete flavour and CP symmetries.
The synergy of JUNO, DUNE and T2HK data can provide an exhaustive test of the considered approach to lepton mixing based on non-Abelian discrete lepton flavour symmetries combined with the CP symmetry.

\end{abstract}
\thispagestyle{empty}
\vfill

\clearpage

{\hypersetup{linkcolor=black}
\tableofcontents
}

\section{Introduction }

The neutrino oscillation data accumulated over many years
of measurements following the discovery of neutrino oscillations\cite{Super-Kamiokande:1998kpq,Kajita:2016cak,SNO:2001kpb,SNO:2002tuh,
McDonald:2016ixn} revealed the existence of three-flavour neutrino mixing
involving three light massive neutrinos (see, e.g.,~\cite{ParticleDataGroup:2024cfk}). In this reference three-neutrino mixing framework the unitary PMNS
neutrino matrix is parametrised by three angles and, depending
on whether the massive neutrinos are Dirac of Majorana particles,
by one Dirac  or one Dirac and two Majorana CP-violation (CPV) phases.
In the so-called ``standard'' parametrisation we will employ in the present study, the Pontecorvo–Maki–Nakagawa–Sakata (PMNS) matrix has the form~\cite{ParticleDataGroup:2018ovx,ParticleDataGroup:2024cfk}:
\begin{equation}
\label{eq:PMNS_def}
U=\begin{pmatrix}
c_{12}c_{13}  &   s_{12}c_{13}   &   s_{13}e^{-i\delta_{CP}}  \\
-s_{12}c_{23}-c_{12}s_{13}s_{23}e^{i\delta_{CP}}   &  c_{12}c_{23}-s_{12}s_{13}s_{23}e^{i\delta_{CP}}  &  c_{13}s_{23}  \\
s_{12}s_{23}-c_{12}s_{13}c_{23}e^{i\delta_{CP}}   & -c_{12}s_{23}-s_{12}s_{13}c_{23}e^{i\delta_{CP}}  &  c_{13}c_{23}
\end{pmatrix}\text{diag}(1,e^{i\frac{\alpha_{21}}{2}},e^{i\frac{\alpha_{31}}{2}})\,,
\end{equation}
where $c_{ij}\equiv \cos\theta_{ij}$, $s_{ij}\equiv \sin\theta_{ij}$, $\delta_{CP}$ is the Dirac CP-violation phase and
$\alpha_{21,31}$ are the two Majorana CPV phases~\cite{Bilenky:1980cx}. As is well known, the three-flavour neutrino oscillation probabilities
depend, in addition to the three neutrino mixing angles,
on the two neutrino mass squared differences
$\Delta m^2_{21}\equiv m^2_2-m^2_1$ and
$\Delta m^2_{31}\equiv m^2_3-m^2_1$, and on the Dirac CP violating (CPV) phase. The five parameters $\theta_{12}$,  $\theta_{13}$,  $\theta_{23}$,
$\Delta m^2_{21}$ and $|\Delta m^2_{31}|$ have been measured in neutrino
oscillation experiments with high precision.
However, the sign of  $\Delta m^2_{31}$ cannot be determined
from the existing data, the two possible signs corresponding to two types of
neutrino mass  spectra: with normal ordering (NO) if  $\Delta m^2_{31} > 0$,
and with inverted ordering (IO) for $\Delta m^2_{31} < 0$.
The data on the Dirac CPV phase is inconclusive having large uncertainties
\cite{ParticleDataGroup:2024cfk}.
There is no experimental information on the values of the Majorana
phases, which play important role, e.g., in the phenomenology of
neutrinoless double beta decay
(see, e.g., \cite{ParticleDataGroup:2018ovx}).

For the solar, reactor and atmospheric neutrino mixing angles
$\theta_{12}$,  $\theta_{13}$ and $\theta_{23}$, it was obtained in the
latest analysis of the global neutrino oscillation data~\cite{Esteban:2024eli}:
\begin{eqnarray}
\label{eq:th12th13exp}
\hskip-0.2in \sin^2\theta_{12}& = &0.308^{+0.012}_{-0.011}\,, \quad
\sin^2\theta_{13}=0.02215^{+0.00056}_{-0.00058}\,,\\
\hskip-0.2in  \label{eq:th23expSK}
\sin^2\theta_{23} &= &0.470^{+0.017}_{-0.013}~{\rm for~NO}\,,~
 \sin^2\theta_{23}=0.550^{+0.012}_{-0.015}~{\rm for~IO}~
{\rm with~SuperKamiokande}\,,\\
\hskip-0.2in  \label{eq:th23expNoSK}
\sin^2\theta_{23} &= &0.561^{+0.012}_{-0.015}~{\rm for~NO}\,,~
\sin^2\theta_{23}=0.562^{+0.012}_{-0.015}~{\rm for~IO}~
{\rm without~SuperKamiokande}\,,
\end{eqnarray}
where the $1\sigma$ C.L. uncertainties are given and ``SuperKamiokande'' refers to SuperKamiokande atmospheric neutrino data. Similar results have been obtained in the global analyses performed by the Valencia group~\cite{deSalas:2020pgw} and
Bari group~\cite{Capozzi:2025wyn,Capozzi:2025ovi}. Thus, the neutrino oscillation experiments have revealed that the neutrino or lepton mixing involves two large and one small angles. This is in strong contrast with the quark mixing exhibiting three small angles.

Understanding the origin of the peculiar pattern of lepton mixing is one of the major problems in today's particle physics. It is part of the more general formidable and still unresolved fundamental problem of uncovering the origin(s) of the lepton and quark flavours. Several possible mechanisms of generating the observed pattern of lepton mixing have been proposed in the literature
(see, e.g.,~\cite{Feruglio:2015jfa} for a review).
Among the most simple and elegant, if not
the most simple and elegant, is arguably the mechanism based on non-Abelian discrete flavour symmetries, which have been extensively explored, see, e.g.,~\cite{Altarelli:2010gt,Ishimori:2010au,King:2013eh,King:2014nza,King:2017guk,Petcov:2017ggy,Xing:2020ijf,Feruglio:2019ybq,Almumin:2022rml,Ding:2023htn,Ding:2024ozt} for reviews. Combining non-Abelian discrete flavour symmetries with the generalized CP (gCP) symmetry proved to be a very effective approach
to address the lepton mixing problem leading to testable predictions. In this approach, the underlying flavour theory is assumed to be invariant
under certain non-Abelian discrete flavour symmetry $G_{f}$ and gCP symmetry denoted as $H_{CP}$, which should be compatible with $G_f$~\cite{Holthausen:2012dk,Chen:2014tpa}. Thus the full symmetry group is $G_{f}\rtimes H_{CP}$~\cite{Feruglio:2012cw}. The three generation of leptons fields are assigned to transform non-trivially as some irreducible representations under the action of $G_f$, so that different entries of
the Yukawa couplings are correlated with each other.
The lepton flavour mixing arises from the non-trivial breaking of the flavour and CP symmetry group $G_{f}\rtimes H_{CP}$. The residual symmetries
$\mathcal{G}_{\ell}=G_{\ell}\rtimes H^{\ell}_{CP}$ and $\mathcal{G}_{\nu}=G_{\nu}\rtimes H^{\nu}_{CP}$~
\footnote{If neutrinos are Majorana particle, the residual flavour and CP symmetries of neutrino sector would be commutable, so that $\mathcal{G}_{\nu}$ would reduce to the direct product
$\mathcal{G}_{\nu}=G_{\nu}\times H^{\nu}_{CP}$.}
are preserved by charged lepton and neutrino sectors, respectively,
where $G_{\ell}$,  $G_{\nu}$ are subgroups of $G_{f}$ and
analogously $H^{\ell}_{CP}$, $H^{\nu}_{CP}$ are subsets of $H_{CP}$.
It is remarkable that the form of the
lepton mixing matrix as well as the values of the mixing angles and
CPV phases are completely fixed by the groups
$\mathcal{G}_{\ell}$, $\mathcal{G}_{\nu}$ and
their relative embedding into $G_{f}\rtimes H_{CP}$,
the details of the symmetry breaking dynamics being irrelevant.
Most importantly, this approach is very predictive since
the lepton mixing matrix can only depend on very few parameters.

Recently the Jiangmen Underground Neutrino Observatory (JUNO) experiment~\cite{JUNO:2025fpc},
considered originally in~\cite{Petcov:2001sy,Choubey:2003qx,Learned:2006wy,Li:2013zyd}, reported a high-precision determination
of the solar oscillation parameters $\sin^2\theta_{12}$ and  $\Delta m^2_{21}$
using the first 59.1 days of data ~\cite{JUNO:2025gmd}:
\begin{equation}
\label{eq:s12sq_JUNO_59day}
\sin^2\theta_{12}=0.3092\pm0.0087,~~~ \Delta m^2_{21}=(7.50\pm0.12)\times 10^{-5} \text{eV}^2\,.
\end{equation}
where we have given also
$1\sigma$ uncertainties in the best fit values.
This remarkable results improve the precision on $\sin^2\theta_{12}$
and $\Delta m^2_{21}$, reached in the global analyses using
data accumulated over at least several years of measurements,
by reducing the respective  $1\sigma$ uncertainties
by the factors of 1.6 and 1.3, respectively.
It is estimated that after six years of data collection,
the JUNO experiment will measure the oscillation parameters
$\Delta m^2_{21}$, $\sin^2\theta_{12}$ and $\Delta m^2_{31}$
with an $1\sigma$ uncertainty respectively of $0.3\%$, $0.5\%$ and $0.2\%$~\cite{JUNO:2022mxj} - unprecedented precision that cannot be matched by any other experiment, and can also determine the
neutrino mass ordering.
Different aspects of the implication of JUNO's first results
have been already discussed
in~\cite{Chao:2025sao,Li:2025hye,Zhang:2025jnn,Ge:2025csr,Huang:2025znh,Xing:2025xte,Chen:2025afg,Jiang:2025hvq,He:2025idv,Luo:2025pqy,Petcov:2025aci,Capozzi:2025ovi,Ding:2025dzc,Goswami:2025wla}.

The approach to the lepton mixing problem based on non-Abelian discrete flavour symmetries combined with the gCP symmetry can lead to sharp predictions for lepton mixing angles and CPV phases. In this work, we shall confront the theoretical predictions of models based on this approach
with the first JUNO results and the results of latest analysis of global neutrino data \texttt{NuFIT}-v6.0~\cite{Esteban:2024eli}, and we will perform  statistical analysis of the predictions of the considered models. We shall focus on the breaking patterns of the non-Abelian discrete flavour and CP symmetries, which enforce the
lepton mixing matrix to depend on only one or two real free parameters, concentrating on models with  massive Majorana neutrinos. In these cases the six observables of the PMNS matrix, the three mixing angles and three CPV phases,
are expressed in terms of one or two continuous real
parameters and, in certain models, on one or two discrete parameters taking limited number of fixed values. As a result, some of the lepton mixing angles and CPV phases
are strongly correlated with each other. These correlations can be tested experimentally. Besides the solar neutrino mixing angle $\theta_{12}$, we study the predictions for
the atmospheric neutrino angle $\theta_{23}$, Dirac CPV phase $\delta_{CP}$ and Majorana CPV phases $\alpha_{21}$ and $\alpha_{31}$, as well as for the CPV rephasing invariant $J_{CP}$ which controls the magnitude of CPV effects in neutrino oscillations~\cite{Krastev:1988yu} and is a leptonic analog of the rephasing invariant introduced by Jarlskog
in the quark sector~\cite{Jarlskog:1985ht}. It is interesting to observe that the JUNO's first result can already
strongly disfavor or even  exclude some mixing patterns predicted by the breaking of the considered flavour and CP symmetries. For example, the lepton mixing matrix with the second column given by $(\frac{1}{\sqrt{3}}, \frac{1}{\sqrt{3}}, \frac{1}{\sqrt{3}})^T$ is ruled out by JUNO, although it is compatible with \texttt{NuFIT}-v6.0 at
$3\sigma$ confidence level. Moreover, we see that the high precision measurement of $\theta_{12}$ by JUNO after six years of data collection, and the precise measurement of $\theta_{23}$ and $\delta_{CP}$ by the future long baseline experiments DUNE and T2HK allow to perform exhaustive test of the approach to lepton mixing based on the non-Abelian discrete flavour and CP symmetries. The JUNO's result has been studied in the context of discrete flavour symmetry in the case of when a $Z_2$ residual symmetry is assumed to be preserved by the neutrino mass term~\cite{Ge:2025csr,Petcov:2025aci}. In this work, we shall perform a comprehensive analysis of possible symmetry
breaking patterns in models in which a combination of
non-Abelian discrete flavour symmetry and CP symmetry
are assumed to hold initially. The potential of testing the non-Abelian discrete symmetry approach to the lepton flavour problem by the prospective data from JUNO, DUNE and
T2HK has been investigated in~\cite{Petcov:2018snn}, where only models based on the $A_4$ ($T^\prime$), $S_4$ and $A_5$ symmetries, but not also on the gCP symmetry, were considered. The capability of the ESSnuSB experiment to test and discriminate between a class of lepton flavor models based on discrete non-Abelian flavor symmetries was studied in~\cite{Blennow:2020snb}. In the models considered in the present study with discrete flavour and gCP symmetries,
one obtains predictions for the Majorana phases
$\alpha_{21}$ and $\alpha_{31}$ as well and
sharper predictions for the Dirac phase  $\delta_{CP}$.

The structure of this paper is organised as follows.
In section~\ref{sec:framework} the framework of determining lepton mixing from non-Abelian discrete flavour and CP symmetry breaking is presented, and we give the master formula of the lepton mixing matrix for different
symmetry breaking patterns. The symmetry breaking pattern with
$\mathcal{G}_{\ell}=Z_n\; (n\geq3)$ or $K_{4}$,
$\mathcal{G}_{\nu} = Z_{2}\times H^{\nu}_{CP}$ and the possible lepton mixing
matrices are studied in section~\ref{sec:onepara_lepton_mixing}, and
 expressions of lepton mixing parameters as well as CP invariants for
each case are presented. In this scenario, the lepton mixing matrix depends
on a single real rotation angle $\theta$,
and one column of the lepton mixing matrix is fixed by the residual symmetry.
We study the symmetry breaking pattern
$\mathcal{G}_{\ell}=Z_{2}\times H_{CP}^{\ell}$, $\mathcal{G}_{\nu}=Z_{2}\times H_{CP}^{\nu}$ in section~\ref{sec:twopara_lepton_mixing}, the resulting lepton
mixing matrix depends on two free rotation angles $\theta_{\ell}$ and
$\theta_{\nu}$, and only one entry of the lepton mixing matrix is fixed
by symmetry consideration. The current neutrino oscillation data can be well
accommodated in this scenario, and we make definite predictions for
$\theta_{23}$ and $\delta_{CP}$.
A statistical analysis is performed for
both NO and IO neutrino mass spectra in
section~\ref{sec:numerical-analysis},
the best fit values and the $3\sigma$ allowed regions of the
mixing parameters are listed in tables for both neutrino mass spectra.
The $\chi^2$ distributions of
$\sin^2\theta_{12}$, $\sin^2\theta_{23}$ and $\delta_{CP}$ are also plotted.
Finally we present our main conclusion in section~\ref{sec:conclusion}.
The details on the flavour groups $A_{5}$, $\Sigma(168)$, $\Delta(6n^{2})$ and $D_n$
are given in Appendix~\ref{sec:group_theory}.
The allowed regions of the lepton mixing parameters for all the viable
one-parameter and two-parameters
models are collected in
Appendix~\ref{sec:3sigma-ranges}. We have explored an alternative symmetry breaking pattern
$\mathcal{G}_{\ell}=Z_2$, $\mathcal{G}_{\nu}=K_{4}\times H^{\nu}_{CP}$ in
Appendix~\ref{app:Z2-K4},
for which the resulting lepton mixing matrix also depends
on two real free parameters. However, no phenomenologically viable lepton mixing matrix can be obtained
in this scenario because the predicted value of the solar neutrino mixing angle $\theta_{12}$ lies significantly outside the $3\sigma$ region allowed by JUNO.

\section{\label{sec:framework}Theoretical approach and master formulas}

It is well known that  in the non-Abelian discrete symmetry approach to the lepton flavour
problem one can predict both lepton mixing angles and CPV phases from the breaking of flavour and gCP symmetries~\cite{Feruglio:2012cw,Feruglio:2019ybq,Ding:2024ozt}. In this approach, the full flavour and CP symmetry described by the group $G_{f}\rtimes H_{CP}$~\footnote{When combining the flavour symmetry with gCP symmetry, a certain consistency condition has to be satisfied and the gCP symmetry corresponds to an automorphism of the flavour symmetry  group $G_{f}$~\cite{Holthausen:2012dk,Chen:2014tpa}. The full symmetry group is the semi-direct product of the flavour and gCP symmetry groups $G_{f}$ and $H_{CP}$, and it is generically denoted as $G_{f}\rtimes H_{CP}$~\cite{Feruglio:2012cw}.},  valid at some high energy scale, is broken down to $\mathcal{G}_{\ell}=G_{\ell}\rtimes H^{\ell}_{CP}$ and $\mathcal{G}_{\nu}=G_{\nu}\times H^{\nu}_{CP}$
in the charged lepton and neutrino sectors, respectively, where $G_{\ell}$,  $G_{\nu}$, with  $G_{\ell} \neq G_{\nu}$, and $H^{\ell}_{CP}$, $H^{\nu}_{CP}$ are subgroups of $G_{f}$ and $H_{CP}$ respectively. The lepton flavour mixing is completely fixed by the group structure  of $G_{f}\rtimes H_{CP}$ and the residual symmetries~(see, e.g., \cite{Chen:2014wxa,Chen:2015nha} and the review articles~\cite{Feruglio:2019ybq,Ding:2024ozt,Petcov:2017ggy}), with the details of the dynamics of the flavour and CP symmetries breaking being irrelevant. In order to accommodate the large lepton mixing angles, the three generations of left-handed leptons are usually assumed to transform as a faithful  irreducible triplet representation $\bm{3}$ of $G_f$.

As is also well known, the residual symmetries significantly constrain the lepton mass matrices and their diagonalization matrices (see, e.g., Refs.~\cite{Feruglio:2019ybq,Ding:2024ozt,Petcov:2017ggy} for details). In the case of the charged lepton mass matrix $m_{\ell}$, assumed to be given in the right-left (R-L) convention  (i.e., $\overline{\psi}_R m_{\ell}\psi_L$),  the residual symmetry $G_{\ell}\rtimes H^{\ell}_{CP}$ constrains the hermitian combination $m^{\dagger}_{\ell}m_{\ell}$ as follows~\footnote{The residual flavour and gCP symmetries of the charged leptons satisfy the following consistency condition:
\begin{eqnarray}\label{eq:consis-conD-Chlep}
X_{\ell\bm{r}}\rho^{*}_{\bm{r}}(g_{\ell})X^{-1}_{\ell\bm{r}}
=\rho_{\bm{r}}(g^{-1}_{\ell}),\qquad g_{\ell}\in G_{\ell}\,.
\end{eqnarray}}
\begin{eqnarray}
\label{eq:flavour-Clepton}
&&\rho^{\dagger}_{\bm{3}}(g_{\ell})m^{\dagger}_{\ell}m_{\ell}\rho_{\bm{3}}(g_{\ell})
=m^{\dagger}_{\ell}m_{\ell},  \qquad g_{\ell}\in G_{\ell}\,, \\
\label{eq:CP-Clepton}
&&X^{\dagger}_{\ell\bm{3}}m^{\dagger}_{\ell}m_{\ell}X_{\ell\bm{3}}
=(m^{\dagger}_{\ell}m_{\ell})^{*},  \qquad X_{\ell\bm{3}} \in H^{\ell}_{CP}\,.
\end{eqnarray}
It follows from Eqs.~\eqref{eq:flavour-Clepton} and \eqref{eq:CP-Clepton} that $X_{\ell\bm{3}}$ and $\rho_{\bm{3}}(g_{\ell})$ are diagonalized  by the same unitary  matrix $U_{\ell}$:
\begin{eqnarray}
\label{eq:Ul-master-formula}
U_{\ell}^{\dagger}\rho_{\bm{3}}(g_{\ell})U_{\ell}
=\widehat{\rho}_{\bm{3}}(g_{\ell})\,,
~~~~~~
U_{\ell}^{\dagger}X_{\ell\bm{3}}U_{\ell}^{*}
=\widehat{X}_{\ell\bm{3}}\,,
\end{eqnarray}
where $U^{\dagger}_{\ell}m^{\dagger}_{\ell}m_{\ell}U_{\ell}
=\text{diag}(m^2_e, m^2_{\mu}, m^2_{\tau})$,  and $\widehat{\rho}_{\bm{3}}(g_{\ell})$, $\widehat{X}_{\ell\bm{3}}$ are diagonal phase matrices. Note that if the residual flavour symmetry $G_{\ell}$ distinguishes the three families, the inclusion of gCP $H^{\ell}_{CP}$ would not add any information as far as lepton mixing is concerned~\cite{Li:2014eia,Ding:2013bpa}

Similarly, the light neutrino  Majorana mass matrix $m_{\nu}$ should be invariant under the action of residual symmetry $G_{\nu}\times H^{\nu}_{CP}$\footnote{The restricted consistency condition in the  case of neutrino Majorana mass term reads:
\begin{eqnarray}
\label{eq:consis-conD-neutrino}
X_{\nu\bm{r}}\rho^*_{\bm{r}}(g_\nu)X_{\nu\bm{r}}^{-1}
=\rho_{\bm{r}}(g_\nu),\qquad g_{\nu}\in G_{\nu},~~X_{\nu\bm{r}}\in H^{\nu}_{CP}\,.
\end{eqnarray}
This implies that the residual flavour and CP symmetries in the neutrino sector are commutable.}
This implies (for $m_\nu$ written in the R-L convention)\footnote{For neutrino Dirac mass matrix the invariance under the action of the residual symmetry leads to constraints similar to those  for the charged lepton mass matrix discussed above.}:
\begin{eqnarray}
\label{eq:flavour-neutrino}
&&\rho^T_{\bm{3}}(g_\nu)m_\nu\rho_{\bm{3}}(g_\nu) =m_\nu, \qquad g_{\nu}\in G_{\nu}\,, \\
\label{eq:CP-neutrino}&&X_{\nu\bm{3}}^Tm_\nu X_{\nu\bm{3}} =\,m_\nu^*,\qquad X_{\nu\bm{3}} \in H^{\nu}_{CP}\,,
\end{eqnarray}
As a consequence, the neutrino diagonalization matrix $U_{\nu}$ is strongly  constrained by the residual symmetry:
\begin{eqnarray}
\label{eq:Unu-master-formula}
U_{\nu}^{\dagger}\rho_{\bm{3}}(g_\nu)U_\nu
=\text{diag}(\pm 1,\pm 1,\pm 1)\,,\qquad
U_\nu^{\dagger}X_{\nu\bm{3}}U_{\nu}^{*}=\text{diag}(\pm 1,\pm 1,\pm 1)\,.
\end{eqnarray}
Once the flavour symmetry group $G_f$ and the residual symmetries $G_{\ell}\rtimes H^{\ell}_{CP}$  and $G_{\nu}\times H^{\nu}_{CP}$ are specified,
$U_{\ell}$ and $U_{\nu}$ can be constructed by solving
Eqs.~\eqref{eq:Ul-master-formula} and \eqref{eq:Unu-master-formula}.
Subsequently the  lepton mixing matrix can be determined as
\begin{equation}
U=U^{\dagger}_{\ell}U_{\nu}\,.
\end{equation}

In the following, we present several typical choices of the residual  symmetries for which the lepton mixing matrix only depends on one to two  free parameters~\footnote{In certain cases of $A_4$, $S_4$ and $A_5$ symmetries one gets predictions either for $\sin^2\theta_{12}$, or for $\sin^2\theta_{23}$, or else for the Dirac CPV phase $\delta_{CP}$ in the PMNS lepton mixing matrix even when the lepton mixing matrix depends on 3 real constant parameters (see, e.g., \cite{Girardi:2015rwa,Petcov:2018snn}). Here we do not consider these cases.}. The residual symmetries $G_{\ell}$ and  $G_{\nu}$  are  Abelian subgroups of the non-Abelian discrete symmetry $G_f$. In the present work, $G_{\ell}$ is restricted to $Z_n$ ($n\geq 2$) or $K_{4}=Z_2\times Z_2$, and $G_{\nu}$ is taken to be  $Z_2$ for Majorana neutrinos~\cite{Chen:2014wxa,Chen:2015nha,Ding:2024ozt}. When $G_{f}\rtimes H_{CP}$ is broken down to $Z_n$ ($n\geq 3$) and $K_{4}\times H^{\nu}_{CP}$ in the charged lepton and neutrino sectors, respectively, the resulting lepton mixing matrix takes a constant form determined entirely by these residual symmetries~\cite{deAdelhartToorop:2011re, Holthausen:2012wt, Fonseca:2014koa, Talbert:2014bda, Yao:2015dwa}. This classification yields 17 sporadic mixing patterns, all of which are now ruled out by experimental data~\cite{Fonseca:2014koa, Yao:2015dwa}. If instead $G_{f}\rtimes H_{CP}$ is broken into $Z_2$ and $K_{4}\times H^{\nu}_{CP}$ in the two sectors, the corresponding lepton mixing matrix depends on two real parameters~\cite{Lu:2018oxc}. However, this breaking pattern is disfavored by the latest global analysis from \texttt{NuFIT}~\cite{Esteban:2024eli}, as detailed in Appendix~\ref{app:Z2-K4}.

\subsection{\label{sec:abelian-subgroup-g_l} $\mathcal{G}_{\ell}=Z_n\; (n\geq3)$ or $K_{4}$ and $\mathcal{G}_{\nu}=Z_2\times H^{\nu}_{CP}$}

If the flavour symmetry $G_{f}$ is finite and the residual symmetries $G_{l}$ generated by $g_{\ell}$ distinguishes the three charged lepton families, then $G_{l}$ can be taken to be a cyclic group $Z_{n}$ with $n\geq3$ (or a Klein $K_{4}$ group). In this case, the representation matrix $\rho_{\bm{3}}(g_{\ell})$ is diagonalized by a unitary matrix $\Sigma_{\ell}$ via
\begin{equation}
 \Sigma_{\ell}^{\dagger} \rho_{\bm{3}}(g_{\ell}) \Sigma_{\ell} = \rho^{\text{diag}}_{\bm{3}}(g_{\ell}),
\end{equation}
where the columns of $\Sigma_{\ell}$ correspond to the eigenvectors of $\rho_{\bm{3}}(g_{\ell})$. The matrix $\Sigma_{\ell}$ is defined up to multiplication by a diagonal unitary matrix $Q_{\ell}$ and a permutation matrix $P_{\ell}$. As a consequence of Eq.~\eqref{eq:Ul-master-formula}, the charged lepton diagonalization matrix can be written as
\begin{equation}\label{eq:Ul1}
U_{\ell} = \Sigma_{\ell} P_{\ell} Q_{\ell}\,.
\end{equation}
It implies that  the unitary diagonalization matrix $U_{\ell}$ of the charged lepton mass matrix is determined entirely, up to column permutations and phases.

Concerning the neutrino sector, since $X_{\nu\bm{3}}$ is a symmetric and unitary matrix, by performing the Takagi factorization $X_{\nu\bm{3}}$ can be written as
\begin{equation}
\label{eq:XnuSigma}
X_{\nu\bm{3}}=\Sigma_{\nu}\Sigma_{\nu}^T\,,~~~\text{with}~~~\Sigma_{\nu}^{\dagger}\rho_{\bm{3}}(g_{\nu})\Sigma_{\nu}=\pm\text{diag}(1, -1, -1)\,.
\end{equation}
The $Z_2$ symmetry constrains $\rho^{diag}_{\bm{3}}(g_{\nu})$ to have two identical eigenvalues which is either $+1$ or $-1$. These can be any two of the three eigenvalues of $\rho_{\bm{3}}(g_{\nu})$. In Eq.~\eqref{eq:XnuSigma} we have chosen them
to be the 2nd and the 3rd ones, but we will account for the other possibilities by adding an appropriate permutation matrix $P_{\nu}$ in the neutrino diagonalization matrix
(see below).

The unitary matrix $\Sigma_{\nu}$ is constructed following the method outlined in Ref.~\cite{Yao:2016zev}. In accordance with the constraints from Eq.~\eqref{eq:Unu-master-formula}, the neutrino diagonalization matrix $U_{\nu}$ takes the form~\cite{Feruglio:2012cw,Yao:2016zev}:
\begin{equation}
\label{eq:Unu-Z2CP}U_{\nu}=\Sigma_{\nu} R_{23}(\theta)P_{\nu}Q_{\nu}\,,
\end{equation}
where $R_{23}(\theta)$ denotes a rotation in the 2–3 plane by an angle $\theta$, and $P_{\nu}$ represents an arbitrary permutation matrix. Here $Q_{\nu}$ in Eq.~\eqref{eq:Unu-Z2CP} is a diagonal matrix with elements equal to $\pm1$ and $\pm i$. Without loss of generality it can be given as $Q_{\nu}=\text{diag}(1,i^{k_{1}},i^{k_{2}})$ with $k_{1,2}=0, 1, 2, 3$. Hence the lepton mixing matrix $U_{}$ is of the form~\cite{Feruglio:2012cw,Chen:2014wxa,Yao:2016zev}
\begin{equation}
\label{eq:Upmns-Zn-Z2CP}
U_{}=U_{\ell}^{\dagger}U_{\nu}=Q_{\ell}^{\dagger}P_{\ell}^T\Sigma_{\ell}^{\dagger}\Sigma_{\nu} R_{23}(\theta)P_{\nu}Q_{\nu}\,,
\end{equation}
where the diagonal phase matrix $Q_{\ell}$ can be absorbed through a redefinition of the charged lepton fields, whereas the diagonal phase matrix $Q_{\nu}$ is to shift the Majorana CP phases $\alpha_{21}$ and $\alpha_{31}$ by integer multiples of $\pi$.  The rotation matrix $R_{23}(\theta)$ in Eq.~\eqref{eq:Upmns-Zn-Z2CP}  and the rotation matrix $R_{13}(\theta)$ to appear later are defined as
\begin{equation}\label{eq:RotationMatrix}
R_{13}(\theta)=\left(
\begin{array}{ccc}
 \cos\theta  & 0 & \sin\theta  \\
 0 & 1 & 0 \\
 -\sin\theta  & 0 & \cos\theta
\end{array}
\right), \qquad
R_{23}(\theta)=\left(
\begin{array}{ccc}
 1 & 0 & 0 \\
 0 & \cos \theta & \sin \theta \\
 0 & -\sin \theta  & \cos \theta
\end{array}
\right)\,.
\end{equation}
Thus the lepton mixing matrix in Eq.~\eqref{eq:Upmns-Zn-Z2CP} depends only on a single parameter $\theta$ which can be determined from the well measured reactor angle $\theta_{13}$. This allows predictions for the solar and atmospheric mixing angles $\theta_{12}$ and $\theta_{23}$, as well as for the CPV phases. Furthermore, since the $Z_2$ residual flavour symmetry can only distinguish one neutrino family from the other two generations, so that only one column of the lepton mixing matrix is fixed through the choice of residual symmetry, as can be seen from Eq.~\eqref{eq:Upmns-Zn-Z2CP}. Note that the lepton mixing matrix is determined up to possible permutations of rows and columns, since both neutrino and charged lepton masses are not constrained in this approach. The permutation matrices $P_{\ell}$ and $P_{\nu}$ in Eq.~\eqref{eq:Upmns-Zn-Z2CP} can take the following six forms:
\begin{equation}
\begin{aligned}
\label{eq:per_matrix}
P_{123}=&\left(\begin{array}{ccc}
1  & 0  &  0 \\
0  & 1  &  0\\
0  & 0  &  1
\end{array}\right),\qquad
P_{231}=\left(\begin{array}{ccc}
0&  1&  0 \\
0&  0&  1  \\
1&  0&  0
\end{array}\right),\qquad
P_{312}=\left(\begin{array}{ccc}
0&  0  &1  \\
1&  0  &0 \\
0&  1  &  0
\end{array}\right)\,,\\
P_{132}=&\left(\begin{array}{ccc}
1  &  0 &  0 \\
0  &  0 &  1 \\
0  &  1 &  0
\end{array}\right),\qquad
P_{213}=\left(\begin{array}{ccc}
0  &  1  &  0 \\
1  &  0  &  0 \\
0  &  0  &  1
\end{array}\right),\qquad
P_{321}=\left(\begin{array}{ccc}
0 &0 &1  \\
0 &1 &0  \\
1 &0 &0
\end{array}\right)\,.
\end{aligned}
\end{equation}
By comparing with the oscillation data in Ref.~\cite{Esteban:2024eli}, one can determine which permutation matrices $P_{\ell}$ and $P_{\nu}$ are phenomenologically allowed. If the residual symmetry of neutrino is $\mathcal{G}_{\nu}=Z_2$ where no residual CP symmetry is preserved, the lepton mixing matrix would be of the form of Eq.~\eqref{eq:Upmns-Zn-Z2CP} by replacing $R_{23}(\theta)$ with a unitary rotation in the (23) plane. As a consequence, more free parameters would be involved and the values of Majorana phases can not be predicted.

\subsection{\label{sec:Z2-g_l} $\mathcal{G}_{\ell}=Z_2\times H^{\ell}_{CP}$ and $\mathcal{G}_{\nu}=Z_2\times H^{\nu}_{CP}$}

When the residual symmetry in the charged lepton sector is assumed to be $Z^{g_{\ell}}_2\times H^{\ell}_{CP}$, the diagonalization matrix  $U_{\ell}$ of the charged lepton mass matrix is constrained through the Takagi factorization described in the preceding section to take the following form~\cite{Yao:2016zev}:
\begin{equation}\label{eq:sigma_l}
U_{\ell}=\Sigma_{\ell}R_{23}(\theta_{\ell})P_{\ell}Q_{\ell}\,,
\end{equation}
where $\Sigma_{\ell}$ is the Takagi factorization of the charged lepton residual CP transformation, $Q_{\ell}=\text{diag}(e^{i\alpha_{e}},e^{i\alpha_{\mu}},e^{i\alpha_{\tau}})$ with arbitrary real parameters $\alpha_{e}$, $\alpha_{\mu}$, $\alpha_{\tau}$, and $P_{\ell}$ denotes a generic three-dimensional permutation matrix chosen from the possibilities listed in Eq.~\eqref{eq:per_matrix}. Consequently, the residual symmetry fixes the lepton mixing matrix as
\begin{equation}\label{eq:gen_PMNS}
U\equiv U^{\dagger}_{\ell}U_{\nu}=Q^{\dagger}_{\ell}P^{T}_{\ell}R^{T}_{23}(\theta_{\ell})\Sigma R_{23}(\theta_{\nu})P_{\nu}Q_{\nu}\,,
\end{equation}
with
\begin{equation}\label{eq:Sigma_matrix}
\Sigma\equiv\Sigma^{\dagger}_{\ell}\Sigma_{\nu}\,.
\end{equation}
The phase matrix $Q_{\ell}$ can be absorbed by the charged lepton fields and the effect of $Q_{\nu}$ is a possible change of the Majorana phases by $\pi$. In the model independent framework, where lepton masses are not predicted, the PMNS matrix is determined up to permutations of its rows and columns, indicated by the left multiplication by $P^{T}_{\ell}$ and right multiplication by $P_{\nu}$. Both permutation matrices $P_{\ell}$ and $P_{\nu}$ may take any of the six forms specified in Eq.~\eqref{eq:per_matrix}. Notably, in this setup, one entry of the PMNS matrix is constrained to a constant value by residual symmetry—specifically, the (11) entry of $\Sigma$.

Furthermore, all mixing angles and CPV phases depend exclusively on two free parameters $\theta_{\ell}$ and $\theta_{\nu}$. The fundamental domain for both parameters is $\left[0, \pi\right)$, since the lepton mixing matrix $U$ from Eq.~\eqref{eq:gen_PMNS} satisfies
\begin{eqnarray}
\nonumber && U(\theta_{\ell}+\pi,\theta_{\nu})=P^{T}_{\ell}\text{diag}(1,-1,-1)P_{\ell}U(\theta_{\ell},\theta_{\nu}), \\
&& U(\theta_{\ell},\theta_{\nu}+\pi)=U(\theta_{\ell},\theta_{\nu})P^{T}_{\nu}\text{diag}(1,-1,-1)P_{\nu}\,,
\end{eqnarray}
where the diagonal matrices $P^{T}_{\ell}\text{diag}(1,-1,-1)P_{\ell}$ and $P^{T}_{\nu}\text{diag}(1,-1,-1)P_{\nu}$ can be absorbed into $Q_{\ell}$ and $Q_{\nu}$, respectively. In addition, the mixing matrix $U$ exhibits the following symmetry properties:
\begin{equation}
\label{eq:PMNS_symmetry} P^{T}_{\ell}P_{132}P_{\ell}U(\theta_{\ell},\theta_{\nu})=Q^{\prime}_{\ell}U(\theta_{\ell}+\frac{\pi}{2},\theta_{\nu})\,,\qquad U(\theta_{\ell},\theta_{\nu})P^{T}_{\nu}P_{132}P_{\nu}=U(\theta_{\ell},\theta_{\nu}+\frac{\pi}{2})Q^{\prime}_{\nu}\,,
\end{equation}
where $Q^{\prime}_{\ell}$ and $Q^{\prime}_{\nu}$ are given by
\begin{equation}\label{eq:re_phase}
Q^{\prime}_{\ell}=P^{T}_{\ell}P_{132}P_{\ell}Q_{\ell}P^{T}_{\ell}P^{T}_{132}\text{diag}(1,-1,1)P_{\ell}Q^{\dagger}_{\ell}, \quad
Q^{\prime}_{\nu}=Q_{\nu}P^{T}_{\nu}\text{diag}(1,-1,1)P^{T}_{132}P_{\nu}Q^{\dagger}_{\nu}P^{T}_{\nu}P_{132}P_{\nu}\,.
\end{equation}
Here $Q^{\prime}_{\ell}$ represents an arbitrary phase matrix, while $Q^{\prime}_{\nu}$ is a diagonal matrix with entries $\pm1$ and $\pm i$. Both can be absorbed into $Q_{\ell}$ and $Q_{\nu}$, respectively. As a result, Eq.~\eqref{eq:PMNS_symmetry} implies that applying the row permutation $P^{T}_{\ell}P_{132}P_{\ell}$ and column permutation $P^{T}_{\nu}P_{132}P_{\nu}$ to $U$ does not produce a new mixing pattern for any specific choice of $P_{\ell}$ and $P_{\nu}$. Therefore, among the 36 possible permutations of rows and columns, only nine yield physically independent mixing patterns. This also means that the matrix element fixed by residual symmetry can occupy any one of nine distinct positions in the mixing matrix.

In the paradigm of discrete flavor and CP symmetries, the CP violation arises from the mismatch between the residual CP symmetries of the charged lepton and neutrino sectors. If some common residual CP transformation is shared by the charged lepton and neutrino sectors, the CP symmetry would be preserved, the Dirac and Majorana CP violation phases would be equal to $0$ or $\pi$~\cite{Feruglio:2012cw,Ding:2024ozt}. On the other hand, if no common residual CP transformation is shared by the charged
lepton and neutrino sectors, the CP symmetry  is not conserved and the Majorana and/or Dirac phases take CP-violating values. 

\section{Lepton  mixing with one free parameter\label{sec:onepara_lepton_mixing}}

In this section, we consider the scenario that the flavour and CP symmetry $G_{f}\rtimes H_{CP}$ is broken down to
$\mathcal{G}_{\ell}=Z_n\; (n\geq3)$ or $K_{4}$ and $\mathcal{G}_{\nu}=Z_2\times H^{\nu}_{CP}$ in the charged lepton and neutrino sectors respectively. The master formula of the lepton mixing matrix is given by Eq.~\eqref{eq:Upmns-Zn-Z2CP}. It is notable that the lepton mixing matrix and all mixing parameters depend on a single rotation angle $\theta$, and one column of the mixing matrix is fixed by residual symmetry. This type of symmetry breaking pattern has also been studied for many flavour symmetry groups combined with gCP, such as $A_4$~\cite{Feruglio:2012cw,Ding:2013bpa,Li:2016nap}, $S_4$~\cite{Feruglio:2012cw,Li:2013jya,Ding:2013hpa,Li:2014eia,Feruglio:2013hia,Penedo:2017vtf}, $T^\prime$~\cite{Girardi:2013sza}, $\Delta(27)$~\cite{Branco:2015hea,Branco:2015gna}, $\Delta(48)$~\cite{Ding:2013nsa,Ding:2014hva}, $A_5$~\cite{Li:2015jxa,DiIura:2015kfa,Ballett:2015wia,Turner:2015uta,DiIura:2018fnk}, $\Delta(96)$~\cite{Ding:2014ssa} and the infinite group series $\Delta(3n^2)=(Z_n\times Z_n)\rtimes Z_3$~\cite{Hagedorn:2014wha,Ding:2015rwa}, $\Delta(6n^2)=(Z_n\times Z_n)\rtimes S_3$~\cite{Hagedorn:2014wha,Ding:2014ora} and $D^{(1)}_{9n, 3n}=(Z_{9n}\times Z_{3n})\rtimes S_3$~\cite{Li:2016ppt}. A systematical analysis of the discrete flavour group $G_f$ up to order 2000 was performed in~\cite{Yao:2016zev} in this scenario. It was found that all viable mixing patterns can be obtained by considering $A_{5} \rtimes H_{CP}$, $\Sigma(168) \rtimes H_{CP}$,  $\Delta(6n^{2}) \rtimes H_{CP}$ and $D^{(1)}_{9n, 3n}\rtimes H_{CP}$. Since the series group $D^{(1)}_{9n, 3n}$ is a subgroup of $\Delta(6(9n)^2)$ group for any positive integer $n$~\cite{Li:2016ppt}, all breaking patterns originating from $D^{(1)}_{9n, 3n} \rtimes H_{CP}$ can be derived from those of $\Delta(6(9n)^2)\rtimes H_{CP}$. Therefore
we will not explicitly discuss the flavour symmetry $D^{(1)}_{9n, 3n}$ combined with CP.

The possible residual subgroups $\mathcal{G}_{\ell}=Z_n$ or $K_{4}$ and $\mathcal{G}_{\nu}=Z_2\times H^{\nu}_{CP}$ have been studied for the symmetries $A_{5} \rtimes H_{CP}$, $\Sigma(168) \rtimes H_{CP}$ and  $\Delta(6n^{2}) \rtimes H_{CP}$~\cite{Yao:2016zev}, and five possible patterns of lepton mixing matrix are found to be compatible with current experimental data of neutrino oscillation~\cite{Yao:2016zev}.
For phenomenological analysis in the next section, we report the explicit form of the mixing matrix together with the residual symmetry in the following. For each independent mixing pattern, we can straightforwardly extract the expressions of the mixing angles $\sin^2\theta_{13}$, $\sin^2\theta_{12}$, $\sin^2\theta_{23}$ and the CP invariants $J_{CP}$, $I_{1}$, $I_{2}$ in the usual way. In the present work, we will adopt the standard parametrization of the lepton mixing matrix in Eq.~\eqref{eq:PMNS_def}. As regards the CPV, three weak basis invariants $J_{CP}$, associated with the Dirac CPV phase  $\delta_{CP}$, which is a leptonic analog of the invariant introduced by Jarlskog in the quark sector~\cite{Jarlskog:1985ht} and controls the magnitude of the CPV effects in neutrino oscillations ~\cite{Krastev:1988yu}, and $I_1$ and $I_{2}$~related to the Majorana phases
$\alpha_{21}$ and $\alpha_{31}$~\cite{Branco:1986gr,Nieves:1987pp,Nieves:2001fc,Jenkins:2007ip,Branco:2011zb}
can be defined as:
\begin{eqnarray}
\nonumber && J_{CP}=\Im{(U_{11}U_{33}U^{*}_{13}U^{*}_{31})}=\frac{1}{8}\sin2\theta_{12}\sin2\theta_{13}\sin2\theta_{23}\cos\theta_{13}\sin\delta_{CP}\,, \\
\nonumber&& I_1=\Im\left(U^{*2}_{11}U^2_{12}\right)=\frac{1}{4}\sin^22\theta_{12}\cos^4\theta_{13}\sin\alpha_{21}\,,\\
\label{eq:def_CP_invariants}&&I_2=\Im\left(U^{*2}_{11}U^2_{13}\right)=\frac{1}{4}\sin^22\theta_{13}\cos^2\theta_{12}\sin(\alpha_{31}-2\delta_{CP})\,.
\end{eqnarray}
We note that a given Majorana phase $\alpha_{i1}$ would violate the CP-invariance if both ${\rm Im} (U^{*2}_{11}U^2_{1i}) \neq 0$ and ${\rm Re} (U^{*2}_{11}U^2_{1i}) \neq 0$, $i=2,3$ (see, e.g., \cite{Bilenky:2001rz}).

\begin{description}[labelindent=-0.8em, leftmargin=0.3em]

\item[~~ (\uppercase\expandafter{\romannumeral1})]
The lepton  mixing matrix compatible with oscillation data is given by
\begin{equation}\label{eq:pmnsIa}
U^{I}=\frac{1}{\sqrt{3}}
\left(\begin{array}{ccc}
\sqrt{2} \sin \varphi _1 ~&~ e^{i \varphi _2} ~&~ \sqrt{2}\cos\varphi_1 \\
\sqrt{2}\cos\left(\varphi _1-\frac{\pi }{6}\right) ~&~ -e^{i\varphi_2} ~&~ -\sqrt{2} \sin \left(\varphi_1-\frac{\pi}{6}\right) \\
\sqrt{2}\cos\left(\varphi_1+\frac{\pi}{6}\right)  ~&~  e^{i\varphi_2} ~&~ -\sqrt{2}\sin\left(\varphi_1+\frac{\pi}{6}\right)
\end{array}
\right)R_{23}(\theta)Q_{\nu}\,,
\end{equation}
where $\varphi_1$ and $\varphi_2$ are rational angles determined by residual symmetries. The mixing matrix $U^{I}$ arises from the flavour symmetry groups $\Delta(6n^2)$ combined with gCP symmetry. The group structure of $\Delta(6n^2)$ is detailed in Appendix~\ref{sec:Delta6n2}. It is generated by four generators $a$, $b$, $c$, and $d$. In our working basis, the matrix representations of these generators are either real symmetric or diagonal across all irreducible representations. As a consequence, the gCP transformation matrices coincide in the same form with the representation matrices of the group generators in this basis.

The symmetry breaking pattern leading to $U^{I}$ is as follows~\cite{Ding:2014ora}:
\begin{equation}
\Delta(6n^2)\rtimes H_{CP}~~:~~  G_{\ell}=\left\langle ac^{s}d^{t}\right\rangle,~~
G_{\nu}=Z^{bc^xd^x}_2,~~
X_{\nu\bm{r}}=\rho_{\bm{r}}(c^{\gamma}d^{-2x-\gamma}) \,,
\end{equation}
with $s,t,x,\gamma=0,1,\cdots n-1$. The parameters $\varphi_1$ and $\varphi_2$ in the PMNS matrix $U^{I}$ is then given by:
\begin{equation}\label{eq:phi12_I}
\varphi_1=\frac{s-x}{n}\pi,\qquad \varphi_2=\frac{2t-s-3(\gamma+x)}{n}\pi\,,
\end{equation}
which are interdependent of each other, and they can take the discrete values
\begin{equation}
\varphi_1=0, \pm\frac{1}{n}\pi, \pm\frac{2}{n}\pi, \ldots \pm\frac{n-1}{n}\pi\,, \quad \varphi_2~(\mathrm{mod}~2\pi)=0, \frac{1}{n}\pi, \frac{2}{n}\pi, \ldots \frac{2n-1}{n}\pi\,,
\end{equation}

In particular, widely studied smaller groups $S_4\cong[24,12]$ ($n=2$) and $\Delta(96)\cong [96,64]$ ($n=4$) can admit a reasonably good fit to the experimental data. This is compatible with the known results in the literature~\cite{Feruglio:2012cw,Ding:2013hpa,Li:2014eia,Li:2013jya,Ding:2014ssa}.
From the PMNS matrix $U^{I}$ in Eq.~\eqref{eq:pmnsIa}, we can read out the lepton mixing angles as follow
\begin{equation}
\begin{aligned}
\sin^2\theta_{13}=&\frac{1}{3}\left(1+\cos^2\theta\cos2\varphi_1+\sqrt{2}\sin2\theta\cos\varphi_1\cos\varphi_2\right),\\
\sin^2\theta_{12}=&\frac{1+\sin^2\theta\cos2\varphi_1-\sqrt{2}\sin2\theta\cos\varphi_1\cos\varphi_2}
{2-\cos^2\theta\cos2\varphi_1-\sqrt{2}\sin2\theta\cos\varphi_1\cos\varphi_2}= 1 + \frac{\cos2\varphi_1 - 1}{3\cos^2\theta_{13}}
\,,\\
\sin^2\theta_{23}=&\frac{1-\cos^2\theta\sin\left(\pi/6+2\varphi_1\right)-\sqrt{2}\sin2\theta\cos\varphi_2\sin\left(\pi/6-\varphi_1\right)}
{2-\cos^2\theta\cos2\varphi_1-\sqrt{2}\sin2\theta\cos\varphi_1\cos\varphi_2}\,.
\end{aligned}
\end{equation}
A sum rule between the solar and reactor mixing angles is given by the relation
\begin{equation}\label{eq:th12th13varph1}
3\cos^2\theta_{12}\cos^2\theta_{13}=2\sin^2\varphi_1\,,
\end{equation}
which leads to the constraint $\cos^2\theta_{12}\cos^2\theta_{13} \leq 2/3$. Using the experimentally determined $3\sigma$ ranges $0.275 \leq \sin^2\theta_{12} \leq 0.345$ and $0.02030 \leq \sin^2\theta_{13} \leq 0.02388$~\cite{Esteban:2024eli}, the allowed values for the parameter $\varphi_1$ are found to be
\begin{equation}
\label{eq:varphi_range} \varphi _1/\pi\in[0.4351, 0.5649]\cup [1.4351, 1.5649]\,.
\end{equation}
This indicates that $\varphi_1$ is approximately $\pi/2$ or $3\pi/2$. Additionally, when $\theta_{13}$ is restricted to its $3\sigma$ experimental bounds, the solar mixing angle $\theta_{12}$ must lie within the following interval:
\begin{equation}
 0.317\leq\sin^{2}\theta_{12}\leq0.345\,.
\end{equation}
This range will be directly tested by the JUNO experiment, which is projected to constrain $\sin^2\theta_{12}$ to $[0.3022, 0.3118]$ at $3\sigma$ after six years of operation~\cite{JUNO:2022mxj}.

The three CP rephasing invariants $J_{CP}$, $I_1$, and $I_2$ are expressed as:
\begin{equation}
\begin{aligned}
&\left|J_{CP}\right|=\frac{1}{6\sqrt{6}}\left|\sin2\theta\sin\varphi_2\sin3\varphi_1\right|\,,\\
&|I_1|=\frac{4}{9} \left|\cos \theta  \sin ^2\varphi _1 \sin \varphi _2 \left(\cos \theta  \cos \varphi _2-\sqrt{2} \sin \theta  \cos \varphi _1\right)\right|\,,\\
&|I_2|=\frac{4}{9} \left|\sin \theta  \sin ^2\varphi _1 \sin \varphi _2 \left(\sin \theta  \cos \varphi _2+\sqrt{2} \cos \theta  \cos \varphi _1\right)\right|\,.
\end{aligned}
\end{equation}
The above three CP invariants are conventionally defined in Eq.~\eqref{eq:def_CP_invariants}. In this work, we shall present the absolute  values of $J_{CP}$, $I_1$ and $I_2$ because the signs of $I_1$ and $I_2$ depend  in the case of CP-invariance on the  relative CP parity of the relevant neutrino states, which is encoded in the matrix $Q_{\nu}$, and the overall signs of all the three CP invariant would be changed if the left-handed lepton doublets are assigned to conjugate triplet $\bar{\bm{3}}$ instead of $\bm{3}$.

Furthermore, an exact sum rule relating the mixing angles and the Dirac CPV phase can be derived as follows:
\begin{equation}\label{eq:angle_phse_caseIa}
\begin{aligned}
\cos\delta_{CP}=&\frac{\cos2\theta_{23}\left(3\cos2\theta_{12}-2\sin^2\varphi_1\right)+\sqrt{3}\sin2\varphi_1}
{3\sin2\theta_{12}\sin\theta_{13}\sin2\theta_{23}}
\\
=&\frac{
\cos2\theta_{23}
 \left(\cos^2\theta_{12}\sin^2\theta_{13} - \sin^2\theta_{12}\right )
+ (2 - 3\cos^2\theta_{12}\cos^2\theta_{13})^{\frac{1}{2}}
\cos\theta_{12}\cos\theta_{13}
}
{\sin2\theta_{12}\sin\theta_{13}\sin2\theta_{23}}\,,
\end{aligned}
\end{equation}
where we have used Eq.~\eqref{eq:th12th13varph1} to derive the second expression for $\cos\delta_{CP}$. This relation can alternatively be derived from the identities $|U_{\mu1}|^2=2\cos^2(\varphi_1-\pi/6)/3$ and $|U_{\tau1}|^2=2\cos^2(\varphi_1+\pi/6)/3$. Because the parameter $\varphi_1$ should be around $\pi/2$ or $3\pi/2$ as shown in Eq.~\eqref{eq:varphi_range}, the sum rule of Eq.~\eqref{eq:angle_phse_caseIa} is approximately
\begin{equation}
\label{eq:sum_app_caseIa}\cos\delta_{CP}\simeq\frac{(3\cos2\theta_{12}-2)\cot2\theta_{23}}{3\sin2\theta_{12}\sin\theta_{13}}\,.
\end{equation}
This implies that $\delta_{CP}$ would be nearly maximal if the atmospheric angle $\theta_{23}$ takes the maximal value $\theta_{23}=\pi/4$. We allow the three mixing angles to freely vary in the experimentally preferred $3\sigma$ ranges~\cite{Esteban:2024eli}, then the sum rule Eq.~\eqref{eq:sum_app_caseIa} leads to
\begin{equation}
-0.3440\leq\cos\delta_{CP}\leq0.4526\,.
\end{equation}
Needless to say, the improved measurement of the mixing angles particularly $\theta_{12}$ and $\theta_{23}$ could help to make more precise prediction for $\delta_{CP}$ in our framework.

Note that taking the fixed column vector
$\sqrt{\frac{2}{3}}\left(\sin\varphi_{1},\cos(\varphi_{1}-\frac{\pi}{6}),\cos(\varphi_{1}+\frac{\pi}{6})\right)^{T}$
as the second or third column of the PMNS matrix leads to a prediction incompatible with experimental observations.

\item[~~ (\uppercase\expandafter{\romannumeral2})]

The lepton mixing matrix for the second case is
\begin{equation}\label{eq:pmnsII}
U^{II}_{1}=\frac{1}{\sqrt{3}}
 \left(
\begin{array}{ccc}
 e^{i \varphi _1} & 1 & e^{i \varphi _2} \\
 \omega  e^{i \varphi _1} & 1 & \omega ^2 e^{i \varphi _2} \\
 \omega ^2 e^{i \varphi _1} & 1 & \omega  e^{i \varphi _2} \\
\end{array}
\right)R_{13}(\theta)Q_{\nu}\,,\qquad
U^{II}_{2}=P_{132}U^{II}_{1}\,,
\end{equation}
where $\omega=e^{2\pi i/3}$. These two forms differ only by the exchange of the second and third rows. Such mixing structures can be derived from flavour symmetry based on $\Delta(6n^2)$ flavour symmetry group combined with CP symmetry. The associated residual symmetry is:
\begin{equation}
\Delta(6n^2)\rtimes H_{CP}~~: \quad   G_{\ell}=\left\langle ac^{s}d^{t}\right\rangle,
\quad G_{\nu}=Z^{c^{n/2}}_{2}, \quad  X_{\nu\bm{r}}=\rho_{\bm{r}}(c^{\gamma}d^{\delta})\,,
\end{equation}
with $n$ being an even integer. The parameters $\varphi_1$ and $\varphi_2$ are defined as
\begin{equation}
\varphi_{1}=\frac{\gamma+\delta+2s}{n}\pi,\qquad \varphi_{2}=\frac{2\delta-\gamma+2t}{n}\pi\,.
\end{equation}
They are interdependent of each other and can take the discrete values
\begin{equation}
\varphi_1, \varphi_2~(\textrm{mod}~2\pi)=0, \frac{1}{n}\pi, \frac{2}{n}\pi,\ldots,\frac{2n-1}{n}\pi\,,
\end{equation}

The smallest symmetry group capable of reproducing the experimental mixing angles for specific values of $\theta$ is $S_4$ (i.e., $\Delta (24)\cong S_4$ with $(\varphi_1,\varphi_2) = (0, \pi)$). The second column of $U^{II}_{1}$ and $U^{II}_{2}$ are $(1, 1, 1)^{T}/\sqrt{3}$ which is the second column of tri-bimaximal mixing matrix, and consequently they are the so-called TM2 pattern. We can extract the following results for the lepton mixing angles
\begin{eqnarray}
\nonumber \sin^2\theta_{13}&=&\frac{1}{3}\left[1+\sin2\theta\cos(\varphi_2-\varphi_1)\right], \qquad
\sin^2\theta_{12}=\frac{1}{2-\sin2\theta\cos(\varphi_2-\varphi_1)}\,,\\
\nonumber \sin^2\theta_{23}&=&\frac{1-\sin2\theta\sin\left(\varphi_2-\varphi_1+\pi/6\right)}{2-\sin2\theta\cos(\varphi_2-\varphi_1)}~~~\mathrm{for}~~~U^{II}_{1}\,,\\
\label{eq:mixing_angles_case_II} \sin^2\theta_{23}&=&\frac{1+\sin2\theta\sin\left(\varphi_2-\varphi_1-\pi/6\right)}{2-\sin2\theta\cos(\varphi_2-\varphi_1)}~~~\mathrm{for}~~~U^{II}_{2}\,.
\end{eqnarray}
Therefore, the solar and reactor mixing angles satisfy the well-known sum rule:
\begin{equation}\label{eq:TM2_sum_rule}
3\cos^2\theta_{13}\sin^2\theta_{12}=1\,.
\end{equation}
This implies a lower bound on the solar mixing angle $\sin^2\theta_{12} > 1/3$. Eliminating the rotation angle $\theta$, we find that the reactor mixing angle and the atmospheric mixing angle are related as follow,
\begin{equation}
\begin{split}
&\frac{3\cos^2\theta_{13}\sin^2\theta_{23}-1}{1-3\sin^2\theta_{13}}=\frac{1}{2}+\frac{\sqrt{3}}{2}\tan\left(\varphi_2-\varphi_1\right),~~~\text{for}~~~U^{II}_{1}\,,\\
&\frac{3\cos^2\theta_{13}\sin^2\theta_{23}-1}{1-3\sin^2\theta_{13}}=\frac{1}{2}-\frac{\sqrt{3}}{2}\tan\left(\varphi_2-\varphi_1\right),~~~\text{for}~~~U^{II}_{2}\,.
\end{split}
\end{equation}
For both the mixing matrices $U^{II}_{1}$ and $U^{II}_{2}$, the three CP invariants take the form
\begin{equation}
\begin{split}
\left|J_{CP}\right|&=\frac{1}{6\sqrt{3}}\left|\cos2\theta\right|,\\ |I_1|&=\frac{2}{9}\left|\left(\cos\theta\cos\varphi_1-\sin\theta\cos\varphi_2\right) \left(\cos\theta\sin\varphi_1-\sin\theta\sin\varphi_2\right)\right|\,,\\
|I_2|&=\frac{1}{9}\left|\cos2\theta\sin\left(2\varphi_1-2\varphi_2\right)\right|\,.
\end{split}
\end{equation}
The $J_{CP}$ invariant can be expressed in terms of $\theta_{13}$ and $\theta_{23}$:
\begin{equation}
\left|J_{CP}\right|
=\frac{1}{6}\left |\cos^4\theta_{13}\sin^22\theta_{23}
- \cos^22\theta_{13}\right|^{\frac{1}{2}}\,.
\end{equation}
Furthermore, the mixing angles and Dirac CPV phase fulfill the following sum rule
\begin{equation}
\label{eq:angle_phase_caseII}\cos\delta_{CP}=\frac{\cos2\theta_{13}\cot2\theta_{23}}{\sqrt{3\cos^2\theta_{13}-1}\,\sin\theta_{13}}\,.
\end{equation}
Therefore, the value of $\delta_{CP}$ is strongly correlated with the deviation of $\theta_{23}$ from maximal mixing. Using the $3\sigma$ ranges of $\sin^2\theta_{13}$ and $\sin^2\theta_{23}$ from the global fit~\cite{Esteban:2024eli}, we find that $\cos\delta_{CP}$ can span nearly the entire interval $[-0.8330, 0.6333]$, indicating that the current sum rule provides only a weak constraint on the Dirac CP phase. Nevertheless, with future neutrino experiments expected to significantly reduce the uncertainty in the atmospheric mixing angle $\theta_{23}$, the sum rule given in Eq.~\eqref{eq:angle_phase_caseII} could place a stringent bound on the allowed values of $\delta_{CP}$.

The flavour symmetry group $G_{f} = S_4$ yields the mixing patterns $U^{II}_{1}$ and $U^{II}_{2}$ for the parameter values $(\varphi_1, \varphi_2) = (0, \pi)$. In this configuration, the atmospheric mixing angle $\theta_{23}$ and the Dirac CPV phase $\delta_{CP}$ are both predicted to be maximal. Furthermore, the Majorana phases are constrained to be either $0$ or $\pi$. Notably, $U^{II}_{1}$ and $U^{II}_{2}$ represent the same physical mixing pattern in this case, as they are connected through a redefinition of the parameter $\theta$ and the matrix $Q_{\nu}$:
\begin{equation}
U^{II}_{2}(\theta,\varphi_1=\pi, \varphi_2=0)=U^{II}_{1}(\frac{\pi}{2}-\theta,\varphi_1=\pi, \varphi_2=0)\text{diag}(1,1,-1)\,.
\end{equation}

Note that the mixing matrices in Eq.~\eqref{eq:pmnsII} can also be obtained from breaking the series symmetry $\Delta(3n^{2}) \rtimes H_{CP}$~\cite{Ding:2015rwa}. Here, gCP transformations can be consistently defined within $\Delta(3n^{2})$, provided that nontrivial singlet representations are absent from the model. Hence, the minimal group that yields the mixing patterns $U^{II}_{1}$ and $U^{II}_{2}$ is $A_{4}$, whose predictions coincide with those of $S_{4}$.

Finally, from the sum rule given in Eq.~\eqref{eq:TM2_sum_rule}, using the $3\sigma$ range of $\sin^2\theta_{13}$ from~\cite{Esteban:2024eli}, we obtain $0.3402 \leq \sin^2\theta_{12} \leq 0.3415$. The JUNO experiment has determined the $3\sigma$ range of the solar mixing angle $\sin^{2}\theta_{12}$ to be $[0.2831, 0.3353]$~\cite{JUNO:2025gmd}. Thus, the solar mixing angle is predicted to be larger than the $3\sigma$ upper limit of JUNO. Therefore the mixing patterns $U^{II}_{1}$ and $U^{II}_{2}$, as well as the TM2 mixing matrix, are excluded by the JUNO data~\cite{Zhang:2025jnn,Jiang:2025hvq,He:2025idv,Petcov:2025aci}, although they are marginally compatible with neutrino global analysis \texttt{NuFIT}-v6.0~\cite{Esteban:2024eli}. In a concrete model in which either the mixing pattern $U^{II}_{1}$ or $U^{II}_{2}$ is realized, the experimental data could possibly be accommodated if the solar mixing angle undergoes appropriate corrections induced by higher dimensional operators etc~\cite{Ding:2024ozt}.

\item[~~(\uppercase\expandafter{\romannumeral3})]
In this case, the lepton mixing matrix is given by
\begin{equation}\label{eq:mixing_III}
U^{III}=\left(
\begin{array}{ccc}
-i\sqrt{\frac{\phi_g}{\sqrt{5}}} & \sqrt{\frac{1}{\sqrt{5}\phi_g}} & 0 \\
i\sqrt{\frac{1}{2\sqrt{5}\phi_g}} & \sqrt{\frac{\phi_g}{2\sqrt{5}}} & -\frac{1}{\sqrt{2}} \\
i\sqrt{\frac{1}{2\sqrt{5}\phi_g}} & \sqrt{\frac{\phi_g}{2\sqrt{5}}} & \frac{1}{\sqrt{2}}
\end{array}
\right)R_{13}(\theta)Q_{\nu}\,,
\end{equation}
where $\phi_g=(\sqrt{5}+1)/2$ is the golden ratio. This the mixing matrix can arise from the breaking of the flavour symmetry group $A_5\rtimes H_{CP}$. In the chosen basis, the consistent gCP transformation matrices coincide structurally with the representation matrices of the flavour symmetry. Then the mixing matrix $U^{III}$ results from the following symmetry-breaking structure~\cite{Li:2015jxa,DiIura:2015kfa,Ballett:2015wia}:
\begin{equation}
A_{5}\rtimes H_{CP}~~:\quad G_{\ell}=Z^{T}_{5}, \quad  G_{\nu}=Z^{T^3ST^2ST^3}_{2}, \quad  X_{\nu \bm{r}}=\rho_{\bm{r}}(S)\,,
\end{equation}
This scenario has previously been identified in studies of $A_5$ flavour symmetry combined with gCP invariance~\cite{Li:2015jxa,DiIura:2015kfa,Ballett:2015wia}, and our results are in full agreement with those earlier findings.

For the mixing matrix $U^{III}$, the mixing angles read
\begin{equation}\label{eq:thij4b}
\sin^2\theta_{13}=\frac{\phi_g}{\sqrt{5}} \sin ^2\theta,\quad
\sin^2\theta_{12}=\frac{4-2\phi_g}{5-2\phi_g+\cos 2\theta}\,,\quad \sin^2\theta_{23}=\frac{1}{2}\,.
\end{equation}
The solar and reactor mixing angles have the correlation
\begin{equation}\label{eq:th12th13_III}
\sin^{2}\theta_{12}\cos^2\theta_{13}=(3-\phi_g)/5\,.
\end{equation}
Using the $3\sigma$ range of the reactor mixing angle $0.02030\leq\sin^{2}\theta_{13}\leq0.02388$~\cite{Esteban:2024eli}, we get
\begin{equation}
0.2821\leq\sin^{2}\theta_{12}\leq0.2832,
\end{equation}
where the theoretical upper limit of $\sin^3\theta_{12}=0.2832$  is only slightly larger than the lower limit of 0.2831 provided by the JUNO experiment. Therefore, this mixing pattern will soon be tested by the JUNO experiment. For the CPV phases, we find $\delta_{CP}$ is exactly maximal while both Majorana phases $\alpha_{21}$ and $\alpha_{31}$ are trivial with
\begin{equation}
|J_{CP}|=\frac{1}{4}\sqrt{\frac{\phi_g}{5\sqrt{5}}}\;|\sin 2\theta|
= \frac{1}{2\sqrt{5}}\left( \frac{\sqrt{5}-1}{\sqrt{5}+1}
\right)^{\frac{1}{2}}\sin\theta_{13}\cot\theta_{12}\,,
~~ \cos\delta_{CP}=0\,,~~ I_1=I_2=0\,.
\end{equation}
where we have used Eq. (\ref{eq:thij4b}) to get the expressions for $|J_{CP}|$ in terms of $\theta_{13}$ and $\theta_{12}$.

\item[~~(\uppercase\expandafter{\romannumeral4}) ]

The lepton mixing matrices for this case are given by
\begin{equation}\label{eq:mixing_IV}
U^{IV}_{1}=\frac{1}{2}\left(
\begin{array}{ccc}
 \phi_g  & 1 & \phi_g -1 \\
 \phi_g -1 & -\phi_g  & 1 \\
 1 & 1-\phi_g  & -\phi_g
\end{array}
\right)R_{23}(\theta)Q_{\nu}\,, \qquad
U^{IV}_{2}=P_{132}U^{IV}_{1}\,.
\end{equation}
These two mixing matrices are related by an exchange of their second and third rows. Similar to case III, this mixing pattern also originates from the flavour symmetry group $A_5$ in combination with gCP and they come from the following breaking pattern:
\begin{equation}
A_{5}\rtimes H_{CP}~~:\quad G_{\ell}=K^{(ST^2ST^3S,TST^4)}_{4}, \quad  G_{\nu}=Z^{S}_{2},  \quad X_{\nu \bm{r}}=\rho_{\bm{r}}(T^3ST^2ST^3) \,.
\end{equation}
Earlier studies of this mixing pattern in the context of $A_5$ flavour symmetry and CP can be found in Refs.~\cite{Li:2015jxa,DiIura:2015kfa,Ballett:2015wia}. We can extract the following results for the mixing angles
\begin{equation}
\begin{split}
&\sin^2\theta_{13}=\frac{(\cos\theta+\phi_g\sin\theta)^2}{4\phi_g^2},\qquad \sin^2\theta_{12}=\frac{(\phi_g\cos\theta-\sin\theta)^2}{4\phi_g^2-(\cos\theta+\phi_g\sin\theta)^2}\,,\\
&\sin^2\theta_{23}=\frac{\phi^2_{g}(\cos\theta-\phi_g\sin\theta)^2}{4\phi_g^2-(\cos\theta+\phi_g\sin\theta)^2}~~~\text{for}~~~U^{IV}_{1},\\
&\sin^2\theta_{23}=\frac{(\sin\theta+\phi^2_{g}\cos\theta)^2}{4\phi_g^2-(\cos\theta+\phi_g\sin\theta)^2}~~~\text{for}~~~U^{IV}_{2}\,.
\end{split}
\end{equation}
After some tedious calculations, we find the following relations between the mixing angles~\footnote{The sum rules relating $\sin^2\theta_{12}$ and $\sin^2\theta_{13}$
in Eqs. (\ref{eq:th12th13_III}) and (\ref{eq:mixing_angles_correlations}) were derived also in~\cite{Girardi:2015rwa} in models with $A_5$ symmetry but without gCP symmetry, which were analysed phenomenologically in~\cite{Petcov:2018snn,Petcov:2025aci}.
These models lead to different predictions for $\delta_{CP}$ and $\sin^2\theta_{23}$, thus are experimentally
distinguishable from the models considered in the present work.}:
{\small\begin{eqnarray}
\nonumber && \hskip-0.35in 4\cos^{2}\theta_{12}\cos^{2}\theta_{13}=1+\phi_g\,, \\
\nonumber &&\hskip-0.35in  \sin^{2}\theta_{23}=1-\left[\frac{\sqrt{5}}{\phi_g}+(1+2\phi_g)\tan^{2}\theta_{13}+2\phi_g\tan\theta_{13}\sqrt{\sqrt{5}\phi_g-(2+3\phi_g)\tan^{2}\theta_{13}}\right]/5, ~\text{for} ~ U^{IV}_{1}\\
\label{eq:mixing_angles_correlations}&&\hskip-0.35in  \sin^{2}\theta_{23}=\left[\frac{\sqrt{5}}{\phi_g}+(1+2\phi_g)\tan^{2}\theta_{13}+2\phi_g\tan\theta_{13}\sqrt{\sqrt{5}\phi_g-(2+3\phi_g)\tan^{2}\theta_{13}}\right]/5,  ~\text{for} ~ U^{IV}_{2}
\end{eqnarray}}
Using the $3\sigma$ range of the reactor mixing angle $\sin^{2}\theta_{13}$, we have
\begin{eqnarray}
\nonumber && 0.3295\leq\sin^{2}\theta_{12}\leq0.3319,\\
\nonumber && 0.5148\leq\sin^{2}\theta_{23}\leq0.5324\quad  \text{for} \quad  U^{IV}_{1}\,, \\
&& 0.4676\leq\sin^{2}\theta_{23}\leq0.4852\quad  \text{for} \quad U^{IV}_{2}\,,
\end{eqnarray}
for NO.
The CP invariants $J_{CP}$, $I_1$ and $I_2$ are found to vanish exactly so that both Dirac and Majorana CP phases take CP conserving values 0 and $\pi$.

\item[~~(\uppercase\expandafter{\romannumeral5}) ]

The lepton mixing matrix is
\begin{equation}\label{eq:mixing_V}
U^{V}=\frac{1}{2\sqrt{3}}\left(
\begin{array}{ccc}
\sqrt{3}-1 &~ 2e^{-i \varphi } &~ -(\sqrt{3}+1)e^{\frac{3\pi i}{4}} \\
-\sqrt{3}-1 &~ 2e^{-i \varphi } &~ (\sqrt{3}-1)e^{\frac{3\pi i}{4}} \\
2 &~ 2e^{-i \varphi } &~ 2e^{\frac{3\pi i}{4}}
\end{array}
\right)R_{13}(\theta)Q_{\nu}\,,
\end{equation}
where $\varphi=\arctan (2-\sqrt{7})$. This form of the mixing matrix arises from the breaking of the flavour symmetry group $\Sigma(168) \cong PSL(2,7) \cong \Gamma_{7}$ and gCP symmetry. In our basis, the gCP transformation consistent with $\Sigma(168)$ flavour symmetry takes the same form as the flavour symmetry transformation itself. The mixing matrix $U^{V}$ results from the symmetry breaking pattern
\begin{equation}
\Sigma(168)\rtimes H_{CP}~~:\quad G_{\ell}=Z^{ST^4ST^2S}_{3}, \qquad  G_{\nu}=Z^{S}_{2}, \qquad  X_{\nu \bm{r}}=\rho_{\bm{r}}(1)\,.
\end{equation}

In this scenario, one column of the PMNS matrix is fixed as $(1, 1, 1)^{T}/\sqrt{3}$, which must correspond to the second column to match experimental lepton mixing data. Among the six mixing matrices generated by row permutations of $U^{V}$, four are compatible with current experimental data:
\begin{equation}\label{eq:Sigma168_four_permutations}
U^{V}_{1}=U^{V}\,,\qquad U^{V}_{2}=P_{132}U^{V}\,,\qquad
U^{V}_{3}=P_{213}U^{V}\,,\qquad U^{V}_{4}=P_{231}U^{V}\,.
\end{equation}
Notably, $U^{V}_{2}$ and $U^{V}_{4}$ are related to $U^{V}_{1}$ and $U^{V}_{3}$ by exchange of their second and third rows. Moreover, the four matrices satisfy the following symmetry relations
\begin{eqnarray}
\nonumber U^{V}_{3}(\theta)&=& e^{\frac{3\pi i}{4}}\left[U^{V}_{1}(\frac{\pi}{2}-\theta)\right]^{*}\text{diag}(-1,-ie^{\frac{i}{2}\arctan{(\sqrt{7}/3)}},1)\,, \\
U^{V}_{4}(\theta)&=& e^{\frac{3\pi i}{4}}\left[U^{V}_{2}(\frac{\pi}{2}-\theta)\right]^{*}\text{diag}(-1,-ie^{\frac{i}{2}\arctan{(\sqrt{7}/3)}},1)\,.
\end{eqnarray}
It implies that the mixing matrices  $U^{V}_{3}$ ($U^{V}_{4}$)  from left-handed leptons assigned to $\bm{\bar{3}}$ and $U^{V}_{1}$ ($U^{V}_{2}$) from left-handed leptons assigned to $\bm{3}$  describe identical physical mixing patterns, up to the Majorana CP phase $\alpha_{21}$. This equivalence follows from the fact that $U^{V}_{3}$ ($U^{V}_{4}$) can be obtained from $\left(U^{V}_{1}\right)^*$ ($\left(U^{V}_{2}\right)^*$) by reparameterizing $\theta$ up to the diagonal phase $Q_{\nu}$ and an overall phase. Consequently, the predictions for lepton mixing parameters from $U^{V}_{3}$ and $U^{V}_{4}$ follow directly from those of $U^{V}_{1}$ and $U^{V}_{2}$, and are therefore not presented separately.

For $U^{V}_{1}$ and $U^{V}_{2}$, the mixing angles and all three $CP$ rephasing invariants are given by
\begin{eqnarray}
\nonumber&&\sin^2\theta_{13}=\frac{1}{12}\left(4+2\sqrt{3}\cos2\theta+\sqrt{2}\sin2\theta\right),\\
\nonumber&&\sin^2\theta_{12}=\frac{4}{8-2\sqrt{3}\cos2\theta-\sqrt{2}\sin2\theta}\,,\\
\nonumber&&\sin^2\theta_{23}=\frac{4-2\sqrt{3}\cos2\theta+\sqrt{2}\sin2\theta}{8-2\sqrt{3}\cos2\theta-\sqrt{2}\sin2\theta}~~~\text{for}~~~U^{V}_{1}\,,\\
\nonumber&&\sin^2\theta_{23}=\frac{4-2\sqrt{2}\sin2\theta}{8-2\sqrt{3}\cos2\theta-\sqrt{2}\sin2\theta}~~~\text{for}~~~U^{V}_{2}\,,\\
\nonumber&& |J_{CP}|=\frac{1}{6\sqrt{6}}\left|\sin2\theta\right|\,,\quad |I_2|=\frac{1}{36}\left|\cos2\theta-\sqrt{6}\sin2\theta\right|\,,\\
\label{eq:V_mix_pars}&&|I_1|=\frac{1}{72}\left|2\sqrt{7}-\sqrt{3}+\left(2-\sqrt{21}\right)\cos2\theta-\sqrt{14}\sin2\theta\right|\,.
\end{eqnarray}
Then we can derive the following sum rules among the mixing angles
\begin{equation}
\label{eq:sum_rule_case_V}\begin{aligned}
  \sin^2\theta_{12}\cos^2\theta_{13}&=\frac{1}{3}\,,\\
  \sin^2\theta_{23}\cos^2\theta _{13}&=\frac{1}{42} \left(9+15\cos2\theta_{13}\pm 2\sqrt{3}\sqrt{12 \cos 2 \theta _{13}-9 \cos 4 \theta _{13}-4}\right)~~\text{for $U^{V}_{1}$}~~\,,\\
  \sin^2\theta_{23}\cos^2\theta _{13}& =\frac{1}{21}\left(6+3\cos2\theta_{13}\pm\sqrt{3}\sqrt{12\cos2\theta_{13}-9\cos4\theta_{13}-4}\right)~~\text{for $U^{V}_{2}$}\,.
\end{aligned}
\end{equation}
In the  $U^{V}_{1}$ and $U^{V}_{2}$ cases, as it is not difficult to show,  the CP invariant $|J_{CP}|$ can be expressed in terms of $\theta_{23}$  and $\theta_{13}$:
\begin{eqnarray}\label{eq:JCP6ab}
\nonumber |J_{CP}|&=&\frac{1}{6\sqrt{3}}\left|1 - 3\cos^2\theta_{23}\cos^2\theta_{13}\right|,~~~{\rm for}~~~U^{V}_{1}\,, \\
|J_{CP}|&=&\frac{1}{6\sqrt{3}}\left|1 - 3\sin^2\theta_{23}\cos^2\theta_{13}\right|,~~~{\rm for}~~~U^{V}_{2}\,.
\end{eqnarray}
From the expression of reactor angle in Eq.~\eqref{eq:V_mix_pars}, we see that the minimum value of the reactor mixing angle fulfill $\sin^2\theta_{13}\simeq 0.02153$ when $\theta\simeq 0.5617\pi$. Requiring this mixing pattern to be compatible with the global analysis \texttt{NuFIT}~\cite{Esteban:2024eli} at $3\sigma$ level, we find that the allowed regions of the three mixing angles for normal ordering as follow, 
\begin{eqnarray}
\nonumber &&0.02153\leq\sin^{2}\theta_{13}\leq0.02388\,,\\
\nonumber&&0.3407\leq\sin^{2}\theta_{12}\leq0.3415,\\
\nonumber && 0.5405\leq\sin^{2}\theta_{23}\leq0.5850,~~\text{for} ~ U^{V}_{1}\,, \\
 \label{eq:estimate_case_V}&& 0.4350\leq\sin^{2}\theta_{23}\leq0.4595~~\text{for} ~ U^{V}_{2}\,,
\end{eqnarray}
Obviously the atmospheric mixing angle $\theta_{23}$ is non-maximal in this case. What is more important, the specific value of the solar mixing
angle $\theta_{12}$ predicted by this mixing model has been ruled out by the first results from the JUNO experiment~\cite{JUNO:2025gmd}, although it is marginally compatible with \texttt{NuFIT}~\cite{Esteban:2024eli}. However, agreement with the experimental data could possibly be achieved in a specific model producing this mixing pattern at leading order, if $\theta_{12}$ acquires moderate corrections. As regards the leptonic CP violation phases,  the sum rule of Eq.~\eqref{eq:angle_phase_caseII} among the mixing angles and Dirac CP phase is satisfied for all the four PMNS matrices~\footnote{The sum rules for $\sin^2\theta_{12}$ and $\cos\delta_{CP}$, Eqs.~\eqref{eq:sum_rule_case_V} and \eqref{eq:angle_phase_caseII},
are present in models based on $A_4$, $S_4$ and $A_5$ symmetries but without gCP symmetry as well~\cite{Girardi:2015rwa}. However, they lead to different predictions for
$\sin^2\theta_{23}$ and thus for $\delta_{CP}$~\cite{Petcov:2018snn,Petcov:2025aci}, and therefore are experimentally distinguishable from the models considered by us.} $U^{V}_{1}$, $U^{V}_{2}$, $U^{V}_{3}$ and $U^{V}_{4}$.

\end{description}

\section{Lepton mixing with two free parameters \label{sec:twopara_lepton_mixing}}

In this section, we shall discuss the breaking patterns in which charged lepton and neutrino sectors preserve $\mathcal{G}_{\ell}=Z^{g_{\ell}}_{2}\times H^{\ell}_{CP}$ and $\mathcal{G}_{\nu}=Z^{g_{\nu}}_{2}\times H^{\nu}_{CP}$, respectively. The master formula of the lepton mixing matrix is given by Eq.~\eqref{eq:gen_PMNS}. We see that the lepton mixing angles and CP phases in this scenario depend on two real parameters $\theta_{\ell}$ and $\theta_{\nu}$, with only one element of the PMNS matrix being fixed by symmetry. This breaking pattern has been studied for the flavor symmetry groups $S_4$~\cite{Lu:2016jit}, $\Delta(6n^2)$~\cite{Li:2017abz} and the Dihedral groups $D_n$~\cite{Lu:2019gqp} in combination with gCP. Similar to section~\ref{sec:onepara_lepton_mixing}, we are also concerned with the flavour and CP symmetries $A_{5}\rtimes H_{CP}$, $\Sigma(168)\rtimes H_{CP}$ and $\Delta(6n^{2})\rtimes H_{CP}$. All independent residual symmetry of the structure $Z_2\times CP$ would be analyzed for each symmetry group, and the corresponding predictions for lepton flavour mixing will be studied in the following.

\subsection{\label{sec:two_pars_from_A5}The breaking patterns of $A_{5}\rtimes H_{CP}$}

\begin{table}[t!]
\centering
        \renewcommand{\tabcolsep}{0.4mm}
\renewcommand{\arraystretch}{1.2}
\begin{tabular}{|c|c|c|c|}
\hline \hline
 \multicolumn{3}{|c|}{The residual $Z_{2}\times CP$ from $A_{5}\rtimes H_{CP}$ and the corresponding $\Sigma_{\ell}(\Sigma_{\nu})$}  \\ \hline
 $Z^{g_{\ell}}_{2}$ ($Z^{g_\nu}_{2}$) &  $X_{\ell\bm{r}}$ ($X_{\nu\bm{r}}$)  &    $\Sigma_{\ell}$ ($\Sigma_{\nu}$) \\ \hline

\multirow{2}{*}{$Z^{S}_{2}$} & $\rho_{\bm{r}}(1)$  &  $U_{GR}$    \\ \cline{2-3}
  & $\rho_{\bm{r}}(T^{3}ST^{2}ST^{3})$
  &  $U_{GR}\,\text{diag}(i,1,i)$ \\ \hline

\multirow{2}{*}{$Z^{TST^{4}}_{2}$} & $\rho_{\bm{r}}(T^{2})$  &  $\text{diag}(1,\omega_{5},\omega_{5}^{4})\,U_{GR}$    \\ \cline{2-3}
  & $\rho_{\bm{r}}(T^{3}ST^{2}ST^{3}S)$
  &  $\text{diag}(1,\omega_{5},\omega_{5}^{4})\,U_{GR}\,\text{diag}(-i,-i,1)$ \\ \hline

\multirow{2}{*}{$Z^{T^{2}ST^{3}}_{2}$} & $\rho_{\bm{r}}(T^{4})$  &  $\text{diag}(1,\omega_{5}^{2},\omega_{5}^{3})\,U_{GR}$    \\ \cline{2-3}
  & $\rho_{\bm{r}}(ST^{2}S)$
  &  $\text{diag}(1,\omega_{5}^{2},\omega_{5}^{3})\,U_{GR}\,\text{diag}(i,1,i)$ \\ \hline

\multirow{2}{*}{$Z^{T^{4}(ST^{2})^{2}}_{2}$} & $\rho_{\bm{r}}(T^{2})$  &  $\text{diag}(1,\omega_{5},\omega_{5}^{4})\,U_{GR}\,P_{213}$    \\ \cline{2-3}
  & $\rho_{\bm{r}}(TST)$
  &  $\text{diag}(1,\omega_{5},\omega_{5}^{4})\,U_{GR}\,\text{diag}(1,-i,i)\,P_{213}$ \\ \hline

\multirow{2}{*}{$Z^{T^{3}ST^{2}ST^{3}}_{2}$} & $\rho_{\bm{r}}(1)$  &  $U_{GR}\,P_{213}$    \\ \cline{2-3}
  & $\rho_{\bm{r}}(S)$
  &  $U_{GR}\,\text{diag}(1,i,i)\,P_{213}$ \\ \hline

\multirow{2}{*}{$Z^{T^{3}ST^{2}ST^{3}S}_{2}$} & $\rho_{\bm{r}}(1)$  &  $\frac{1}{\sqrt{2}}\left(
\begin{array}{ccc}
 0 & 0 & \sqrt{2} \\
 -1 & 1 & 0 \\
 1 & 1 & 0 \\
\end{array}
\right)$    \\ \cline{2-3}
  & $\rho_{\bm{r}}(S)$
  &  $U_{GR}\,\text{diag}(1,i,i)\,P_{321}$ \\ \hline \hline

\multicolumn{3}{|c|}{The independent breaking patterns from $A_{5}\rtimes H_{CP}$ and the corresponding $\Sigma$ matrix}  \\ \hline

 ($Z^{g_{\ell}}_{2}$, $Z^{g_\nu}_{2}$) &  ($X_{\ell\bm{r}}$, $X_{\nu\bm{r}}$)  &   $\Sigma=\Sigma^{\dagger}_{\ell}\Sigma_{\nu}$ \\ \hline

\multirow{2}{*}{($Z^{S}_{2}$, $Z^{TST^{4}}_{2}$)} & ($\rho_{\bm{r}}(1)$, $\rho_{\bm{r}}(T^{3}ST^{2}ST^{3}S)$)  &  $\Sigma^{VI}_{1}=\text{diag}(i,-i,1)\,U_{RC}\,\text{diag}(-1, 1, -1)$   \\ \cline{2-3}
 & ($\rho_{\bm{r}}(T^{3}ST^{2}ST^{3})$, $\rho_{\bm{r}}(T^{3}ST^{2}ST^{3}S)$)  &  $\Sigma^{VI}_{2}=\text{diag}(i,1,1)\,U_{RC}\,\text{diag}(1,-1,1)$   \\ \hline

\multirow{2}{*}{($Z^{S}_{2}$, $Z^{T^{2}ST^{3}}_{2}$)} & ($\rho_{\bm{r}}(1)$, $\rho_{\bm{r}}(ST^{2}S)$)  &  $\Sigma^{VI}_{3}=\text{diag}(1,-1,-i)\,P_{312}U_{RC}\,\text{diag}(i,1,-1)$   \\ \cline{2-3}
 & ($\rho_{\bm{r}}(T^{3}ST^{2}ST^{3})$, $\rho_{\bm{r}}(ST^{2}S)$)  &  $\Sigma^{VI}_{4}=\text{diag}(i,1,1)\,P_{312}U_{RC}\,\text{diag}(i,1,-1)$   \\ \hline

 \multirow{2}{*}{($Z^{S}_{2}$, $Z^{T^{4}(ST^{2})^{2}}_{2}$)} & ($\rho_{\bm{r}}(1)$, $\rho_{\bm{r}}(T^{3}ST^{2}ST^{3}S)$)  &  $\Sigma^{VI}_{5}=\text{diag}(1,-1,-i)\,U_{RC}P_{213}\,\text{diag}(i,1,1)$   \\ \cline{2-3}
 & ($\rho_{\bm{r}}(T^{3}ST^{2}ST^{3})$, $\rho_{\bm{r}}(T^{3}ST^{2}ST^{3}S)$)  &  $\Sigma^{VI}_{6}=\text{diag}(i,1,1)\,U_{RC}P_{213}\,\text{diag}(i,1,1)$   \\ \hline \hline

\end{tabular}
\caption{\label{tab:sum_A5_Sigma}
The possible residual subgroups of the structure $Z_2\times CP$ and the corresponding Takagi factorization matrices $\Sigma_{\ell}(\Sigma_{\nu})$, and possible six viable breaking patterns and the corresponding $\Sigma=\Sigma^{\dagger}_{\ell}\Sigma_{\nu}$ matrices from breaking the symmetry $A_{5}\rtimes H_{CP}$. The matrices $U_{GR}$ is the golden ratio mixing matrix in Eq.~\eqref{eq:def_UGR}, the matrix $U_{RC}$ is defined in Eq.~\eqref{eq:def_URC}, and  the permutation matrices $P_{213}$ and $P_{321}$ are given in Eq.~\eqref{eq:per_matrix}. In each $\Sigma$ matrix, the (11) entry is fully constrained by combinations of residual symmetries of the charged lepton and neutrino sectors.}
\end{table}

We now classify the distinct symmetry breaking patterns that arise from breaking the flavour group $A_{5}\rtimes H_{CP}$. If a pair of residual flavour subgroups $\{Z^{g^{\prime}_{\ell}}_{2},Z^{g^{\prime}_{\nu}}_{2}\}$  is related to $\{Z^{g_{\ell}}_{2},Z^{g_{\nu}}_{2}\}$ by conjugation under some element $h\in A_{5}$, i.e.
\begin{equation}
Z^{g^{\prime}_{\ell}}_{2}=hZ^{g_{\ell}}_{2}h^{-1}, \qquad Z^{g^{\prime}_{\nu}}_{2}=hZ^{g_{\nu}}_{2}h^{-1}\,,
\end{equation}
then these two breaking patterns will yield identical phenomenological predictions for mixing
parameters. As all fifteen $Z_{2}$ subgroups are conjugate to each other.  As a consequence, it is sufficient to only consider the residual symmetries  $Z^{S}_{2}\times H^{\ell}_{CP}$ in the charged lepton sector. A systematic analysis of the breaking of $A_{5}\rtimes H_{CP}$ shows that the neutrino residual symmetry $Z^{g_{\nu}}_{2}\times H^{\nu}_{CP}$ can be chosen from the following five subgroups of $A_{5}\rtimes H_{CP}$:
\begin{equation}\label{eq:A5_nu_Z2subs}
Z_{2}^{TST^4}\times H^{\nu}_{CP}, ~~ Z_{2}^{T^2ST^3}\times H^{\nu}_{CP}, ~~ Z_{2}^{T^4(ST^2)^2}\times H^{\nu}_{CP}, ~~ Z_{2}^{ T^3ST^2T^3}\times H^{\nu}_{CP}, ~~ Z_{2}^{ T^3ST^2T^3S}\times H^{\nu}_{CP}\,.
\end{equation}
Following the procedures presented in section~\ref{sec:framework},  the corresponding Takagi factorization for each residual symmetry can be calculated, and all these results are summarized in table~\ref{tab:sum_A5_Sigma}, where $U_{GR}$ denotes the golden ratio mixing matrix, given by
\begin{equation}\label{eq:def_UGR}
U_{GR}=\left(
\begin{array}{ccc}
-\sqrt{\frac{\phi_g}{\sqrt{5}}} & \sqrt{\frac{1}{\sqrt{5}\phi_g}} & 0\\
\sqrt{\frac{1}{2\sqrt{5}\phi_g}} & \sqrt{\frac{\phi_g}{2\sqrt{5}}} & -\frac{1}{\sqrt{2}} \\
\sqrt{\frac{1}{2\sqrt{5}\phi_g}} & \sqrt{\frac{\phi_g}{2\sqrt{5}}} & \frac{1}{\sqrt{2}}
\end{array}
\right)\,.
\end{equation}
We compute the absolute of the fixed elements of the PMNS matrices for the five independent breaking patterns (see Eq.~\eqref{eq:A5_nu_Z2subs}), yielding the following values, respectively:
\begin{equation}
\frac{\phi_{g}}{2}, \quad \frac{1}{2}, \quad \frac{1}{2\phi_{g}}, \quad 0, \quad 0\,.
\end{equation}
A comparison with current experimental constraints indicates that only the patterns containing $\frac{\phi_{g}}{2}$, $\frac{1}{2}$, and $\frac{1}{2\phi_{g}}$ may be phenomenologically acceptable. Moreover, the positions of these elements within the mixing matrix are strongly constrained: $\frac{\phi_{g}}{2}$ must be the $(11)$ entry, $\frac{1}{2\phi_{g}}$ must occupy either the $(21)$ or $(31)$ position, and $\frac{1}{2}$ can be assigned to the $(21)$, $(22)$, $(31)$ or $(32)$ entry.

Considering all independent residual CP choices yields twelve distinct breaking patterns; a numerical scan shows that only six of them are compatible with present data. These six viable breakings and their mixing matrices are summarized in table~\ref{tab:sum_A5_Sigma}. Here we define the matrix
\begin{equation}\label{eq:def_URC}
U_{RC}=\frac{1}{2} \left(
\begin{array}{ccc}
 \phi_{g} & 1/\phi_{g} & 1 \\
 1/\phi_{g} & 1 & -\phi_{g} \\
 1 & -\phi_{g} & -1/\phi_{g} \\
\end{array}
\right)\,.
\end{equation}
From the six breaking patterns summarized in table~\ref{tab:sum_A5_Sigma}, we obtained the following 11 viable mixing matrices:
\begin{eqnarray}
\nonumber && U^{VI}_{1}=R^{T}_{23}(\theta_{\ell})\Sigma^{VI}_{1}R_{23}(\theta_{\nu}), \quad U^{VI}_{2}=R^{T}_{23}(\theta_{\ell})\Sigma^{VI}_{2}R_{23}(\theta_{\nu}), \quad U^{VI}_{3,1}=P_{213}R^{T}_{23}(\theta_{\ell})\Sigma_{3}R_{23}(\theta_{\nu}),\\
\nonumber &&U^{VI}_{3,2}=P_{231}R^{T}_{23}(\theta_{\ell})\Sigma^{VI}_{3}R_{23}(\theta_{\nu}), \qquad U^{VI}_{4,1}=P_{213}R^{T}_{23}(\theta_{\ell})\Sigma^{VI}_{4}R_{23}(\theta_{\nu}),  \\
\nonumber &&U^{VI}_{4,2}=P_{231}R^{T}_{23}(\theta_{\ell})\Sigma^{VI}_{4}R_{23}(\theta_{\nu}), \qquad  U^{VI}_{4,3}=P_{213}R^{T}_{23}(\theta_{\ell})\Sigma^{VI}_{4}R_{23}(\theta_{\nu})P_{213}, \\
\nonumber && U^{VI}_{4,4}=P_{231}R^{T}_{23}(\theta_{\ell})\Sigma^{VI}_{4}R_{23}(\theta_{\nu})P_{213},  \qquad U^{VI}_{5}=P_{213}R^{T}_{23}(\theta_{\ell})\Sigma^{VI}_{5}R_{23}(\theta_{\nu}), \\
\label{eq:PMNS_caseVI}&& U^{VI}_{6,1}=P_{213}R^{T}_{23}(\theta_{\ell})\Sigma^{VI}_{6}R_{23}(\theta_{\nu}),
\qquad  U^{VI}_{6,2}=P_{231}R^{T}_{23}(\theta_{\ell})\Sigma^{VI}_{6}R_{23}(\theta_{\nu}),
\end{eqnarray}
where we have omitted the diagonal phase matrices $Q_{\ell}$ and $Q_{\nu}$ here. As $Q_{\ell}$ can be absorbed by the charged lepton fields, while $Q_{\nu}$ is a diagonal matrix with entries $\pm1$ or $\pm i$. Its effect is to change the Majorana phases by $\pi$. Observe that the mixing matrices $U^{VI}_{3,2}$, $U^{VI}_{4,2}$, $U^{VI}_{4,4}$ and $U^{VI}_{6,2}$ are derived from $U^{VI}_{3,1}$, $U^{VI}_{4,1}$, $U^{VI}_{4,3}$ and $U^{VI}_{6,1}$ by interchanging the second and third rows of the PMNS matrix, respectively. As established, such an exchange transforms the atmospheric mixing angle via $\theta_{23}\rightarrow\pi/2-\theta_{23}$, shifts the Dirac CP phase as $\delta_{CP}\rightarrow\pi+\delta_{CP}$, and leaves all other mixing parameters unchanged. For each independent mixing pattern, the mixing angles $\sin^2\theta_{13}$, $\sin^2\theta_{12}$, $\sin^2\theta_{23}$ and the CP invariants $J_{CP}$, $I_{1}$, $I_{2}$ are computed in the standard manner.

Note that the diagonal phase matrix $\text{diag}(i,1,1)$ in $\Sigma^{VI}_{2}$, $\Sigma^{VI}_{4}$, and $\Sigma^{VI}_{6}$ commutes with both rotation matrices $R_{23}(\theta_{\ell})$ and $R_{23}(\theta_{\nu})$, and can therefore be absorbed into the phase matrices $Q_{\ell}$ and $Q_{\nu}$. As a result, the mixing matrices $U^{VI}_{2}$, $U^{VI}_{4,i}$ and $U^{VI}_{6,i}$ in Eq.~\eqref{eq:PMNS_caseVI} are real up to the overall phase matrices $Q_{\ell}$ and $Q_{\nu}$, leading to trivial CP phases.

We see that the mixing parameters depend on the continuous parameters $\theta_{\ell}$ and $\theta_{\nu}$. As a consequence, sum rules among the mixing angles and the Dirac CP phase $\delta_{CP}$ can be found as follows~\footnote{The sum rule relating $\cos^{2}\theta_{12}$ and $\cos^2\theta_{13}$
in cases $U^{VI}_{1}$ and $U^{VI}_{2}$ coincides with the sum rule in Eq.~\eqref{eq:mixing_angles_correlations}
since $\phi_{g}^{2}=1+\phi_{g}$. It holds also in an $A_5$ model with no gCP symmetry having three real parameters~\cite{Girardi:2015rwa}, in which $\sin^2\theta_{23}$ and $\delta_{CP}$ are unconstrained \cite{Petcov:2018snn}.}:
\begin{eqnarray}
\nonumber U^{VI}_{1}, U^{VI}_{2}~&:&~~~4\cos^{2}\theta_{12}\cos^2\theta_{13}=\phi_{g}^{2}\,, \\
\nonumber  U^{VI}_{4,1}~&:&~~~1=\frac{1-4\sin^2\theta_{12}\cos^2\theta_{23}-4\sin^2\theta_{13}\cos^2\theta_{12}\sin^2\theta_{23}}
{2\sin2\theta_{12}\sin\theta_{13}\sin2\theta_{23}}\,, \\
\nonumber U^{VI}_{4,2}~&:&~~~1=\frac{1-4\sin^2\theta_{12}\sin^2\theta_{23}-4\sin^2\theta_{13}\cos^2\theta_{12}\cos^2\theta_{23}}
{2\sin2\theta_{12}\sin\theta_{13}\sin2\theta_{23}}\,, \\
\nonumber U^{VI}_{4,3}~&:&~~~1=-\frac{1-4\cos^2\theta_{12}\cos^2\theta_{23}-4\sin^2\theta_{13}\sin^2\theta_{12}\sin^2\theta_{23}}
{2\sin2\theta_{12}\sin\theta_{13}\sin2\theta_{23}}\,, \\
\nonumber  U^{VI}_{4,4}~&:&~~~1=-\frac{1-4\cos^2\theta_{12}\sin^2\theta_{23}-4\sin^2\theta_{13}\sin^2\theta_{12}\cos^2\theta_{23}}
{2\sin2\theta_{12}\sin\theta_{13}\sin2\theta_{23}}\,, \\
\nonumber U^{VI}_{6,1}~&:&~~~1=-\frac{1/\phi_{g}^{2}-4\sin^2\theta_{12}\cos^2\theta_{23}-4\sin^2\theta_{13}\cos^2\theta_{12}\sin^2\theta_{23}}
{2\sin2\theta_{12}\sin\theta_{13}\sin2\theta_{23}}\,, \\
\nonumber  U^{VI}_{6,2}~&:&~~~1=-\frac{1/\phi_{g}^2-4\sin^2\theta_{12}\sin^2\theta_{23}-4\sin^2\theta_{13}\cos^2\theta_{12}\cos^2\theta_{23}}
{2\sin2\theta_{12}\sin\theta_{13}\sin2\theta_{23}}\,, \\
\nonumber U^{VI}_{3,1}~&:&~~~\cos\delta_{CP}=\frac{1-4\sin^2\theta_{12}\cos^2\theta_{23}-4\sin^2\theta_{13}\cos^2\theta_{12}\sin^2\theta_{23}}
{2\sin2\theta_{12}\sin\theta_{13}\sin2\theta_{23}}\,, \\
\nonumber U^{VI}_{3,2}~&:&~~~\cos\delta_{CP}=-\frac{1-4\sin^2\theta_{12}\sin^2\theta_{23}-4\sin^2\theta_{13}\cos^2\theta_{12}\cos^2\theta_{23}}
{2\sin2\theta_{12}\sin\theta_{13}\sin2\theta_{23}}\,, \\
\label{eq:correlation_VI}
U^{VI}_{5}~&:&~~~\cos\delta_{CP}=\frac{1/\phi_{g}^{2}-4\sin^2\theta_{12}\cos^2\theta_{23}-4\sin^2\theta_{13}\cos^2\theta_{12}\sin^2\theta_{23}}
{2\sin2\theta_{12}\sin\theta_{13}\sin2\theta_{23}}\,,
\end{eqnarray}
Using the $3\sigma$ range of $\theta_{13}$ as input, we find $\sin^{2}\theta_{12}\in[0.3295,0.3319]$ for mixing matrices $U^{VI}_{1}$ and $U^{VI}_{2}$. This prediction will be directly tested by the JUNO experiment after six years of operation~\cite{JUNO:2022mxj}. If all the three mixing angles freely vary within their $3\sigma$  intervals, the last three relations lead to $[0.6699,1]$, $[-1,-0.5853]$ and $[-1,-0.4346]$, respectively. 

\subsection{\label{sec:two_pars_from_Sigma168}The breaking patterns of $\Sigma(168)\rtimes H_{CP}$}

Let us systematically analyze the distinct breaking patterns arising from the flavour symmetry group $\Sigma(168)\rtimes H_{CP}$ with $\mathcal{G}_{\ell}=Z^{g_{\ell}}_{2}\times H^{\ell}_{CP}$ and $\mathcal{G}_{\nu}=Z^{g_{\nu}}_{2}\times H^{\nu}_{CP}$. The group $\Sigma(168)$ has 21 $Z_{2}$ shbroups, and all them are conjugate to each other.  Without loss of generality, we consider the residual symmetries  $\mathcal{G}_{\ell}=Z^{S}_{2}\times H^{\ell}_{CP}$ in the charged lepton sector. After performing  a comprehensive analysis of lepton mixing patterns which arise from the breaking of $\Sigma(168)\rtimes H_{CP}$, we find that the residual symmetry $Z^{g_{\nu}}_{2}\times H^{\nu}_{CP}$ of neutrino sector can be any one of the following five subgroups:
\begin{equation}\label{eq:Sigma168_nu_Z2subs}
Z_{2}^{TST^6}\times H^{\nu}_{CP}, ~~ Z_{2}^{T^2ST^5}\times H^{\nu}_{CP}, ~~ Z_{2}^{T^4ST^3}\times H^{\nu}_{CP}, ~~ Z_{2}^{ TST^3ST^2}\times H^{\nu}_{CP}, ~~ Z_{2}^{ TST^5ST^5}\times H^{\nu}_{CP}\,.
\end{equation}
It is easy to check that the absolute of the fixed elements of the PMNS matrices for the five  breaking patterns yielding the following values, respectively:
\begin{equation}
\frac{1}{\sqrt{2}}, \quad \frac{1}{2}, \quad \frac{1}{\sqrt{2}}, \quad 0, \quad 0\,.
\end{equation}
A comparison with current experimental data on lepton mixing parameters shows that only the pattern with fixed elements $\frac{1}{\sqrt{2}}$ and $\frac{1}{2}$ are potentially viable. Furthermore, the allowed positions of these elements within the mixing matrix are highly constrained: $\frac{1}{\sqrt{2}}$ must be the $(23)$ or $(33)$ entry, and $\frac{1}{2}$ can be assigned to the $(21)$, $(22)$, $(31)$ or $(32)$ entry. A systematic scan of the parameter space for all independent breaking patterns and their mixing matrices shows that only six matrices can accommodate the experimental data. These viable candidates arise from two breaking patterns:
\begin{eqnarray}
\nonumber && G_{\ell}=Z^{S}_{2}, \quad  G_{\nu}=Z^{T^{2}ST^{5}}_{2},  \quad X_{l \bm{r}}=\rho_{\bm{r}}(1), \quad  X_{\nu \bm{r}}=\rho_{\bm{r}}(T^4) \,, \\
\label{eq:Sigma168_2VBPs}&& G_{\ell}=Z^{S}_{2}, \quad  G_{\nu}=Z^{T^{2}ST^{5}}_{2},  \quad X_{l \bm{r}}=\rho_{\bm{r}}(T^{2}ST^{5}ST^{2}), \quad  X_{\nu \bm{r}}=\rho_{\bm{r}}(T^{4}ST^{5}ST^{4}) \,.
\end{eqnarray}
Following the procedures presented in section~\ref{sec:framework}, the $\Sigma\equiv\Sigma_{\ell}^{\dagger}\Sigma_{\nu}$ matrices corresponding to the two breaking patterns in Eq.~\eqref{eq:Sigma168_2VBPs}  can be taken to be
\begin{eqnarray}
\nonumber &&\Sigma^{VII}_{1}=\text{diag}(1,e^{-\frac{\pi i}{4}},e^{\frac{\pi i}{4}})\,\Sigma^{VII}_{2}\,\text{diag}(1,e^{\frac{\pi i}{4}},e^{-\frac{\pi i}{4}})\,,\\
&&  \label{eq:sigma_Sigma168}\Sigma^{VII}_{2}=\frac{1}{4}\left(
\begin{array}{ccc}
 -2 &~ \sqrt{2 \left(\sqrt{7}+3\right)} &~ \sqrt{2 \left(3-\sqrt{7}\right)} \\
 \sqrt{2 \left(\sqrt{7}+3\right)} &~ \sqrt{7}-1 &~ \sqrt{2} \\
 \sqrt{2 \left(3-\sqrt{7}\right)} &~ \sqrt{2} &~ -\sqrt{7}-1 \\
\end{array}
\right)\,.
\end{eqnarray}
Then the six viable mixing matrices corresponding those two viable breaking patterns take the following form:
\begin{eqnarray}
\nonumber && U^{VII}_{1,1}=P_{213}R^{T}_{23}(\theta_{\ell})\Sigma^{VII}_{1}R_{23}(\theta_{\nu})P_{213}, \qquad U^{VII}_{1,2}=P_{231}R^{T}_{23}(\theta_{\ell})\Sigma^{VII}_{1}R_{23}(\theta_{\nu})P_{213}\,,  \\
\nonumber && U^{VII}_{2,1}=P_{213}R^{T}_{23}(\theta_{\ell})\Sigma^{VII}_{2}R_{23}(\theta_{\nu}), \qquad U^{VII}_{2,2}=P_{231}R^{T}_{23}(\theta_{\ell})\Sigma^{VII}_{2}R_{23}(\theta_{\nu})\,, \\
\label{eq:PMNS_caseVII} &&U^{VII}_{2,3}=P_{213}R^{T}_{23}(\theta_{\ell})\Sigma^{VII}_{2}R_{23}(\theta_{\nu})P_{213}, \qquad  U^{VII}_{2,4}=P_{231}R^{T}_{23}(\theta_{\ell})\Sigma^{VII}_{2}R_{23}(\theta_{\nu})P_{213}\,.
\end{eqnarray}
Note that the four lepton mixing matrices $U^{VII}_{2,i}$ are real, therefore implying trivial CPV phases. We see that $U^{VII}_{1,2}$, $U^{VII}_{2,2}$ and $U^{VII}_{2,4}$ are related to $U^{VII}_{1,1}$, $U^{VII}_{2,1}$ and $U^{VII}_{2,3}$ through an exchange of the second and third rows, respectively. Consequently, the mixing angles and CP phases predicted by $U^{VII}_{1,2}$, $U^{VII}_{2,2}$ and $U^{VII}_{2,4}$ can be directly inferred from the corresponding results for $U^{VII}_{1,2}$, $U^{VII}_{2,2}$ and $U^{VII}_{2,4}$, respectively. Given that all mixing parameters are functions of the free parameters $\theta_{l, \nu}$, they exhibit strong mutual correlations. In particular, each mixing pattern yields a specific sum rule that connects the mixing angles and the Dirac CP phase \begin{eqnarray}
\nonumber U^{VII}_{1,1}~&:&~~\cos\delta_{CP}=-\frac{1-4\cos^2\theta_{12}\cos^2\theta_{23}-4\sin^2\theta_{13}\sin^2\theta_{12}\sin^2\theta_{23}}
{2\sin2\theta_{12}\sin\theta_{13}\sin2\theta_{23}}\,, \\
\nonumber  U^{VII}_{1,2}~&:&~~\cos\delta_{CP}=\frac{1-4\cos^2\theta_{12}\sin^2\theta_{23}-4\sin^2\theta_{13}\sin^2\theta_{12}\cos^2\theta_{23}}
{2\sin2\theta_{12}\sin\theta_{13}\sin2\theta_{23}}\,, \\
\nonumber U^{VII}_{2,1}~&:&~~1=\frac{1-4\sin^2\theta_{12}\cos^2\theta_{23}-4\sin^2\theta_{13}\cos^2\theta_{12}\sin^2\theta_{23}}
{2\sin2\theta_{12}\sin\theta_{13}\sin2\theta_{23}}\,,\\
\nonumber U^{VII}_{2,2}~&:&~~1=\frac{1-4\sin^2\theta_{12}\sin^2\theta_{23}-4\sin^2\theta_{13}\cos^2\theta_{12}\cos^2\theta_{23}}
{2\sin2\theta_{12}\sin\theta_{13}\sin2\theta_{23}}\,, \\
\nonumber  U^{VII}_{2,3}~&:&~~1=-\frac{1-4\cos^2\theta_{12}\cos^2\theta_{23}-4\sin^2\theta_{13}\sin^2\theta_{12}\sin^2\theta_{23}}
{2\sin2\theta_{12}\sin\theta_{13}\sin2\theta_{23}}\,, \\
\label{eq:sum_rules_caseVII} U^{VII}_{2,4}~&:&~~1=-\frac{1-4\cos^2\theta_{12}\sin^2\theta_{23}-4\sin^2\theta_{13}\sin^2\theta_{12}\cos^2\theta_{23}}
{2\sin2\theta_{12}\sin\theta_{13}\sin2\theta_{23}}\,.
\end{eqnarray}
The sum rules for $\cos\delta_{CP}$ in terms of the mixing angles, as presented above, are also obtained in Ref.~\cite{Penedo:2017vtf}. This agreement arises because these relations are derived by setting the (22), or (32) entries of the PMNS matrix equal to $1/2$, respectively. However, while the lepton mixing matrix in our framework is determined by two free parameters, the model in Ref.~\cite{Penedo:2017vtf} involves three. As a result, the correlations among neutrino mixing parameters and their allowed ranges differ between the two approaches. When the three mixing angles are varied freely within their $3\sigma$ bounds from Ref.~\cite{Esteban:2024eli}, the first two relations yield $\cos\delta_{CP}$ in the intervals  $[0.3829,1]$, and $[-1,-0.5537]$, respectively. Distinct allowed intervals of $\cos\delta_{CP}$ provide a means to discriminate among the two mixing matrices $U^{VII}_{1,1}$ and $U^{VII}_{1,2}$. Should future measurements determine $\theta_{12}$ and $\theta_{23}$ with significantly improved precision, these sum rules could be used to predict the Dirac CP phase $\delta_{CP}$ based on experimentally measured mixing angles. The derived sum rules thus serve as sensitive probes for testing the validity of this class of mixing patterns~\cite{Ballett:2013wya,Petcov:2014laa}.

\subsection{\label{sec:two_pars_from_D6n2}The breaking patterns of $\Delta(6n^{2})\rtimes H_{CP}$}

\begin{table}[t!]
\centering
\begin{tabular}{|c|c|c|c|}
\hline \hline
\multicolumn{3}{|c|}{The residual $Z_{2}\times CP$ from $\Delta(6n^{2})\rtimes H_{CP}$ and the corresponding $\Sigma_{\ell}(\Sigma_{\nu})$}  \\ \hline
 $Z^{g_{\ell}}_{2}$ ($Z^{g_\nu}_{2}$) &  $X_{\ell\bm{r}}$ ($X_{\nu\bm{r}}$)  &    $\Sigma_{\ell}$ ($\Sigma_{\nu}$) \\

  &   &      \\ [-0.16in]\hline
  &   &     \\[-0.15in]
$Z^{bc^{x}d^{x}}_{2}$  & $\rho_{\bm{r}}(c^{\alpha}d^{-2x-\alpha})$
  &  $\frac{1}{\sqrt{2}}\left(
\begin{array}{ccc}
 -e^{\frac{i \pi  \alpha }{n}} &~ e^{\frac{i \pi  \alpha }{n}} ~& 0 \\
 0 &~ 0 ~& -\sqrt{2} e^{-\frac{2 i \pi  (x+\alpha )}{n}} \\
 e^{\frac{i \pi  (2 x+\alpha )}{n}} &~ e^{\frac{i \pi  (2 x+\alpha )}{n}} ~& 0 \\
\end{array}
\right)$ \\[0.3in] \hline

  &   &     \\[-0.15in]
$Z^{abc^{x}}_{2}$  & $\rho_{\bm{r}}(c^{\alpha}d^{2x+2\alpha})$
  &  $\frac{1}{\sqrt{2}}\left(
\begin{array}{ccc}
 -e^{\frac{i \pi  \alpha }{n}} &~ 0 ~& e^{\frac{i \pi  \alpha }{n}} \\
 e^{\frac{i \pi  (2 x+\alpha )}{n}} &~ 0 ~& e^{\frac{i \pi  (2 x+\alpha )}{n}} \\
 0 &~ -\sqrt{2} e^{-\frac{2 i \pi  (x+\alpha )}{n}} ~& 0 \\
\end{array}\right)$ \\ [0.3in] \hline

&   &     \\[-0.15in]
 $Z^{c^{n/2}}_{2}$  &  $c^{\alpha}d^{\beta}$
  &  $\left(
\begin{array}{ccc}
 0 & 0 & e^{\frac{i \pi  \alpha }{n}} \\
 0 & -e^{-\frac{i \pi  (\alpha -\beta )}{n}} & 0 \\
 e^{-\frac{i \pi  \beta }{n}} & 0 & 0 \\
\end{array}
\right)$ \\[0.30in] \hline\hline

\multicolumn{3}{|c|}{The independent breaking patterns from $\Delta(6n^{2})\rtimes H_{CP}$ and the corresponding $\Sigma$ matrix}  \\ \hline

($Z^{g_{\ell}}_{2}$, $Z^{g_\nu}_{2}$) &  ($X_{\ell\bm{r}}$, $X_{\nu\bm{r}}$)  &   $\Sigma=\Sigma^{\dagger}_{\ell}\Sigma_{\nu}$ \\ \hline

($Z^{bc^{x}d^{x}}_{2}$, $Z^{bc^{y}d^{y}}_{2}$) & ($\rho_{\bm{r}}(c^{\alpha}d^{-2x-\alpha})$, $\rho_{\bm{r}}(c^{\beta}d^{-2y-\beta})$)  &  $\Sigma^{VIII}=\left(
\begin{array}{ccc}
 \cos \varphi_{1} & -i \sin \varphi_{1} & 0 \\
 -i \sin \varphi_{1} & \cos \varphi_{1} & 0 \\
 0 & 0 & e^{i \varphi_{2}} \\
\end{array}
\right)$    \\ \hline

($Z^{bc^{x}d^{x}}_{2}$, $Z^{abc^{y}}_{2}$) & ($\rho_{\bm{r}}(c^{\alpha}d^{-2x-\alpha})$, $\rho_{\bm{r}}(c^{\beta}d^{2y+2\beta})$)  &  $\Sigma^{IX}=\frac{1}{2}\left(
\begin{array}{ccc}
 1 &~ -\sqrt{2} e^{i \varphi_{2}} ~& -1 \\
 -1 &~ -\sqrt{2} e^{i \varphi_{2}} ~& 1 \\
 -\sqrt{2} e^{i \varphi_{1}} &~ 0 ~& -\sqrt{2} e^{i \varphi_{1}} \\
\end{array}
\right)$    \\ \hline

($Z^{bc^{x}d^{x}}_{2}$, $Z^{c^{n/2}}_{2}$) & ($\rho_{\bm{r}}(c^{\alpha}d^{-2x-\alpha})$, $\rho_{\bm{r}}(c^{\beta}d^{\gamma})$)  &  $\Sigma^{X}=\frac{1}{\sqrt{2}}\left(
\begin{array}{ccc}
 e^{i \varphi_{2}} & 0 & -1 \\
 e^{i \varphi_{2}} & 0 & 1 \\
 0 & \sqrt{2} e^{i \varphi_{1}} & 0 \\
\end{array}\right)$    \\ \hline \hline

\end{tabular}
\caption{\label{tab:sum_D6n2_Sigma}
Residual subgroups of the structure $Z_2\times CP$ along with the corresponding Takagi factorization matrices, and the form of $\Sigma=\Sigma^{\dagger}_{\ell}\Sigma_{\nu}$ for the three viable breaking patterns derived from $\Delta(6n^{2})\rtimes H_{CP}$. Parameters $x$, $y$, $\alpha$, $\beta$ and $\gamma$ take integer values from 0 to $n-1$.
The discrete parameters $\varphi_{1}$ and $\varphi_{2}$ that appear in $\Sigma^{VIII}$, $\Sigma^{IX}$, and $\Sigma^{X}$ are determined by the chosen residual symmetries and take the forms $\varphi_{1}=\frac{x-y}{n}\pi$, $\varphi_{2}=\frac{3(x-y+\alpha-\beta)}{n}\pi$ for $\Sigma^{VIII}$, $\varphi_{1}=\frac{3\alpha+2(x+y)}{n}\pi$, $\varphi_{2}=-\frac{3\beta+2(x+y)}{n}\pi$ for $\Sigma^{IX}$, and $\varphi_{1}=\frac{2x+3\alpha-2\beta+\gamma}{n}\pi$, $\varphi_{2}=-\frac{2x+\beta+\gamma}{n}\pi$ for $\Sigma^{X}$.
In each $\Sigma$ matrix, the (11) entry is fully constrained by combinations of residual symmetries of the charged lepton and neutrino sectors. }
\end{table}

Now we shall perform a comprehensive analyze of the lepton mixing patterns arising from the $\Delta(6n^2)$ flavour group and CP symmetries which are broken down to $Z_2\times CP$ in the neutrino and charged lepton sectors. All possible admissible residual subgroups of the structure $Z_2\times CP$ would be considered, and the phenomenological predictions for lepton mixing matrix would be discussed.  Note that $D^{(1)}_{9n, 3n}$ is a subgroup of  $\Delta(6(9n)^2)$ group. We shall not consider that the breaking patterns arise from $D^{(1)}_{9n, 3n}\rtimes H_{CP}$. After performing a comprehensive analysis of lepton mixing patterns derived from the flavour group $\Delta(6n^2)\rtimes H_{CP}$, we find that only three symmetry breaking patterns are found to be phenomenologically viable, with the residual subgroups given by $\mathcal{G}_{\ell} = Z^{bc^{x}d^{x}}_{2} \times H^{\ell}_{CP}$ and $\mathcal{G}_{\nu}$ equal to either $Z^{bc^{y}d^{y}}_{2} \times H^{\nu}_{CP}$, $Z^{abc^{y}}_{2} \times H^{\nu}_{CP}$ and $Z^{c^{n/2}}_{2} \times H^{\nu}_{CP}$. These patterns were fully analyzed in Ref.~\cite{Li:2017abz}. Here, we present updated predictions for the lepton mixing parameters using the latest experimental data.

Following the framework outlined in section~\ref{sec:framework}, we compute the Takagi factorization for the $Z_{2} \times CP$ symmetry in each viable residual symmetry configuration. The results are summarized in table~\ref{tab:sum_D6n2_Sigma}. The corresponding lepton mixing matrix for the three phenomenologically allowed breaking patterns are then derived from the symmetry breaking patterns given in Eq.~\eqref{eq:gen_PMNS}. For each breaking pattern, the matrix $\Sigma = \Sigma_{\ell}^{\dagger} \Sigma_{\nu}$, which is fixed by the residual symmetry, is explicitly listed in table~\ref{tab:sum_D6n2_Sigma}. The fixed element (the $(11)$ entry of each $\Sigma$ matrix) is uniquely determined by the residual flavour symmetry. It is given by the inner product of the eigenvectors corresponding to non-degenerate eigenvalues of the generators $Z^{g_{\ell}}_{2}$ and $Z^{g{\nu}}_{2}$. Then we proceed to study these three cases and their predictions for lepton mixing angles and CPV phases one by one.

\begin{description}[labelindent=-0.8em, leftmargin=0.3em]

\item[~~(\uppercase\expandafter{\romannumeral8})]
 $G_{\ell}=Z^{bc^xd^x}_2$, $X_{\ell\bm{r}}=\rho_{\bm{r}}(c^{\alpha}d^{-2x-\alpha})$, $G_{\nu}=Z^{bc^yd^y}_2$, $X_{\nu\bm{r}}=\rho_{\bm{r}}(c^{\beta}d^{-2y-\beta})$

For this breaking pattern, the $\Sigma^{VIII}$ matrix in table~\ref{tab:sum_D6n2_Sigma} gives rise to the lepton mixing matrix
\begin{equation}\label{eq:PMNS_caseVIII}
U^{VIII}=\left(
\begin{array}{ccc}
 \cos\varphi_{1} &~ s_{\nu}\sin \varphi_{1}  ~& -c_{\nu}\sin \varphi_{1}  \\
 -s_{\ell}\sin \varphi_{1}  &~ c_{\ell} c_{\nu}e^{i \varphi_{2}} +s_{\ell} s_{\nu}\cos \varphi_{1}  ~& c_{\ell} s_{\nu}e^{i \varphi_{2}} -c_{\nu} s_{\ell}\cos \varphi_{1}  \\
 c_{\ell}\sin \varphi_{1}  &~ c_{\nu} s_{\ell}e^{i \varphi_{2}} -c_{\ell} s_{\nu}\cos \varphi_{1}  ~& s_{\ell} s_{\nu}e^{i \varphi_{2}}+c_{\ell} c_{\nu}\cos \varphi_{1}   \\
\end{array}
\right)\,,
\end{equation}
up to row and column permutations, and with phase matrices $Q_{\ell}$ and $Q_{\nu}$ omitted.  Here the parameters are defined as
\begin{equation}
s_{\ell}\equiv\sin\theta_{\ell}, \quad c_{\ell}\equiv\cos\theta_{\ell}, \quad s_{\nu}\equiv\sin\theta_{\nu}, \quad  c_{\nu}\equiv\cos\theta_{\nu}\,.
\end{equation}
The independent parameters $\varphi_1$ and $\varphi_2$ are fixed by the remnant symmetries:
\begin{equation}\label{eq:phi12_VIII}
\varphi_1=\frac{x-y}{n}\pi, \quad  \varphi_2=\frac{3(x-y+\alpha-\beta)}{n}\pi\,.
\end{equation}
They take discrete values:
\begin{eqnarray}
\nonumber&&\varphi_1~(\mathrm{mod}~2\pi)=0, \frac{1}{n}\pi, \frac{2}{n}\pi, \ldots, \frac{2n-1}{n}\pi\,,\\
\nonumber &&\varphi_2~(\mathrm{mod}~2\pi)=0, \frac{3}{n}\pi, \frac{6}{n}\pi, \ldots, \frac{2n-3}{n}\pi,  \quad 3 \mid n\,,\\
\label{eq:para_values_I}&&\varphi_2~(\mathrm{mod}~2\pi)=0, \frac{1}{n}\pi, \frac{2}{n}\pi, \ldots, \frac{2n-1}{n}\pi, \quad 3 \nmid n\,.
\end{eqnarray}
It is easy to check that the mixing matrix in Eq.~\eqref{eq:PMNS_caseVIII} has the following symmetry properties,
\begin{eqnarray}
\nonumber
U^{VIII}(\pi+\varphi_{1},\varphi_{2},\theta_{\ell},\theta_{\nu})&=&\text{diag}(-1,1,-1)U^{VIII}(\varphi_{1},\varphi_{2},\pi-\theta_{\ell},\theta_{\nu})\,,\\
\nonumber
U^{VIII}(\pi-\varphi_{1},\varphi_{2},\theta_{\ell},\theta_{\nu})&=&\text{diag}(-1,1,1)U^{VIII}(\varphi_{1},\varphi_{2},\theta_{\ell},\pi-\theta_{\nu})\text{diag}(1,1,-1)\,,\\
\nonumber U^{VIII}(\varphi_{1}, \pi+\varphi_{2},\theta_{\ell},\theta_{\nu})&=&U^{VIII}(\varphi_{1},\varphi_{2},\theta_{\ell},\pi-\theta_{\nu})\text{diag}(1,-1,1)\,,\\
\label{eq:PMNS_relations_I}
U^{VIII}(\varphi_{1},\pi-\varphi_{2},\theta_{\ell},\theta_{\nu})&=&(U^{VIII})^{*}(\varphi_{1},\varphi_{2},\theta_{\ell},\pi-\theta_{\nu})\text{diag}(1,-1,1)\,.
\end{eqnarray}
The diagonal matrices mentioned above can be incorporated into the redefinition of the lepton fields. In general, it is sufficient to consider the allowed values $\varphi_1$ and $\varphi_2$ within the intervals $0\leq\varphi_1\leq\pi/2$ and $0\leq\varphi_2\leq\pi/2$. As outlined in section~\ref{sec:framework}, the fixed element $\cos\varphi_{1}$ may correspond to any of the nine entries of the PMNS mixing matrix. The 36 possible arrangements resulting from permutations of rows and columns generally yield nine distinct mixing patterns, which can be taken as the independent cases for analysis.
\begin{equation}\label{eq:permu_caseVIII}
\begin{array}{lll}
 U^{VIII}_{1}=U^{VIII}, ~~~~&~~~~ U^{VIII}_{2}=U^{VIII}P_{213},  ~~~~&~~~~ U^{VIII}_{3}=U^{VIII}P_{321}\,, \\
U^{VIII}_{4}=P_{213}U^{VIII}, ~~~~&~~~~  U^{VIII}_{5}=P_{213}U^{VIII}P_{213},~~~~&~~~~ U^{VIII}_{6}=P_{213}U^{VIII}P_{321}\,, \\
U^{VIII}_{7}=P_{231}U^{VIII}, ~~~~&~~~~ U^{VIII}_{8}=P_{231}U^{VIII}P_{213}, ~~~~&~~~~ U^{VIII}_{9}=P_{231}U^{VIII}P_{321}\,.
\end{array}
\end{equation}
It is observed that the mixing matrices $U^{VIII}_{7}$, $U^{VIII}_{8}$ and $U^{VIII}_{9}$ are derivable from $U^{VIII}_{4}$, $U^{VIII}_{5}$ and $U^{VIII}_{6}$  by interchanging their second and third rows. Therefore they lead to the same solar mixing angle, reactor mixing angle and Majorana CP phases while atmospheric angle changes from $\theta_{23}$ to $\pi/2-\theta_{23}$ and the Dirac CP phase changes from $\delta_{CP}$ to $\pi+\delta_{CP}$.  We see that the mixing parameters depend on the continuous parameters $\theta_{\ell}$ and $\theta_{\nu}$ as well as the discrete parameters $\varphi_{1}$ and $\varphi_{2}$. This leads to the following sum rules connecting the mixing angles and the Dirac CP phase $\delta_{CP}$:
\begin{eqnarray}
\nonumber&& U^{VIII}_{1}~:~~~\cos^{2}\theta_{12}\cos^2\theta_{13}=\cos^2\varphi_{1}\,,\qquad  U^{VIII}_{2}~:~~~\sin^{2}\theta_{12}\cos^2\theta_{13}=\cos^2\varphi_{1}\,, \\
\nonumber && U^{VIII}_{6}~:~~~\sin^{2}\theta_{23}\cos^2\theta_{13}=\cos^2\varphi_{1}\,,\qquad  U^{VIII}_{9}~:~~~\cos^{2}\theta_{23}\cos^2\theta_{13}=\cos^2\varphi_{1}\,, \\
\nonumber&& U^{VIII}_{4}~:~~~\cos\delta_{CP}=\frac{2(\cos^2\varphi_{1}-\sin^2\theta_{12}\cos^2\theta_{23}-\sin^2\theta_{13}\cos^2\theta_{12}\sin^2\theta_{23})}
{\sin2\theta_{12}\sin\theta_{13}\sin2\theta_{23}}\,, \\
\nonumber&& U^{VIII}_{5}~:~~~\cos\delta_{CP}=-\frac{2(\cos^2\varphi_{1}-\cos^2\theta_{12}\cos^2\theta_{23}-\sin^2\theta_{13}\sin^2\theta_{12}\sin^2\theta_{23})}
{\sin2\theta_{12}\sin\theta_{13}\sin2\theta_{23}}\,, \\
\nonumber&& U^{VIII}_{7}~:~~~\cos\delta_{CP}=-\frac{2(\cos^2\varphi_{1}-\sin^2\theta_{12}\sin^2\theta_{23}-\sin^2\theta_{13}\cos^2\theta_{12}\cos^2\theta_{23})}
{\sin2\theta_{12}\sin\theta_{13}\sin2\theta_{23}}\,, \\
\label{eq:correlation_VIII}&& U^{VIII}_{8}~:~~~\cos\delta_{CP}=\frac{2(\cos^2\varphi_{1}-\cos^2\theta_{12}\sin^2\theta_{23}-\sin^2\theta_{13}\sin^2\theta_{12}\cos^2\theta_{23})}
{\sin2\theta_{12}\sin\theta_{13}\sin2\theta_{23}}\,.
\end{eqnarray}
For a given value of $\varphi_{1}$, the allowed intervals for $\cos\delta_{CP}$ can be derived from the aforementioned correlations by scanning the mixing angles across their experimentally allowed $3\sigma$ intervals~\cite{Esteban:2024eli}. As a result, such sum rules may be tested and potentially differentiated in upcoming neutrino oscillation experiments, provided the Dirac CPV phase $\delta_{CP}$ is measured. Under this framework, the choice of residual symmetry fixes one element of the mixing matrix to $\cos\varphi_1$, so that constraints on $\varphi_{1}$ can be inferred by requiring $\cos\varphi_1$ to lie within the experimentally favored $3\sigma$ range. Additionally, we conduct a detailed numerical scan of the $\Delta(6n^2)$ group with $n\leq 4$, showing that only the mixing matrices labeled $U^{VIII}_{4}$ through $U^{VIII}_{9}$ are consistent with current mixing angle data for specific choices of the parameters $\theta_{\ell}$ and $\theta_{\nu}$. 

Note that if the left-handed lepton doublets are assigned to a \emph{reducible} three-dimensional representation, given by the direct sum of a singlet and a doublet representation of $G_{f}$, the lepton mixing pattern $U^{VIII}$ in Eq.~\eqref{eq:PMNS_caseVIII} can also arise from the breaking of the dihedral group $D_{n}$ combined with generalized CP symmetry~\cite{Lu:2019gqp}.
The relevant group-theoretical properties of $D_{n}$ are summarized in Appendix~\ref{sec:Dn}.
In this scenario, the residual symmetries in the lepton sector can be chosen as
\begin{equation}
 G_{\ell}=Z_{2}^{SR^{z_{e}}},\quad X_{\ell}=\{R^{-z_{e}+x},\, SR^{x}\},\qquad
 G_{\nu}=Z_{2}^{SR^{z_{\nu}}},\quad X_{\nu}=\{R^{-z_{\nu}+y},\, SR^{y}\}\,,
\end{equation}
where $z_{e}, z_{\nu}=0,1,\ldots,n-1$, and $x=y=0$ for odd $n$, while $x,y=0, n/2$ for even $n$. In this case, the explicit expressions of the discrete parameters $\varphi_{1}$ and $\varphi_{2}$ entering the mixing pattern $U^{VIII}$ are rather involved (see~\cite{Lu:2019gqp} for details). Nevertheless, $\varphi_{1}$ and $\varphi_{2}$ are restricted to the following discrete sets of values:
\begin{equation}\label{eq:phi12_VIII_Dn}
\varphi_{1}~(\text{mod}~2\pi)=0, \frac{1}{n}\pi, \frac{2}{n}\pi, \dots, \frac{2n-1}{n}\pi\,,\qquad  \varphi_{2}~(\text{mod}~2\pi)=0, \frac{1}{2}\pi, \pi,  \frac{3}{2}\pi\,,
\end{equation}
Moreover, $\varphi_{2}$ is further constrained to take only the values $0$ or $\pi$ if the group index $n$ of $D_{n}$ is odd, or if $x=y=0, n/2$ with even $n$.

The fixed entry of $U^{VIII}$ is $\cos\varphi_{1}$, which depends only on $\varphi_{1}$; consequently, the phenomenology is directly affected by the value of $\varphi_{1}$.
By comparing Eq.~\eqref{eq:phi12_VIII} with Eq.~\eqref{eq:phi12_VIII_Dn}, we observe that for the same value of $n$, the parameter $\varphi_{2}$ can generally take a larger set of discrete values in the case of $\Delta(6n^{2})$ than in the case of $D_{n}$, while the allowed values of $\varphi_{1}$ remain identical for both groups.
It is worth emphasizing, however, that the group order of $D_{n}$ is $2n$, which is substantially smaller than the order $6n^{2}$ of $\Delta(6n^{2})$.
From the perspective of maintaining analytical tractability while avoiding excessively large group orders, the dihedral group $D_{n}$ can in fact provide a richer and more flexible phenomenology than $\Delta(6n^{2})$.
For this reason, it is necessary to also consider the realization of the mixing pattern $U^{VIII}$ originating from $D_{n}\rtimes H_{CP}$ in the following discussion.

\item[~~(\uppercase\expandafter{\romannumeral9})] $G_{\ell}=Z^{bc^xd^x}_2$, $X_{\ell\bm{r}}=\rho_{\bm{r}}(c^{\alpha}d^{-2x-\alpha})$, $G_{\nu}=Z^{abc^y}_2$, $X_{\nu\bm{r}}=\rho_{\bm{r}}(c^{\beta}d^{2y+2\beta})$

Using the $\Sigma^{IX}$ matrix in table~\ref{tab:sum_D6n2_Sigma}, we obtain that the lepton mixing matrix is given by
{\small\begin{equation}
\label{eq:UPMNS_caseIX}
U^{IX}=\frac{1}{2}\left(
\begin{array}{ccc}
 1 &~ c_{\nu}+\sqrt{2} e^{i \varphi_{2}} s_{\nu} ~& s_{\nu}-\sqrt{2} e^{i \varphi_{2}} c_{\nu} \\
 s_{\ell}+\sqrt{2} e^{i \varphi_{1}} c_{\ell} &~ s_{\ell} c_{\nu}-\sqrt{2} (e^{i \varphi_{1}} c_{\ell} c_{\nu}+e^{i \varphi_{2}} s_{\ell}s_{\nu}) ~&s_{\ell} s_{\nu}-\sqrt{2}(e^{i \varphi_{1}} c_{\ell} s_{\nu} -e^{i \varphi_{2}} s_{\ell}c_{\nu} )  \\
 c_{\ell}-\sqrt{2} e^{i \varphi_{1}} s_{\ell} &~ c_{\ell}c_{\nu}+\sqrt{2} (e^{i \varphi_{1}} s_{\ell}c_{\nu} -e^{i \varphi_{2}}c_{\ell} s_{\nu}) ~& c_{\ell} s_{\nu}+\sqrt{2} (e^{i \varphi_{1}} s_{\ell} s_{\nu}+ e^{i \varphi_{2}} c_{\ell}c_{\nu}) \\
\end{array}
\right)\,.
\end{equation}}
with
\begin{equation}
 \varphi_1=\frac{3\alpha+2(x+y)}{n}\pi, \qquad \varphi_2=-\frac{3\beta+2(x+y)}{n}\pi   \,.
\end{equation}
Our analysis reveals that the parameters $\varphi_1$ and $\varphi_2$ are not entirely independent. Instead, they are restricted to the following discrete sets:
\begin{eqnarray}
\nonumber&&\varphi_1~(\mathrm{mod}~2\pi)=0, \frac{1}{n}\pi, \frac{2}{n}\pi, \ldots, \frac{2n-1}{n}\pi\,,
\\
\nonumber &&\varphi_1+\varphi_{2}~(\mathrm{mod}~2\pi)=0, \frac{3}{n}\pi, \frac{6}{n}\pi, \ldots, \frac{2n-3}{n}\pi,  \quad 3 \mid n\,, \\
\label{eq:para_values_II}&&\varphi_1+\varphi_{2}~(\mathrm{mod}~2\pi)=0, \frac{1}{n}\pi, \frac{2}{n}\pi, \ldots, \frac{2n-1}{n}\pi, \quad 3 \nmid n\,.
\end{eqnarray}
It is adequate to restrict attention to the fundamental domains $0 \leq \varphi_1 < \pi$ and $0 \leq \varphi_2 < \pi$. According to Eq.~\eqref{eq:UPMNS_caseIX}, we see that one element of the PMNS mixing matrix equals $1/2$. To align with experimental observations, this constant entry may correspond to the (21), (22), (31), or (32) elements of the matrix. Consequently, the PMNS matrix can assume four distinct configurations:
\begin{equation}\label{eq:permu_caseIX}
U^{IX}_{1}=P_{213}U^{IX}, \quad  U^{IX}_{2}=P_{213}U^{IX}P_{213}, \quad
U^{IX}_{3}=P_{231}U^{IX}, \quad  U^{IX}_{4}=P_{231}U^{IX}P_{213}\,.
\end{equation}
We see that $U^{IX}_{3}$ and $U^{IX}_{4}$ are related to $U^{IX}_{1}$ and $U^{IX}_{2}$ through an exchange of the second and third rows of the PMNS mixing matrix, respectively. The strong correlations among the mixing parameters stem from their common dependence on $\theta_{\ell}$ and $\theta_{\nu}$. For these four mixing matrices, the sum rule takes the form of the last four correlations in Eq.~\eqref{eq:correlation_VIII} with $\cos\varphi_{1}$ replaced by $1/2$, then we obtain the corresponding $\cos\delta_{CP}$ intervals: $[0.6699,1]$, $[0.3829,1]$, $[-1,-0.5853]$ and $[-1,-0.5537]$.

\item[~~(\uppercase\expandafter{\romannumeral10})] $G_{\ell}=Z^{bc^xd^x}_2$, $X_{\ell\bm{r}}=\rho_{\bm{r}}(c^{\alpha}d^{-2x-\alpha})$,
 $G_\nu=Z^{c^{n/2}}_2$, $X_{\nu\bm{r}}=\rho_{\bm{r}}(c^{\beta}d^{\gamma})$

We obtain the lepton mixing matrix from the $\Sigma^{X}$ matrix in table~\ref{tab:sum_D6n2_Sigma}. This matrix can be taken to the form
\begin{equation}\label{eq:PMNS_X}
U^{X}=\frac{1}{\sqrt{2}}\left(
\begin{array}{ccc}
 c_{\nu} &~ s_{\nu} ~& -e^{i \varphi_{2}} \\
 s_{\ell}c_{\nu} +\sqrt{2} e^{i \varphi_{1}} c_{\ell} s_{\nu} &~ s_{\ell} s_{\nu}- \sqrt{2} e^{i \varphi_{1}} c_{\ell} c_{\nu} ~& e^{i \varphi_{2}} s_{\ell} \\
 c_{\ell} c_{\nu}-\sqrt{2} e^{i \varphi_{1}} s_{\ell} s_{\nu} &~ c_{\ell} s_{\nu}+\sqrt{2} e^{i \varphi_{1}} s_{\ell}c_{\nu}  ~& e^{i \varphi_{2}} c_{\ell} \\
\end{array}
\right)\,,
\end{equation}
where the parameters $\varphi_1$ and $\varphi_2$ are defined as
\begin{equation}
    \varphi _{5}=\frac{2 x+3 \alpha-2 \beta  +\gamma }{n}\pi, \qquad  \varphi _{6}=-\frac{2 x+\beta +\gamma }{n}\pi\,,
\end{equation}
where the group index $n$ has to be even number. We see that the discrete values of $\varphi_1$ and $\varphi_2$ are correlated in the case that $n$ is divisible by 3. To be specific, their values could be
\begin{eqnarray}
\nonumber&&\varphi_{1}~(\mathrm{mod}~2\pi)=0, \frac{1}{n}\pi, \frac{2}{n}\pi, \ldots, \frac{2n-1}{n}\pi\,, \\
\nonumber &&\varphi_{1}+\varphi_{2}~(\mathrm{mod}~2\pi)=0, \frac{3}{n}\pi, \frac{6}{n}\pi, \ldots, \frac{2n-3}{n}\pi,  \qquad 3 \mid n\\
\label{eq:para_values_X}&&\varphi_{1}+\varphi_{2}~(\mathrm{mod}~2\pi)=0, \frac{1}{n}\pi, \frac{2}{n}\pi, \ldots, \frac{2n-1}{n}\pi, \qquad 3 \nmid n\,.
\end{eqnarray}
It is easy to check that $U^{X}$ has the following symmetry properties:
\begin{eqnarray}
\nonumber && U^{X}(\varphi_{1}+\pi,\varphi_{2},\theta_{\ell},\theta_{\nu})=U^{X}(\varphi_{1},\varphi_{2},\theta_{\ell},\pi-\theta_{\nu})\text{diag}(-1,1,1)\,,\\
\nonumber &&U^{X}(\varphi_{1},\varphi_{2}+\pi,\theta_{\ell},\theta_{\nu})=U^{X}(\varphi_{1},\varphi_{2},\theta_{\ell},\theta_{\nu})\text{diag}(1,1,-1)\,,\\
\nonumber && U^{X}(\varphi_{1},\varphi_{2}+\frac{\pi}{2},\theta_{\ell},\theta_{\nu})=U^{X}(\varphi_{1},\varphi_{2},\theta_{\ell},\theta_{\nu})\text{diag}(1,1,i)\,,\\
\label{eq:caseX_complex}&&U^{X}(\pi-\varphi_{1},\pi-\varphi_{2},\theta_{\ell},\theta_{\nu})=\left(U^{X}(\varphi_{1},\varphi_{2},\theta_{\ell},\pi-\theta_{\nu})\right)^{*} \text{diag}(-1,1,-1)\textcolor{red}{\,.}
\end{eqnarray}
Hence it is  sufficient to focus on the fundamental interval of $0\leq\varphi_1<\pi/2$ and $0\leq\varphi_2~ (\text{mod}\, \pi/2)<\pi/2$~\cite{Li:2017abz}. Furthermore, the impact of $\varphi_{2}$ is confined to the Majorana phase $\alpha_{31}$. Its effect amounts to a simple mapping $\alpha_{31} \rightarrow \alpha_{31} + 2\varphi_{2}$ of the $\varphi_{2}=0$ results, leaving all other parameters unaffected. We see the fixed element is $1/\sqrt{2}$ which can only be the $(32)$ or $(33)$ entry in order to achieve agreement with experimental data. As a consequence, we have two phenomenologically viable mixing patterns after the permutations of rows and columns of the mixing matrix $U^{X}$ are considered,
\begin{equation}
\label{eq:permu_caseX}
 U^{X}_{1}=P_{213}U^{X}\,, \qquad U^{X}_{2}=P_{231}U^{X}\,.
\end{equation}
For the two mixing matrices, all mixing parameters are strongly correlated such that the following sum rules among the mixing angles and Dirac CP phase are found to be satisfied,
\begin{equation}\label{eq:sum_rules_X}
 U^{X}_{1}~:~~~\cos^2\theta_{13}\sin^2\theta_{23}=\frac{1}{2}\,,     \qquad U^{X}_{2}~:~~~\cos^2\theta_{13}\cos^2\theta_{23}=\frac{1}{2}\,,\
\end{equation}
Given the measured reactor mixing angle $\sin^2\theta_{13}\in[0.02030,0.02388]$~\cite{Esteban:2024eli}, we find the atmospheric mixing angle
\begin{equation}\label{eq:sintheta23_caseX}
\sin^{2}\theta_{23}\in
[0.5104,0.5122] ~~ \text{for}~~U^{X}_{1}, \qquad \sin^{2}\theta_{23}\in
[0.4878,0.4896] ~~ \text{for}~~U^{X}_{2},
\end{equation}
which deviates slightly from maximal mixing. These predictions may be tested at future long-baseline experiments DUNE~\cite{DUNE:2020ypp} and T2HK~\cite{Hyper-Kamiokande:2018ofw}, and at the discussed ESS$\nu$SB experiment~\cite{Alekou:2022emd}. A remarkably good fit to experimental data can generally be obtained for any $\Delta(6n^2)$ flavour group with even $n$ in this framework. In the present work, we consider the next group with $n = 4$. Then the allowed values of $\varphi_1$ and $\varphi_2$ include $(0,0)$, $(0,\pi/4)$, $(\pi/4,0)$, $(\pi/4,\pi/4)$, $(\pi/2,0)$, and $(\pi/2,\pi/4)$\footnote{The pairs $(3\pi/4,0)$ and $(3\pi/4,\pi/4)$ are also allowed. The resulting mixing matrices are the complex conjugates of those for $(\pi/4,0)$ and $(\pi/4,\pi/4)$, respectively, up to redefinition of $\theta_\nu$ and $Q_\nu$, yielding identical mixing angles but reversed overall signs of the CP phases.}. From the mixing matrix in Eq.~\eqref{eq:PMNS_X}, for a fixed $\varphi_1$, the cases $\varphi_2 = 0$ and $\varphi_2 = \pi/4$ give identical mixing angles and Dirac CP phase. However, for pattern $U^{X}_{1}$, the Majorana phase $\alpha_{21}$ (or $\alpha_{31}$) shifts by $\pi/2$. Thus, predictions for $\varphi_2 = \pi/4$ can be directly inferred from those with $\varphi_2 = 0$, and it suffices to analyze only the $\varphi_2 = 0$ cases. When $\varphi_1 = \varphi_2 = 0$, all permutations in Eq.~\eqref{eq:permu_caseX} can reproduce the measured mixing angles for appropriate choices of $\theta_\ell$ and $\theta_\nu$. In this scenario, the mixing matrix is real, so both Dirac and Majorana CP phases are trivial.

\end{description}

\section{\label{sec:numerical-analysis}Numerical analysis}

\begin{table}[t!]
\centering
\begin{tabular}{|c|c|c|c|c|} \hline \hline
Observable & bf $\pm~1\sigma$ & $3\sigma$ range \\ \hline
$\sin^2\theta_{12}$ $(\text{NO} \;\&\; \text{IO})$ & $0.308^{+0.012}_{-0.011}$ & $[0.275, 0.345]$   \\
$\sin^2\theta_{13}$(NO) & $0.02215^{+0.00056}_{-0.00058}$ & $[0.02030, 0.02388]$  \\
$\sin^2\theta_{13}$(IO) & $0.02231^{+0.00056}_{-0.00056}$ & $[0.02060, 0.02409]$  \\
$\sin^2\theta_{23}$(NO) & $0.470^{+0.017}_{-0.013}$ & $[0.435, 0.585]$  \\
$\sin^2\theta_{23}$(IO) & $0.550^{+0.012}_{-0.015}$ & $[0.440, 0.584]$  \\
$\delta_{CP}/\pi$(NO) & $1.18^{+0.14}_{-0.23}$ & $[0.69, 2.02]$  \\
$\delta_{CP}/\pi$(IO) & $1.52^{+0.12}_{-0.14}$ & $[1.12, 1.86]$  \\
\hline \hline
\end{tabular}
\caption{\label{tab:lepton-data}The central values, the $1\sigma$ errors and $3\sigma$ ranges of the mixing angles and Dirac CPV phase for leptons. We adopt the values of the lepton mixing parameters from \texttt{NuFIT}-v6.0 with Super-Kamiokanda atmospheric data for normal ordering (NO) and inverted ordering (IO)~\cite{Esteban:2024eli}. }
\end{table}

In this section we outline the statistical framework used to test the mixing–parameter predictions of the models discussed above against current neutrino oscillation data, including the latest results from the global analysis \texttt{NuFIT} and the JUNO experiment. The input values of the lepton mixing parameters are taken from the \texttt{NuFIT}-v6.0 with inclusion of the data on atmospheric neutrinos provided by the Super-Kamiokande and IceCube collaborations~\cite{Esteban:2024eli}, and they are summarized in table~\ref{tab:lepton-data}. To quantify the compatibility of each breaking pattern with experimental measurements, we perform a $\chi^{2}$ analysis for every mixing pattern for both NO and IO neutrino mass spectrum. We construct the global test statistic as~\cite{Marzocca:2013cr,Girardi:2014faa,Girardi:2015vha,Girardi:2015zva,Petcov:2018snn}
\begin{equation}\label{eq:chisq_def}
\chi^{2}(\vec{o})=\sum_{i=1}^{4}\chi_{i}^{2}(o_i)\,, \qquad
\vec{o}=(\sin^2\theta_{12}\,,\sin^2\theta_{13}\,,\sin^2\theta_{23}\,,\delta_{CP})\,,
\end{equation}
where $\chi_i^{2}(o_i)$ denotes the one–dimensional projection provided by \texttt{NuFIT}.
We remark that the global fit of \texttt{NuFIT}-v6.0~\cite{Esteban:2024eli} shows an overall preference for normal ordering, with $\chi^{2}_{\min,\mathrm{IO}} - \chi^{2}_{\min,\mathrm{NO}} = 6.1$ if the Super-Kamiokande data is included.
In our analysis, however, we adopt a conservative treatment and consider the two mass orderings on equal footing by setting $\chi^{2}_{\min,\mathrm{NO}} = \chi^{2}_{\min,\mathrm{IO}}=0$. For a given mixing pattern, the observables in $\vec{o}$ are not independent but satisfy sum rules since they depend on one or two free real parameters. The best-fit values of these parameters, together with the predicted mixing observables, are obtained by minimizing the global $\chi^{2}$. Besides, to determine the one–dimensional profile for a specific observable $o_i$ (e.g. $\sin^2\theta_{12}$), we minimize the global function while fixing $o_i$.
This defines
\begin{equation}
\chi^2(o_{i}) =
\min\left[\chi^2(\vec{o})\Big|_{o_{i}}\right].
\label{eq:chi2alpha}
\end{equation}
The corresponding likelihood is then given by
\begin{equation}
L(o_{i})=\exp\left[-\chi^{2}(o_i)/2\right].
\label{eq:likelihood}
\end{equation}

In the following, we will first confront the predictions of the one- and two-parameter mixing patterns discussed in sections~\ref{sec:onepara_lepton_mixing} and~\ref{sec:twopara_lepton_mixing} with the latest \texttt{NuFIT} results.
A given lepton mixing pattern is regarded as phenomenologically viable if its predictions for the mixing parameters at the minimum of the $\chi^{2}$ function lie within the corresponding $3\sigma$ intervals summarized in table~\ref{tab:lepton-data}.
After that, we confront the predicted values of $\sin^{2}\theta_{12}$ with the most recent 59.1-day JUNO measurement, as well as with the projected precision expected after 6 years of JUNO data taking. We also examine the predictions for $\sin^{2}\theta_{23}$ and $\delta_{CP}$ in light of the anticipated sensitivity reaches of the future long-baseline experiments DUNE and T2HK. 

We next examine the predictions of the considered models for $\sin^{2}\theta_{12}$ with the projected precision of JUNO after six years of data taking of $0.5\%$ as $1\sigma$ uncertainty, assuming that the current best fit value of $\sin^{2}\theta_{12}$ will not change significantly. We further confront the model predictions for $\sin^{2}\theta_{23}$
and $\delta_{CP}$ with the precision on these parameters foreseen to be achieved by the future long baseline neutrino oscillation experiments T2HK~\cite{Hyper-Kamiokande:2018ofw} and DUNE~\cite{DUNE:2020ypp} after 15 years of measurements. In this analysis we assume again that the best values of $\sin^{2}\theta_{23}$ and $\delta_{CP}$ obtained in~\cite{Esteban:2024eli}
and shown in table~\ref{tab:lepton-data}
will not change significantly. The projected precisions of T2HK and  DUNE on $\sin^{2}\theta_{23}$ and $\delta_{CP}$ depend strongly on the found respective best fit values. For $\sin^{2}\theta_{23}$, as illustrated, e.g., in Ref.~\cite{Ballett:2016daj},
the projected precision of T2HK and DUNE ranges from about $1\%$ when the true value lies near the edges of the current $3\sigma$ interval, to roughly $7\%$ around the maximal-mixing point $\sin^{2}\theta_{23}=0.5$.
For the present \texttt{NuFIT} best-fit values,
$\sin^{2}\theta_{23}=0.470$ (NO) and $\sin^{2}\theta_{23}=0.550$ (IO), the expected uncertainty is approximately $3\%$, which we adopt as a representative estimate in the present analysis. In what concerns the measurements of $\delta_{CP}$ in these experiments, the projected precisions vary strongly with the found best fit value $\delta^{\text{bfv}}_{CP}$ of $\delta_{CP}$. For $\delta^{\text{bfv}}_{CP} = 0~(2\pi)$, which for IO
is ruled out by the \texttt{NuFIT} results at $3\sigma$ C.L., both the T2HK and DUNE collaborations report $6^\circ$ as prospective $1\sigma$ uncertainty. At the same time the projected $1\sigma$ uncertainties in the case of $\delta^{\text{bfv}}_{CP} = 270^\circ$ are for
T2HK and DUNE $20^\circ$ and $16^\circ$, respectively.
We do not have a precise information about the precision
at $\delta^{\text{bfv}}_{CP} = 180^\circ$, which is predicted by some of the considered models and which in the case of IO is ruled out at $3\sigma$ C.L. by the \texttt{NuFIT} analysis. We guess that it could correspond to $1\sigma$ uncertainties
of $\sim 10^\circ$ and $\sim 8^\circ$.
We note that the proposed ESS$\nu$B experiment
~\cite{Alekou:2022emd} is foreseen to be able to measure $\delta_{CP}$ with $1\sigma$ uncertainty of $8^\circ$ at $\delta^{\text{bfv}}_{CP}=270^\circ$. In the analysis which follows we will assume that in the future long baseline neutrino oscillation experiments $\delta_{CP}$ will be determined with $1\sigma$ error of $12^\circ$.
Thus, in the view of all assumptions listed above,
the conclusions based on the prospective data on
$\sin^2\theta_{23}$ and $\delta_{CP}$ should probably be considered as indicative only.

Finally, we note that among the one- and two-parameter mixing patterns, several patterns---namely $U^{I}$, $U^{II}$, $U^{VIII}$, $U^{XI}$ and $U^{X}$---arise from the series of flavour groups $\Delta(6n^{2})\rtimes H_{CP}$ (with $U^{VIII}$ also obtainable from $D_{n}\rtimes H_{CP}$). To keep the analysis tractable and avoid excessively large group orders, we restrict our attention to groups $\Delta(6n^{2})$ with $n \leq 4$ and $D_{n}$ with $n \leq 8$ in what follows.

\subsection{\label{sec:num_one}Lepton mixing matrix depending on one free parameter}

\begin{table}[t!]\tabcolsep=0.11cm
\begin{center}
\scalebox{1.0}{\begin{tabular}{|c|c|c|c|c|c|c|c|c|c|c|}
\hline\hline
\multirow{2}{*}{Order} & \multirow{2}{*}{Case} & \multirow{2}{*}{$(\varphi_1,\varphi_2)$} & \multirow{2}{*}{$\theta_{\text{bf}}/\pi$} & \multirow{2}{*}{$\chi^2_{\text{min}}$} & \multirow{2}{*}{$\sin^2\theta_{12}$} & \multirow{2}{*}{$\sin^2\theta_{13}$} & \multirow{2}{*}{$\sin^2\theta_{23}$} & \multirow{2}{*}{$\delta_{CP}/\pi$} & $\alpha_{21}/\pi$ & $\alpha_{31}/\pi$  \\
&&&&&&&&& (mod 1) & (mod 1) \\
\hline
\multirow{8}{*}{NO} & $U^{I}$  & $(-\frac{\pi}{2},\frac{\pi}{2})$ & $0.9170$ & $5.308$ & \cellcolor{red!30}$0.318$ & $0.02215$ & $0.5$ & $1.5$ & $0$ & $0$\\  \cline{2-11} & $U^{II}_{1}$  & $(0,\pi)$ & $0.1918$ & $11.862$ & $0.341$ & $0.02208$ & $0.5$ & $1.5$ & $0$ & $0$\\  \cline{2-11} & $U^{II}_{2}$ & $(0,\pi)$ & $0.3083$ & $11.859$ & $0.341$ & $0.02214$ & $0.5$ & $1.5$ & $0$ & $0$\\  \cline{2-11} & $U^{III}$ & --- & $0.0560$ & $9.600$ & $0.283$ & $0.02217$ & $0.5$ & $1.5$ & $0$ & $0$\\  \cline{2-11} & $U^{IV}_{1}$ & --- & $0.9053$ & $6.414$ & \cellcolor{red!30}$0.331$ & $0.02214$ & $0.523$ & $1$ & $0$ & $0$\\  \cline{2-11} & $U^{IV}_{2}$ & --- & $0.9053$ & $11.856$ & \cellcolor{red!30}$0.331$ & $0.02214$ & $0.477$ & $0$ & $0$ & $0$\\  \cline{2-11} & $U^{V}_{1}$ & --- & $0.5715$ & $8.921$ & $0.341$ & $0.02212$ & $0.554$ & $1.331^{*}$ & $0.161^{*}$ & $0.559^{*}$\\  \cline{2-11} & $U^{V}_{2}$ & --- & $0.5746$ & $13.427$ & $0.341$ & $0.02256$ & $0.450$ & $1.652$ & $0.837$ & $0.439$\\  \cline{2-11}
\hline\hline
\multirow{6}{*}{IO} & $U^{I}$ & $(-\frac{\pi}{2},\frac{\pi}{2})$ & $0.9166$ & $3.887$ & \cellcolor{red!30}$0.318$ & $0.02235$ & $0.5$ & $1.5$ & $0$ & $0$\\  \cline{2-11} & $U^{II}_{1}$ & $(0,\pi)$ & $0.1915$ & $10.471$ & $0.341$ & $0.02227$ & $0.5$ & $1.5$ & $0$ & $0$\\ \cline{2-11} & $U^{II}_{2}$ & $(0,\pi)$ & $0.3085$ & $10.471$ & $0.341$ & $0.02227$ & $0.5$ & $1.5$ & $0$ & $0$\\  \cline{2-11} & $U^{III}$ & --- & $0.0563$ & $8.188$ & $0.283$ & $0.02236$ & $0.5$ & $1.5$ & $0$ & $0$\\  \cline{2-11} & $U^{V}_{1}$ & --- & $0.5736$ & $9.012$ & $0.341$ & $0.02241$ & $0.551$ & $1.343^{*}$ & $0.162^{*}$ & $0.561^{*}$\\  \cline{2-11} & $U^{V}_{2}$ & --- & $0.5758$ & $13.416$ & $0.341$ & $0.02274$ & $0.452$ & $1.646$ & $0.837$ & $0.438$\\ \cline{2-11}
\hline \hline
\end{tabular}}
\caption {\label{tab:best-fit-onepara}
The best-fit values of the free parameter and the corresponding lepton mixing observables for all one-parameter mixing patterns in both normal and inverted mass orderings.
The discrete parameters $(\varphi_{1},\varphi_{2})$ specify the residual symmetries associated with flavour groups of type $\Delta(6n^{2})$ $(n \leq 4)$.
For each pattern, the global minimum of the test statistic is obtained at $\theta=\theta_{\text{bf}}$, where the $\chi^{2}$ function reaches $\chi^{2}_{\text{min}}$.
At the best-fit point, we present the predictions for the neutrino mixing angles $\theta_{12}$, $\theta_{13}$, $\theta_{23}$, the Dirac CPV phase $\delta$, and the Majorana phases $\alpha_{21}$ and $\alpha_{31}$.
Values of CPV phases marked with a superscript $*$ correspond to scenarios in which the left-handed lepton doublets transform as the conjugate triplet $\bar{\mathbf{3}}$.
Entries in the $\sin^{2}\theta_{12}$ column highlighted in red color indicate predictions lying within the $3\sigma$ interval of the JUNO 59.1-day measurement.
}
\end{center}
\end{table}

 For the one-parameter mixing patterns, $U^{I}$, $U^{II}$, $U^{III}$, $U^{IV}$ and $U^{V}$, the best-fit value of the free parameter $\theta$, the minimum $\chi^{2}_{\text{min}}$ and the resulting predictions for the lepton mixing angles $\theta_{12}$, $\theta_{13}$, $\theta_{23}$ as well as the CPV phases $\delta_{CP}$, $\alpha_{21}$ and $\alpha_{31}$ for both NO and IO neutrino mass spectra are collected in table~\ref{tab:best-fit-onepara}. We note that the entries in the $\sin^{2}\theta_{12}$ column highlighted in red indicate predictions that fall within the $3\sigma$ interval of the JUNO 59.1-day measurement~\cite{JUNO:2025gmd},
\begin{equation}\label{eq:s12sq_3sigma_JUNO}
0.2831 \leq \sin^{2}\theta_{12} \leq 0.3353\,,
\end{equation}
where this interval is obtained by taking the central value and adding/subtracting three times the $1\sigma$ uncertainty given in Eq.~\eqref{eq:s12sq_JUNO_59day}. It is worth noting that the quoted $3\sigma$ interval of $\sin^{2}\theta_{12}$ from JUNO is obtained under the assumption of normal mass ordering. In our comparison, we also apply this interval to the IO case as a working assumption, since the JUNO collaboration concludes in
\cite{JUNO:2025gmd}
that the results on $\sin^{2}\theta_{12}$ and $\Delta m^2_{21}$
for the inverted ordering are ``fully compatible'' with those reported
for the normal ordering. The corresponding allowed regions of lepton observables, which are obtained by scanning the parameter space,  requiring that three mixing angles and the Dirac CPV phase $\delta_{CP}$ fall within their experimentally preferred $3\sigma$ ranges given in table~\ref{tab:lepton-data}, are summarized in table~\ref{tab:3sigma-onepara}, in Appendix~\ref{sec:3sigma-ranges}.

Confronting the results in table~\ref{tab:best-fit-onepara}
with the \texttt{NuFIT} $3\sigma$ allowed regions
given in table~\ref{tab:lepton-data},  
we find that a total of 8 (6) one-parameter mixing patterns are phenomenologically viable for NO (IO), once all inequivalent row and column permutations of the lepton mixing matrix are taken into account. For the mixing patterns $U^{I}$ and $U^{II}$ arising from $\Delta(6n^{2})\rtimes H_{CP}$, the smallest group that can reproduce the \texttt{NuFIT} data is $\Delta(24)\cong S_{4}$ with $n=2$.
In case of $(\varphi_{1},\varphi_{2})=\left(-\frac{\pi}{2},\,\frac{\pi}{2}\right)$, the pattern $U^{I}$ reduces to the TM1 form~\cite{Xing:2006ms,Lam:2006wm} with its first column fixed as $\left(-\frac{2}{\sqrt{6}},\, -\frac{1}{\sqrt{6}},\, \frac{1}{\sqrt{6}}\right)^{T}$, as can seen from Eq.~\eqref{eq:pmnsIa}, and it agrees with the \texttt{NuFIT} limits for both NO and IO. The TM2-type patterns, namely $U^{II}$ with $(\varphi_{1},\varphi_{2})=(0,\pi)$ and $U^{V}$ featuring the trimaximal column $\left(\frac{1}{\sqrt{3}},\,\frac{1}{\sqrt{3}},\,\frac{1}{\sqrt{3}}\right)^{T}$ (see Eq.~\eqref{eq:pmnsII} and Eq.~\eqref{eq:mixing_V}), don't agree with JUNO's latest results at $3.6\sigma$ level,
although it is close to the $3\sigma$ upper limit of \texttt{NuFIT} for both NO and IO. Among the one-parameter patterns originating from $A_{5}\rtimes H_{CP}$, the GR2 structure $U^{III}$—which preserves the second column of the Golden Ratio mixing pattern, $\left(\sqrt{\frac{1}{\sqrt{5}\,\phi_{g}}},\,\sqrt{\frac{\phi_{g}}{2\sqrt{5}}},\,\sqrt{\frac{\phi_{g}}{2\sqrt{5}}}\right)^{T}$ (see Eq.~\eqref{eq:mixing_III})—remains compatible with the \texttt{NuFIT} constraints in both NO and IO.
By contrast, the real mixing pattern $U^{IV}$, characterized by the fixed first column $\left(\frac{\phi_{g}}{2},\,\frac{\phi_{g}-1}{2},\,\frac{1}{2}\right)^{T}$ (see Eq.~\eqref{eq:mixing_IV}), is only viable in NO.

\begin{figure}[htbp]
\centering
\includegraphics[scale=0.55]{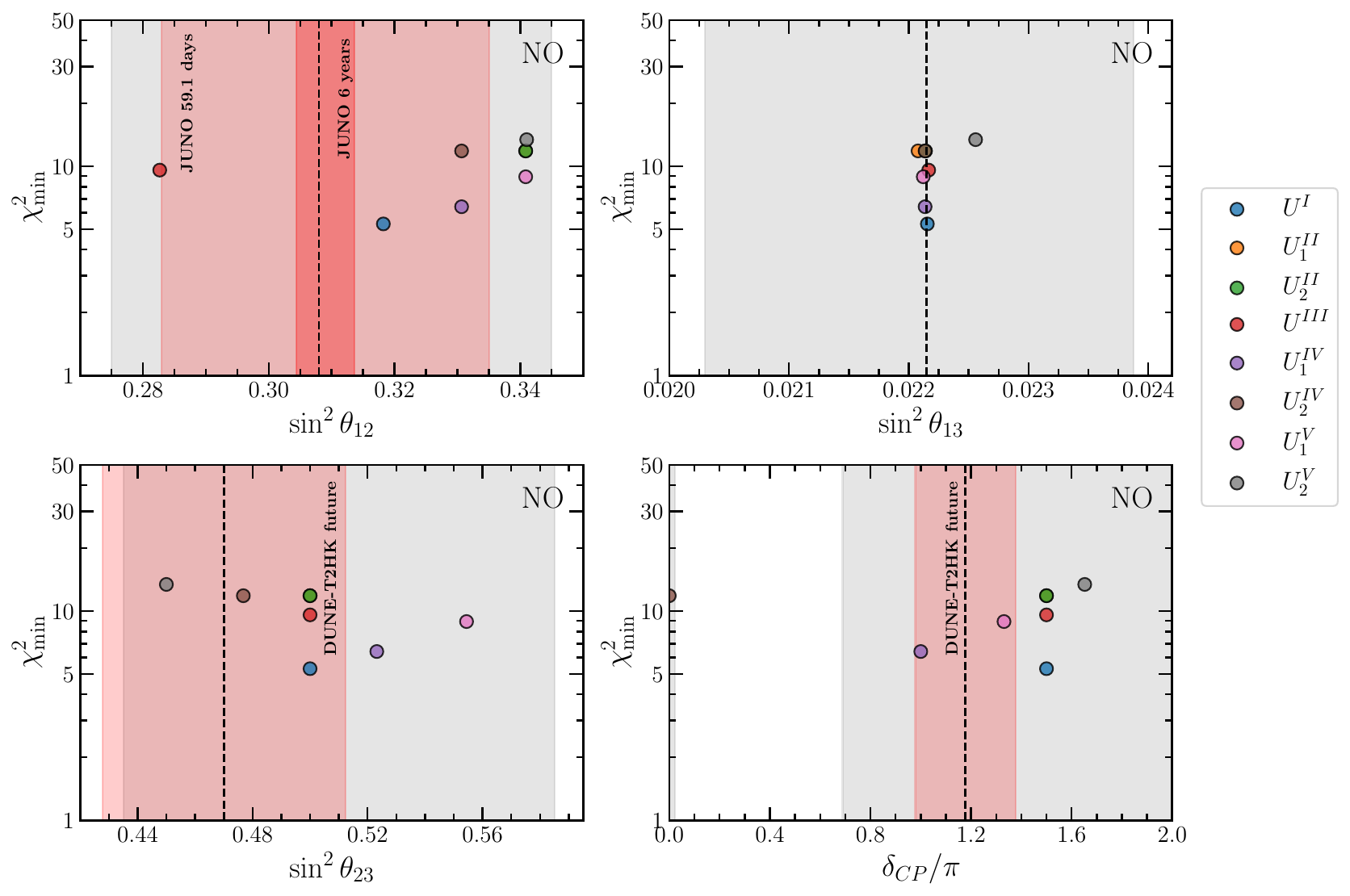}
\includegraphics[scale=0.55]{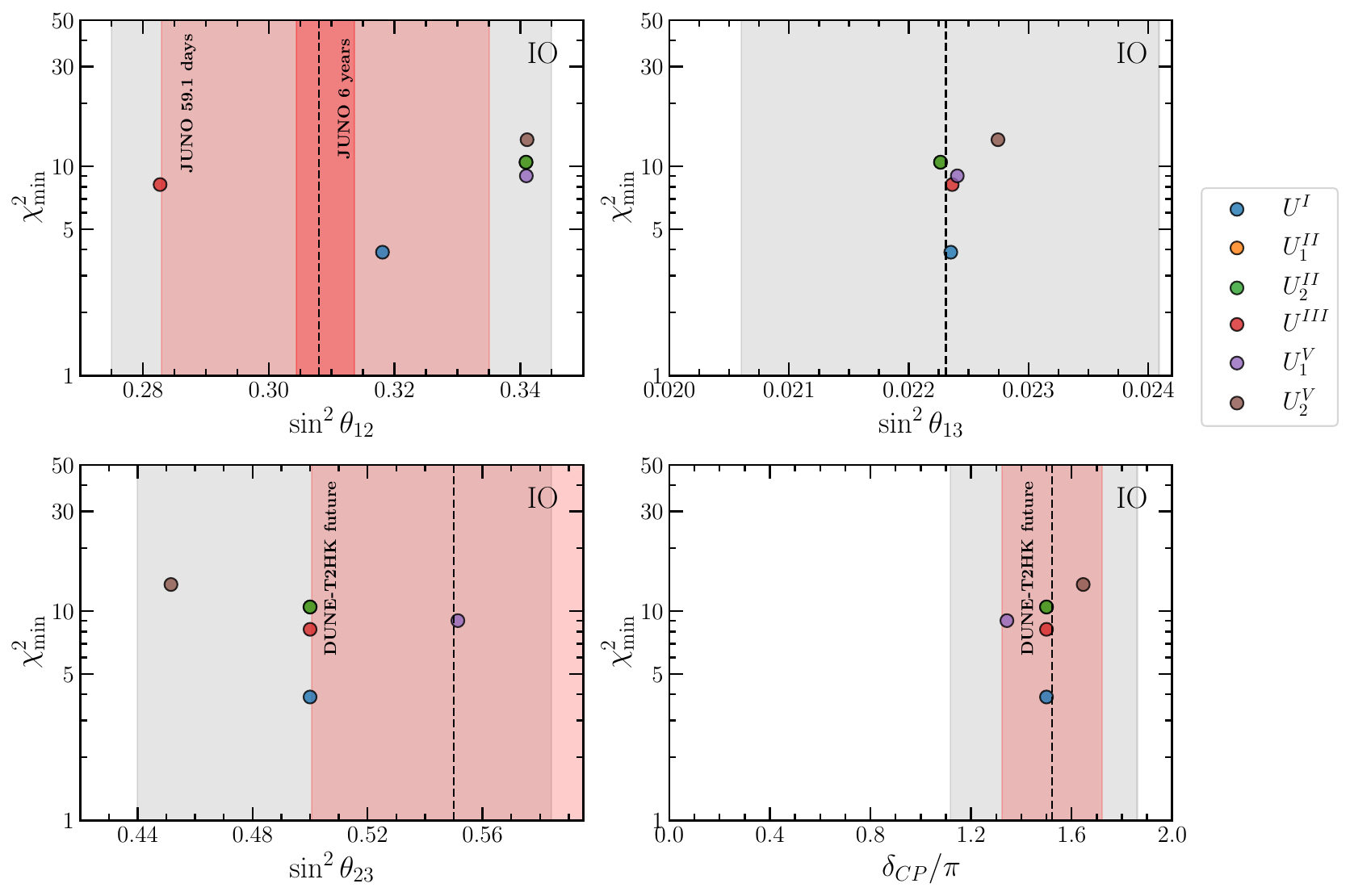}
\caption{Best-fit predictions of the neutrino mixing angles $\theta_{12}$, $\theta_{13}$, $\theta_{23}$ and the Dirac CPV phase $\delta_{CP}$ for lepton mixing patterns with one parameter in case of NO and IO neutrino masses spectrum. The gray regions in all plots are the current $3\sigma$ ranges of the corresponding observables for NO and IO from~\cite{Esteban:2024eli}. The dashed line is the current best-fit value. The light red and red region for $\sin^{2}\theta_{12}$ are the latest $3\sigma$ range obtained from the first 59.1 days of JUNO data~\cite{JUNO:2025gmd} and the prospective $3\sigma$ range of after 6 years of JUNO running with precision of $0.5\%$~\cite{JUNO:2022mxj}. The red regions shown for $\sin^{2}\theta_{23}$ and $\delta_{CP}$ represent the anticipated future $3\sigma$ sensitivities of next-generation long-baseline experiments such as DUNE~\cite{DUNE:2020ypp} and T2HK~\cite{Hyper-Kamiokande:2018ofw}, assuming the current NuFit best-fit values together with prospective precisions of about $3\%$ for $\sin^{2}\theta_{23}$ and $12^\circ$ for $\delta_{CP}$.}
\label{fig:onepara_bfpoints_plots_NO_IO}
\end{figure}

We display in figure~\ref{fig:onepara_bfpoints_plots_NO_IO} the best-fit values of the three mixing angles and $\delta_{CP}$ for all viable one-parameter models under both NO and IO neutrino mass spectrum. The dashed lines indicate the central values of the observables from the \texttt{NuFIT} global fit. The light-red band in the $\sin^{2}\theta_{12}$ panel corresponds to the $3\sigma$ interval derived from the recent JUNO measurement, see Eq.~\eqref{eq:s12sq_3sigma_JUNO}, while the darker and narrower red band indicates the projected $3\sigma$ precision after six years of data taking, assuming the current JUNO central value $\sin^{2}\theta_{12}=0.3092$ and the prospective $1\sigma$ uncertainty of $0.5\%$~\cite{JUNO:2022mxj}. For NO, three one-parameter mixing patterns --- $U^{I}$ (TM1), $U^{IV}_{1}$ and $U^{IV}_{2}$ --- yield predictions for $\sin^{2}\theta_{12}$ that fall within the current JUNO constraint. In contrast, the patterns $U^{II}$ (TM2), $U^{III}$ (GR2), and $U^{V}$ (TM2) are disfavored by the JUNO data at the $3\sigma$ level. None of the one-parameter patterns, however, would remain compatible with the much better sensitivity expected
after six years of operation
if the current best fit value would not change significantly.
For IO, only the pattern $U^{I}$ (TM1)
lies in the present JUNO $3\sigma$ allowed interval, and all patterns
would be excluded once the
planned future precision
is reached. We also observe that the predicted values of $\sin^{2}\theta_{13}$ for nearly all patterns lie very close to the \texttt{NuFIT} best-fit line, reflecting the excellent experimental precision already achieved for this observable.

Finally, the red bands shown for $\sin^{2}\theta_{23}$ and $\delta_{CP}$ represent the anticipated future $3\sigma$ sensitivities of next-generation long-baseline experiments such as DUNE~\cite{DUNE:2020ypp} and T2HK~\cite{Hyper-Kamiokande:2018ofw}. As we have discussed, the future experimental uncertainties on $\sin^{2}\theta_{23}$ and $\delta_{CP}$ foreseen to be achieved by T2HK and DUNE depend significantly on the true (i.e., best fit) values of $\sin^{2}\theta_{23}$ and $\delta_{CP}$. For the present \texttt{NuFIT} best-fit values, $\sin^{2}\theta_{23}=0.47$ (NO) and $\sin^{2}\theta_{23}=0.55$ (IO), the prospective $1\sigma$ uncertainty is approximately $3\%$, which we adopt as a representative estimate in the present study,
as we have indicated. In the case of $\delta_{CP}$, as explained earlier, we adopt $12^\circ$ as prospective $1\sigma$ uncertainty in our analysis.

Combining the numerical results in table~\ref{tab:best-fit-onepara}, we observe that the patterns $U^{I}$, $U^{II}$, and $U^{III}$ predict maximal atmospheric neutrino mixing and CP-violation, $\theta_{23}= \pi/4$ and $\delta_{CP} = 3\pi/2$. In contrast, the real mixing patterns $U^{IV}_{1}$ and $U^{IV}_{2}$ imply CP conservation, but with different values of $\delta_{CP} = 0$ and $\pi$. Neither maximal CP-violating nor CP-conserving $\delta_{CP}$ is predicted by the $U^{V}_{1}$ and $U^{V}_{2}$ patterns, for which also $\theta_{23} \neq \pi/4$. Hence, the precise measurements of $\theta_{23}$ and $\delta_{CP}$ will play an important role in discriminating between these viable mixing schemes.

\begin{figure}[htbp]
\centering
\includegraphics[scale=0.75]{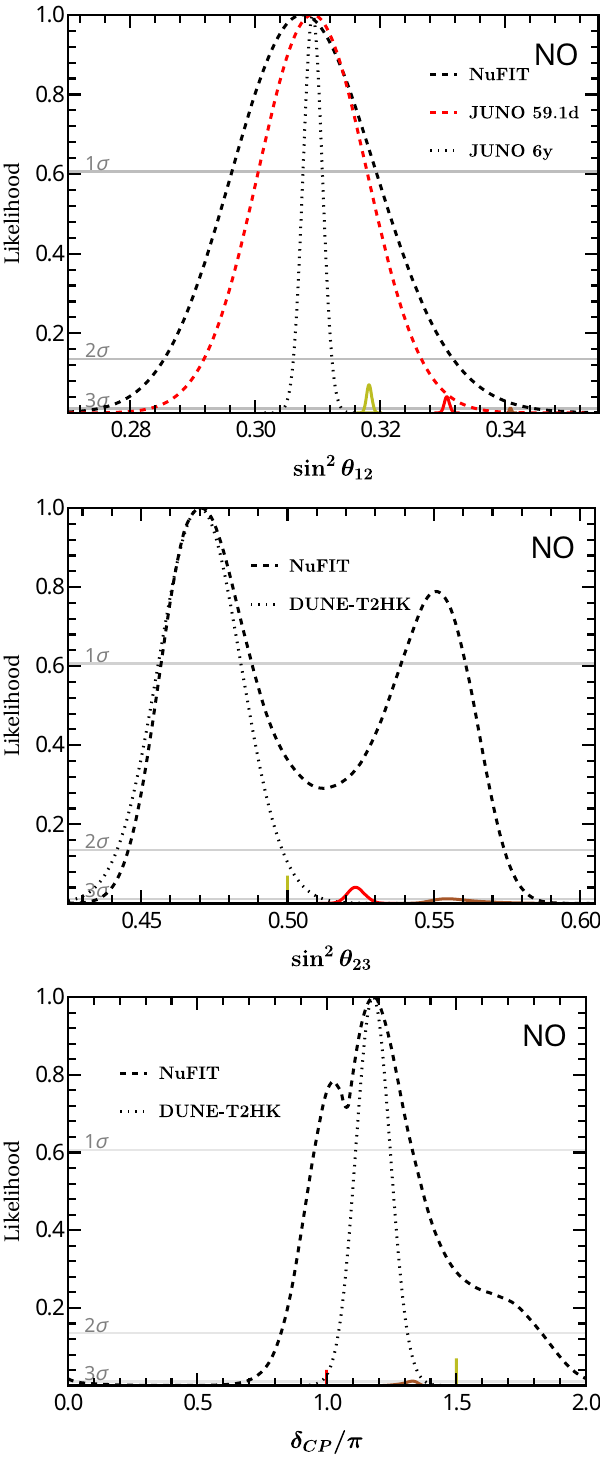}
\includegraphics[scale=0.75]{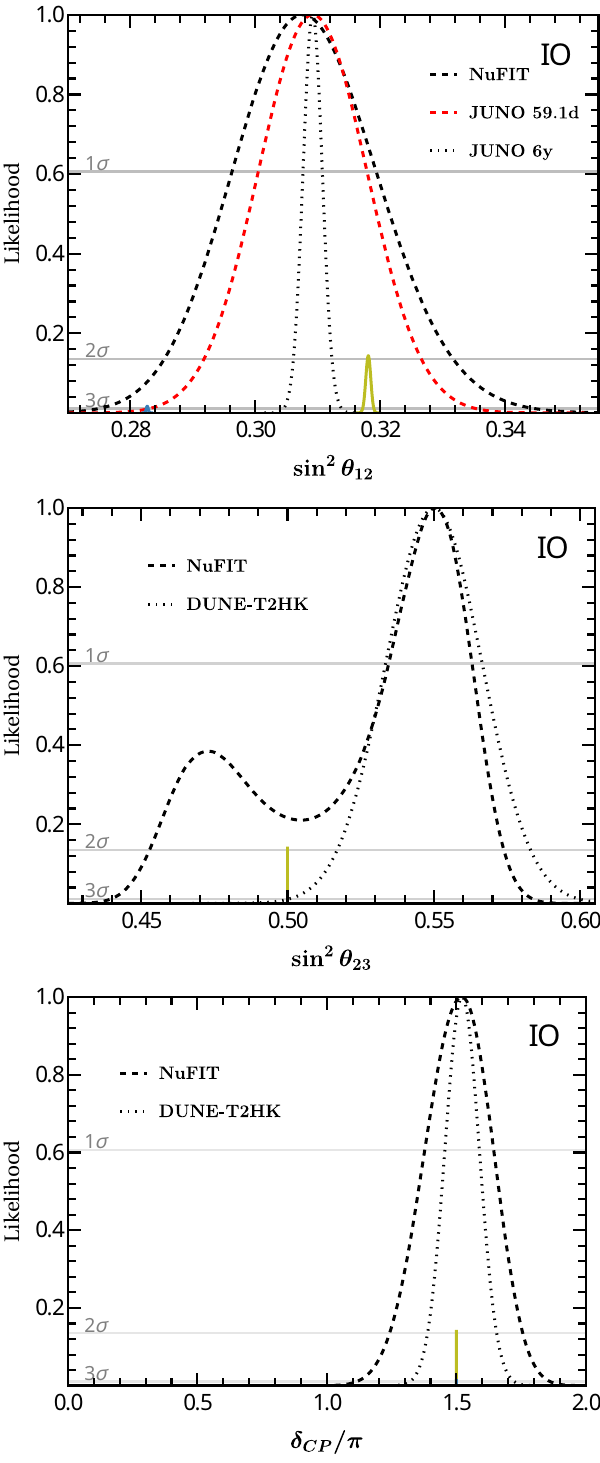}
\includegraphics[scale=0.75]{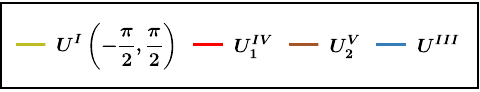}
\caption{\label{fig:one_para_likelihood_NO_IO}The predictions of likelihood profiles for $\sin^{2}\theta_{12}$, $\sin^{2}\theta_{23}$ and $\delta_{CP}$ obtained using the current global data on the neutrino mixing parameters for lepton mixing patterns with one parameter. The dashed curves denote the likelihoods obtained from the \texttt{NuFIT}~\cite{Esteban:2024eli}. In the $\sin^{2}\theta_{12}$ panel, the red dashed curve shows the likelihood from the recent JUNO 59.1-day data~\cite{JUNO:2025gmd}, while the dotted line indicates the projected $0.5\%$ precision expected after six years of JUNO operation~\cite{JUNO:2022mxj}. The dotted curves in the $\sin^{2}\theta_{23}$ and $\delta_{CP}$ panels correspond to the anticipated sensitivities of DUNE~\cite{DUNE:2020ypp} and T2HK~\cite{Hyper-Kamiokande:2018ofw}, assuming the current NuFit best-fit values together with prospective precisions of about $3\%$ for $\sin^{2}\theta_{23}$ and $12^\circ$ for $\delta_{CP}$. The solid colored lines represent the likelihoods predicted by the individual mixing patterns. The left (right) panels employ the one-dimensional projections $\chi^{2}_{i}(o_{i})$ for NO (IO).}
\end{figure}

To further assess the predictive power of the mixing patterns, we evaluate the likelihood profile defined in Eq.~\eqref{eq:likelihood} for $\sin^{2}\theta_{12}$, $\sin^{2}\theta_{23}$ and $\delta_{CP}$ across all one-parameter patterns with $\chi^{2}_{\min}\leq 9$, since models with $\chi^{2}_{\min}>9$ are disfavored at more than $3\sigma$. Under this criterion, only three patterns $U^{I}$ (TM1), $U^{IV}_{1}$ and $U^{V}_{1}$ satisfy $\chi^{2}_{\min}\leq 9$ in the NO case, while for IO only $U^{I}$ and $U^{III}$ remain viable, as summarized in table~\ref{tab:best-fit-onepara}. Figure~\ref{fig:one_para_likelihood_NO_IO} displays the corresponding likelihood profiles. The dashed curves denote the likelihoods for $\sin^{2}\theta_{12}$, $\sin^{2}\theta_{23}$ and $\delta_{CP}$ obtained from the \texttt{NuFIT} analysis. In the $\sin^{2}\theta_{12}$ panel, the red dashed curve shows the likelihood from the recent JUNO 59.1-day data, while the dotted line indicates the projected $0.5\%$ precision expected after six years of JUNO operation. The dotted curves in the $\sin^{2}\theta_{23}$ and $\delta_{CP}$ panels correspond to the anticipated sensitivities of DUNE and T2HK. The solid colored lines represent the likelihoods predicted by the individual mixing patterns. The left (right) panels employ the one-dimensional projections $\chi^{2}_{i}(o_{i})$ for NO (IO).

For $\sin^{2}\theta_{12}$, the predicted profiles are exceedingly narrow because $\sin^{2}\theta_{12}$ is tightly constrained by the small uncertainties in $\sin^{2}\theta_{13}$, as implied by Eqs.~\eqref{eq:th12th13varph1}, \eqref{eq:th12th13_III}, \eqref{eq:mixing_angles_correlations} and \eqref{eq:sum_rule_case_V} for the $U^{I}$,  $U^{III}$, $U^{IV}_{1}$ and $U^{V}_{1}$ patterns, respectively. In the NO case, the prediction of $U^{V}_{1}$ for $\sin^{2}\theta_{12}$ is disfavored by the latest JUNO data, whereas $U^{I}$ and $U^{IV}_{1}$ remain viable at $3\sigma$. In the IO case, $U^{III}$ is excluded by the current JUNO data, while $U^{I}$ is consistent with the data. If JUNO achieves the projected $0.5\%$ precision and the current best fit values would not change significantly, all these patterns could be tested in both NO and IO cases. These conclusions regarding $\sin^{2}\theta_{12}$ for the one-parameter mixing patterns are fully consistent with the earlier analysis based on the best-fit results. 

For $\sin^{2}\theta_{23}$, the patterns $U^{I}$ and $U^{III}$ predict $\sin^{2}\theta_{23}=0.5$, while $U^{IV}_{1}$ and $U^{V}_{1}$ yield sharply peaked predictions with $\sin^{2}\theta_{23}>0.5$, due to correlations with $\sin^{2}\theta_{13}$ encoded in Eqs.~\eqref{eq:mixing_angles_correlations} and \eqref{eq:sum_rule_case_V}. Taking the future projected sensitivities into account, the predicted ranges of $U^{IV}_{1}$ and $U^{V}_{1}$ could be probed in the NO case, while in the IO case both $U^{I}$ and $U^{III}$ could be tested. Regarding the Dirac phase, $U^{I}$ and $U^{III}$ predict $\delta_{CP}=1.5\pi$, while $U^{IV}_{1}$ predicts $\delta_{CP}=\pi$. i.e., absence of CP-violation in neutrino oscillations. In the NO case, the prediction $\delta_{CP}=1.5\pi$ lies outside the projected $3\sigma$ allowed range of the prospective DUNE and T2HK data, whereas $\delta_{CP}=\pi$ remains within it. In contrast, for the IO case, $\delta_{CP}=1.5\pi$ falls within the $3\sigma$ allowed interval.

In summary, the JUNO 59.1-day data imposes significant restrictions on the landscape of one-parameter mixing patterns. The $3\sigma$ interval of $\sin^{2}\theta_{12}$ from JUNO excludes several patterns that would otherwise remain compatible with the \texttt{NuFIT} global fit. We have focused on the flavour symmetry groups $A_{5}$, $\Sigma(168)$ and $D(6n^{2})$ with $n\leq 4$. For NO, we find that only $U^{I}$ (TM1), originating from $S_{4}\rtimes H_{CP}$, and $U^{IV}_{1,2}$, originating from $A_{5}\rtimes H_{CP}$, predict $\sin^{2}\theta_{12}$ values consistent with the current JUNO constraint. For IO, the viable space shrinks further, leaving $U^{I}$ (TM1) as the sole surviving option. Patterns such as $U^{II}$ (TM2), $U^{III}$ (GR2) and $U^{V}$ (TM2), although acceptable under \texttt{NuFIT} alone, become disfavored once the JUNO likelihood is taken into account. It should be emphasized that, for the mixing pattern $U^{I}$, additional phenomenologically viable textures—compatible with the current JUNO data—will emerge other than the TM1 form once higher values of $n$ in the group $\Delta(6n^{2})$ are considered. The future precision measurements of $\sin^{2}\theta_{12}$, $\sin^{2}\theta_{23}$, and $\delta_{CP}$ will provide additional discriminating power for testing these one-parameter mixing patterns.

\subsection{\label{sec:num_two}Lepton mixing matrix depending on two free parameters}

\begin{figure}[htbp]
    \centering
    \includegraphics[scale=0.55]{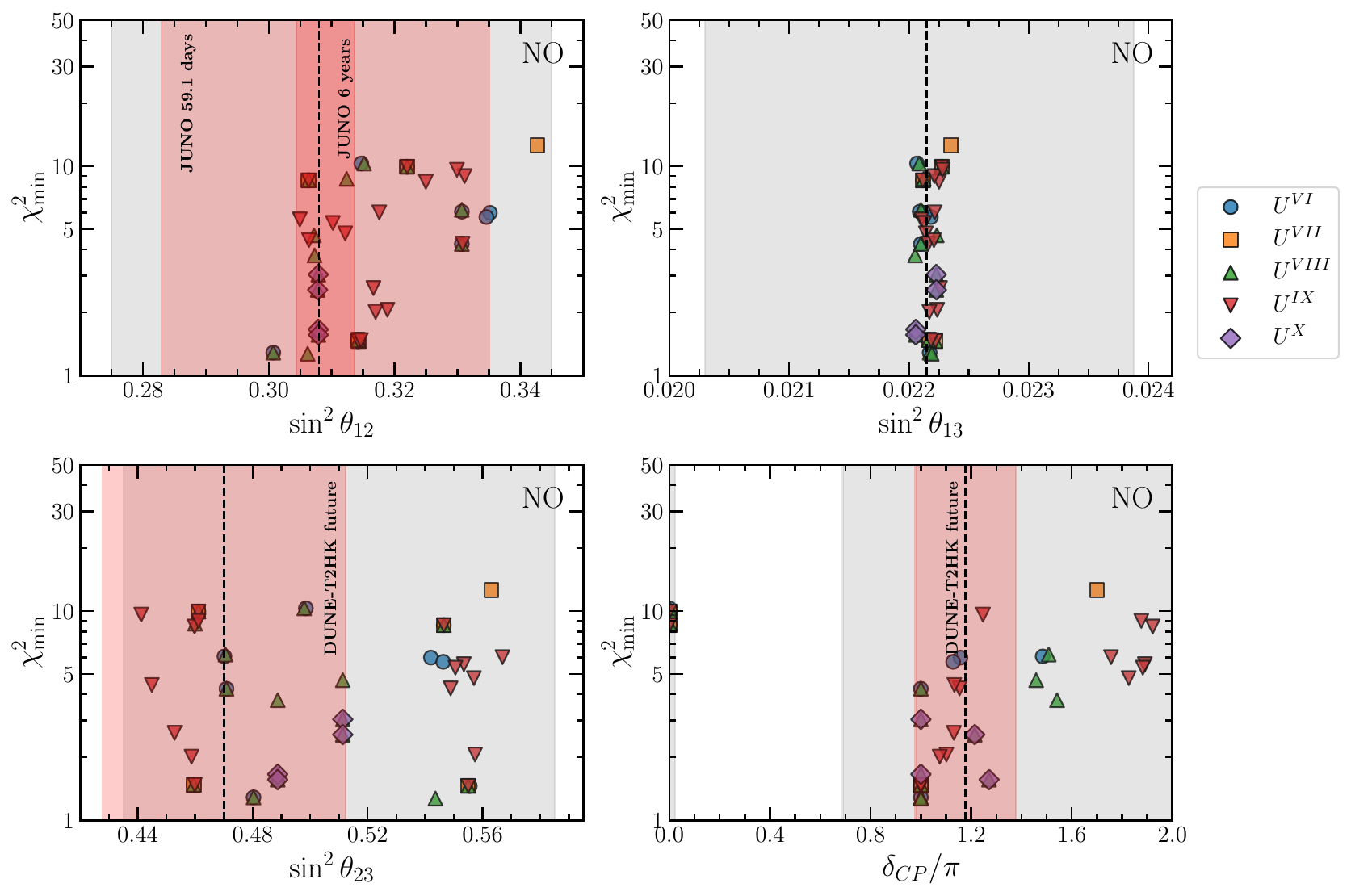}
    \includegraphics[scale=0.55]{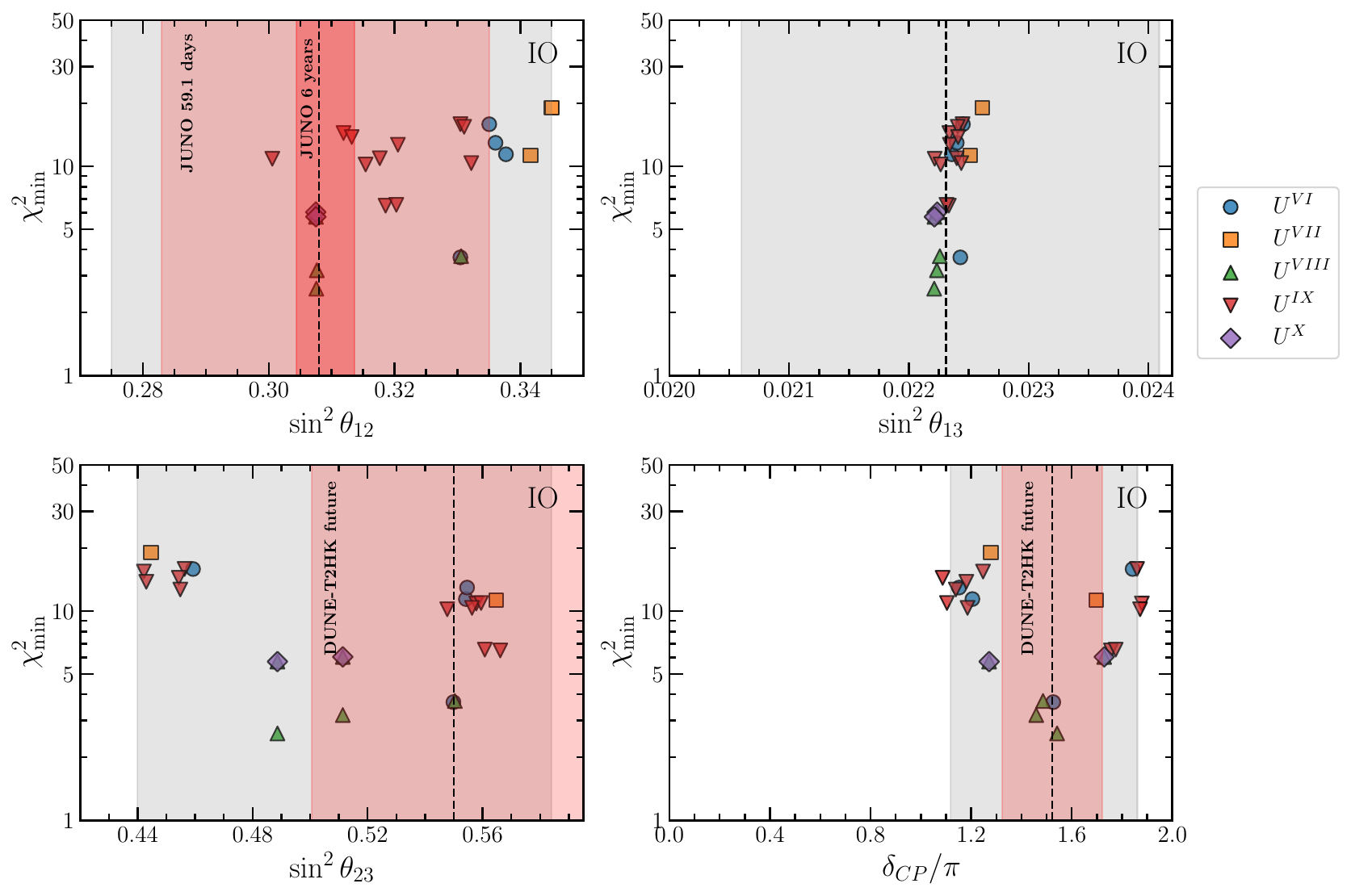}
    \caption{Best-fit predictions of the neutrino mixing angles $\theta_{12}$, $\theta_{13}$, $\theta_{23}$ and the Dirac CPV phase $\delta_{CP}$ for lepton mixing patterns with two parameter in case of NO and IO neutrino masses spectra. The remaining conventions follow those used in figure~\ref{fig:onepara_bfpoints_plots_NO_IO}.\label{fig:twopara_bfpoints_plots_NO_IO}}
\end{figure}

We now turn to discuss the numerical results for the two-parameter mixing patterns $U^{VI}$, $U^{VII}$, $U^{VIII}$, $U^{IX}$ and $U^{X}$. The best-fit values of the two real input parameters $(\theta_{\ell}^{\text{bf}},\theta_{\nu}^{\text{bf}})$, together with the corresponding predictions for the lepton mixing angles and CPV phases, are listed in table~\ref{tab:best-fit-twopara_A5_Sigma168} for $U^{VI}$ and $U^{VII}$, in table~\ref{tab:twopara_bf_CaseVIII_Delta6nsq} for $U^{VIII}$, and in table~\ref{tab:twopara_bf_CaseIX_X_Delta6nsq} for $U^{IX}$ and $U^{X}$. The resulting allowed regions of the lepton observables are summarized in table~\ref{tab:3sigma-twopara_A5_Sigma168}, table~\ref{tab:twopara_3sigma_CaseI_Delta6nsq}, and table~\ref{tab:twopara_3sigma_CaseII_Delta6nsq}, in Appendix~\ref{sec:3sigma-ranges}.
Unlike the one-parameter case, where only 8 (6) patterns are viable for NO (IO), the two-parameter constructions provide substantially greater flexibility: in total, 51 patterns for NO and 24 patterns for IO are compatible with the \texttt{NuFIT} data. We display the best-fit values of the three mixing angles and $\delta_{CP}$ for all viable two-parameter models for both NO and IO neutrino mass spectra in figure~\ref{fig:twopara_bfpoints_plots_NO_IO}.

\begin{figure}[!htbp]
    \centering
    \includegraphics[scale=0.75]{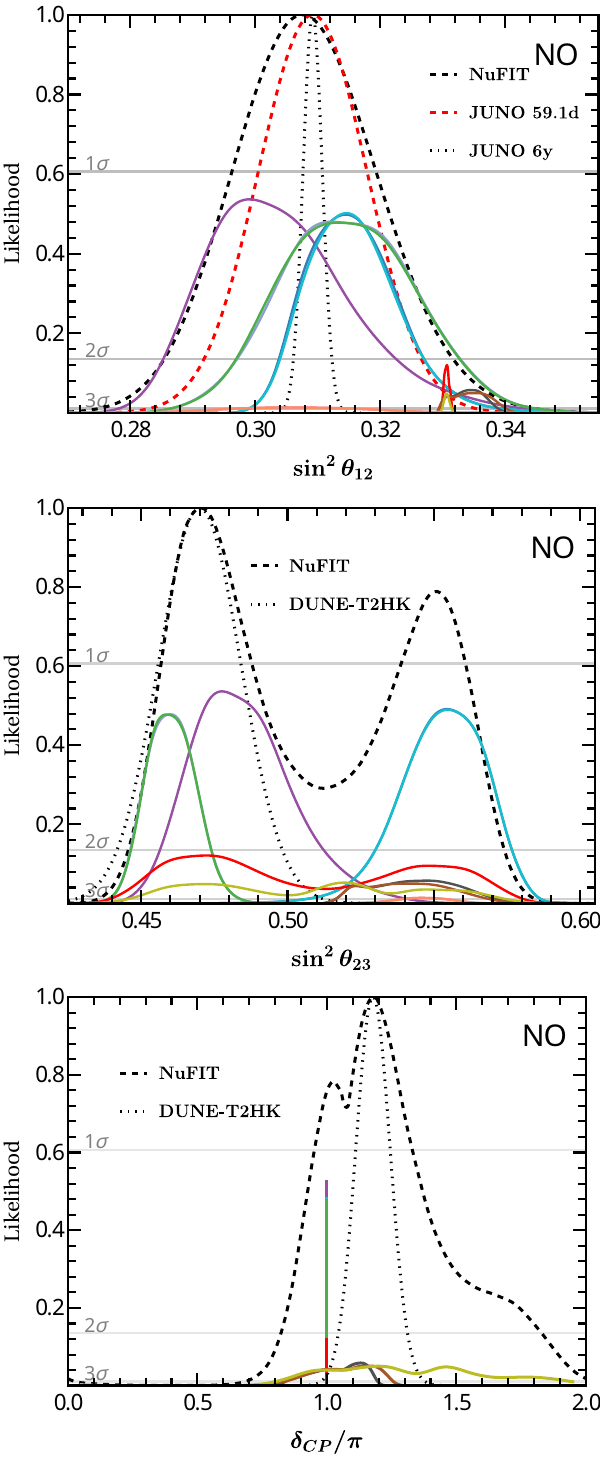}
    \includegraphics[scale=0.75]{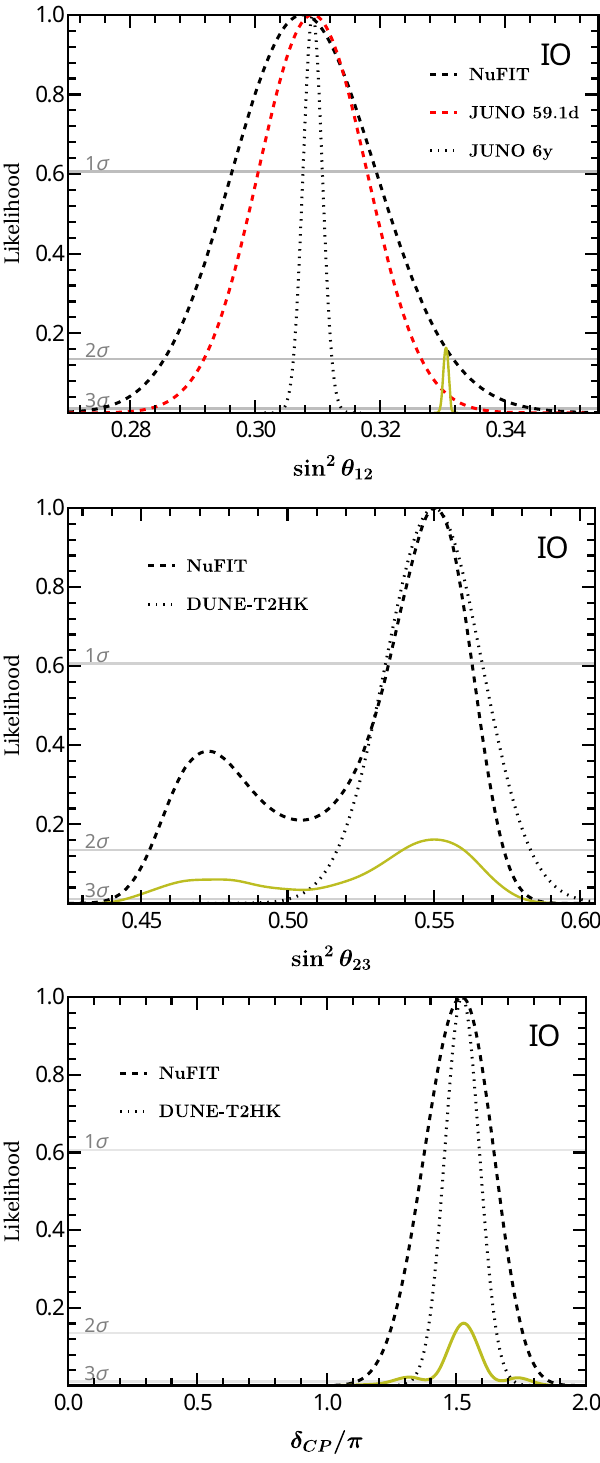}
    \includegraphics[scale=0.9]{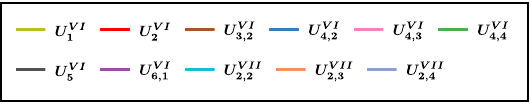}
    \caption{
    The predictions of likelihood profiles for $\sin^{2}\theta_{12}$, $\sin^{2}\theta_{23}$ and $\delta_{CP}$ obtained using the current global data on the neutrino mixing parameters for two-parameter mixing patterns that originate from $A_{5}\rtimes H_{CP}$ and $\Sigma(168)\rtimes H_{CP}$. The remaining conventions follow those used in figure~\ref{fig:one_para_likelihood_NO_IO}.}   \label{fig:two_para_A5_Sigma168}
\end{figure}

For the two-parameter mixing patterns derived from $A_{5}\rtimes H_{CP}$, we find 10 and 4 phenomenologically viable realizations of $U^{VI}$ for NO and IO, respectively, as listed in table~\ref{tab:best-fit-twopara_A5_Sigma168}. The patterns $U^{VI}_{2}$, $U^{VI}_{4,1}$, $U^{VI}_{4,2}$, $U^{VI}_{4,3}$, $U^{VI}_{4,4}$, $U^{VI}_{6,1}$ and $U^{VI}_{6,2}$ predict CP-conserving Dirac and Majorana phases, consistent with the general argument presented in section~\ref{sec:two_pars_from_A5}. Consequently, these patterns
are compatible with present data in the NO case, while they are disfavored for IO mass spectrum. By contrast, $U^{VI}_{1}$, $U^{VI}_{3,2}$ and $U^{VI}_{5}$ yield nontrivial CP-violating phases and can accommodate the current experimental data for both NO and IO. $U^{VI}_{3,1}$ can provide viable predictions only for the IO scenario. Under the sum rule constraints in Eq.~\eqref{eq:correlation_VI}, the allowed regions of the lepton observables given in table~\ref{tab:3sigma-twopara_A5_Sigma168} become narrower than the corresponding $3\sigma$ intervals in table~\ref{tab:lepton-data}. The likelihood profiles for $\sin^{2}\theta_{12}$, $\sin^{2}\theta_{23}$ and $\delta_{CP}$ are shown in figure~\ref{fig:two_para_A5_Sigma168} for selected viable patterns $U^{VI}$ with $\chi^{2}_{\min}\leq 9$. The patterns $U^{VI}_{4,1}$ and $U^{VI}_{6,2}$ for NO, as well as $U^{VI}_{3,1}$, $U^{VI}_{3,2}$ and $U^{VI}_{5}$ for IO, are not shown in the figure.

In the case of the predictions for $\sin^{2}\theta_{12}$, figure~\ref{fig:two_para_A5_Sigma168} shows that the patterns $U^{VI}_{4,2}$, $U^{VI}_{4,4}$, and $U^{VI}_{6,1}$ are consistent with the \texttt{NuFIT} data at the $2\sigma$ level, whereas $U^{VI}_{1}$, $U^{VI}_{2}$, $U^{VI}_{3,2}$, $U^{VI}_{4,3}$, and $U^{VI}_{5}$ remain compatible at $3\sigma$ for the NO scenario. In case of IO, only $U^{VI}_{1}$ is consistent at $2\sigma$. The patterns $U^{VI}_{1}$ and $U^{VI}_{2}$ yield very sharp predictions for $\sin^{2}\theta_{12}$, since $\sin^{2}\theta_{12}$ is correlated with $\sin^{2}\theta_{13}$, see Eq.~\eqref{eq:correlation_VI}. The predictions of the patterns $U^{VI}_{1}$, $U^{VI}_{2}$, $U^{VI}_{3,2}$, and $U^{VI}_{5}$ partially overlap with the $3\sigma$ range of the JUNO 59.1-day measurement, but they are disfavored at the $2\sigma$ level and are expected to be excluded once JUNO reaches its projected six-year precision of $0.5\%$. The remaining four cases $U^{VI}_{4,2}$, $U^{VI}_{4,3}$, $U^{VI}_{4,4}$ and $U^{VI}_{6,1}$ are consistent with both the current and projected JUNO sensitivities for both the NO and IO scenarios.

In what concerns the predictions for $\sin^{2}\theta_{23}$, the patterns $U^{VI}_{3,2}$, $U^{VI}_{4,2}$, $U^{VI}_{4,3}$ and $U^{VI}_{5}$ could be excluded by the projected DUNE and T2HK sensitivities in the NO case, as their predicted values of $\theta_{23}$ lie in second octant. For IO, the pattern $U^{VI}_{1}$  would remain consistent with the data with the anticipated experimental precision. Moreover, all patterns that predict CP-conserving Dirac phase can be tested by future measurements, whereas those yielding nontrivial CP-violating $\delta_{CP}$ would remain viable according to the considered projected sensitivities.

For the two-parameter mixing patterns originating from $\Sigma(168)\rtimes H_{CP}$, we identify 5 viable realizations for NO and 2 for IO, as summarized in table~\ref{tab:best-fit-twopara_A5_Sigma168}. In the NO case, given the fact that $U^{VII}_{2,i}$ is a real matrix (see Eq.~\eqref{eq:sigma_Sigma168}), the four patterns $U^{VII}_{2,i}$ with $i=1,2,3,4$ predict CP-conserving Dirac phase --- 0 by  $U^{VII}_{2,1}$ and $U^{VII}_{2,3}$ and and $\pi$ by $U^{VII}_{2,2}$ and $U^{VII}_{2,4}$. The pattern $U^{VII}_{1,1}$ instead yields CP-violating Dirac phase and it is compatible with global oscillation data for both mass orderings, whereas $U^{VII}_{1,2}$ is viable only for IO. Figure~\ref{fig:two_para_A5_Sigma168} displays the likelihood profiles of 3 patterns $U^{VII}_{2,2}$, $U^{VII}_{2,3}$ and $U^{VII}_{2,4}$. All of them predict $\sin^{2}\theta_{12}$ values lying within both the current JUNO 59.1-day and the projected six-year precision $3\sigma$ intervals. Their predictions for $\sin^{2}\theta_{23}$, however, show different levels of compatibility with future accelerator experiments sensitivities provided the current \texttt{NuFIT} best fit value including the SK data would not change drastically: while the $U^{VII}_{2,2}$ and $U^{VII}_{2,4}$ predictions fall outside the expected DUNE and T2HK ranges, the allowed region of $U^{VII}_{2,3}$ partially overlaps with the projected one. Since all these patterns predict CP-conserving Dirac phase, $\delta_{CP} = 0$ or $\pi$, they will be critically tested once DUNE and T2HK achieve their anticipated precision on $\delta_{CP}$.

\begin{table}[t!]
\centering
\renewcommand{\arraystretch}{1.1}
\scalebox{0.85}{\begin{tabular}{|c|c|c|c|c|c|c|c|c|c|c|}
\hline \hline
\multicolumn{11}{|c|}{$A_{5}\rtimes H_{CP}$} \\ \hline
\multirow{2}{*}{Order} & \multirow{2}{*}{Case} & \multirow{2}{*}{$\theta^{\text{bf}}_{\ell}/\pi$} & \multirow{2}{*}{$\theta^{\text{bf}}_{\nu}/\pi$} & \multirow{2}{*}{$\chi^2_{\text{min}}$}   & \multirow{2}{*}{$\sin^2\theta_{12}$} & \multirow{2}{*}{$\sin^2\theta_{13}$} & \multirow{2}{*}{$\sin^2\theta_{23}$}  & \multirow{2}{*}{$\delta_{CP}/\pi$} & $\alpha_{21}/\pi$ & $\alpha_{31}/\pi$ \\
& & & & & & & & & (mod 1) & (mod 1) \\ \hline
\multirow{10}{*}{NO} & $U^{VI}_{1}$ & $0.2623$ & $0.2424$ & $6.076$ & \
\cellcolor{red!30}$0.331$ & $0.02209$ & $0.470$ & $1.484$ & $0$ & \
$0.969$\\ \cline{2-11} & $U^{VI}_{2}$ & $0.6309$ & $0.2424$ & $4.258$ \
& \cellcolor{red!30}$0.331$ & $0.02210$ & $0.471$ & $1$ & $0$ \
& $0$\\ \cline{2-11} & $U^{VI}_{3,2}$ & $0.9717$ & $0.3973$ & \
$5.998$ & \cellcolor{red!30}$0.335$ & $0.02213$ & $0.542$ & $1.157$ & \
$0.082$ & $0.050$\\ \cline{2-11} & $U^{VI}_{4,1}$ & $0.0054$ & \
$0.3986$ & $9.952$ & \cellcolor{red!30}$0.322$ & $0.02227$ & $0.461$ \
& $0$ & $0$ & $0$\\ \cline{2-11} & \
$U^{VI}_{4,2}$ & $0.0107$ & $0.3915$ & $1.458$ & \
\cellcolor{red!30}$0.314$ & $0.02219$ & $0.556$ & $1$ & $0$ & \
$0$\\ \cline{2-11} & $U^{VI}_{4,3}$ & $0.3984$ & $0.7962$ & \
$8.575$ & \cellcolor{red!30}$0.306$ & $0.02212$ & $0.547$ & \
$0$ & $0$ & $0$\\ \cline{2-11} & \
$U^{VI}_{4,4}$ & $0.8372$ & $0.4333$ & $1.479$ & \
\cellcolor{red!30}$0.314$ & $0.02217$ & $0.459$ & $1$ & $0$ & \
$0$\\ \cline{2-11} & $U^{VI}_{5}$ & $0.4688$ & $0.1028$ & $5.725$ \
& \cellcolor{red!30}$0.335$ & $0.02218$ & $0.546$ & $1.128$ & $0.901$ \
& $0.927$\\ \cline{2-11} & $U^{VI}_{6,1}$ & $0.8407$ & $0.5677$ & \
$1.287$ & \cellcolor{red!30}$0.301$ & $0.02217$ & $0.480$ & $1$ & \
$0$ & $0$\\ \cline{2-11} & $U^{VI}_{6,2}$ & $0.5062$ & \
$0.0872$ & $10.326$ & \cellcolor{red!30}$0.315$ & $0.02207$ & $0.499$ \
& $0$ & $0$ & $0$\\ \cline{2-11}
  \hline \hline
\multirow{4}{*}{IO} & $U^{VI}_{1}$ & $0.2294$ & $0.2418$ & $3.672$ & \cellcolor{red!30}$0.330$ & $0.02243$ & $0.550$ & $1.526$ & $0$ & $0.052$\\ \cline{2-11} & $U^{VI}_{3,1}$ & $0.9714$ & $0.3977$ & $15.911$ & \cellcolor{red!30}$0.335$ & $0.02245$ & $0.459$ & $1.842^{*}$ & $0.917^{*}$ & $0.950^{*}$\\ \cline{2-11} & $U^{VI}_{3,2}$ & $0.0358$ & $0.3921$ & $11.428$ & $0.338$ & $0.02236$ & $0.554$ & $1.205^{*}$ & $0.102^{*}$ & $0.062^{*}$\\ \cline{2-11} & $U^{VI}_{5}$ & $0.5370$ & $0.1057$ & $12.969$ & $0.336$ & $0.02240$ & $0.555$ & $1.152^{*}$ & $0.882^{*}$ & $0.914^{*}$\\ \cline{2-11}  \hline \hline
\multicolumn{11}{|c|}{$\Sigma(168)\rtimes H_{CP}$} \\ \hline
\multirow{2}{*}{Order} & \multirow{2}{*}{Case} & \multirow{2}{*}{$\theta^{\text{bf}}_{\ell}/\pi$} & \multirow{2}{*}{$\theta^{\text{bf}}_{\nu}/\pi$} & \multirow{2}{*}{$\chi^2_{\text{min}}$}   & \multirow{2}{*}{$\sin^2\theta_{12}$} & \multirow{2}{*}{$\sin^2\theta_{13}$} & \multirow{2}{*}{$\sin^2\theta_{23}$}  & \multirow{2}{*}{$\delta_{CP}/\pi$} & $\alpha_{21}/\pi$ & $\alpha_{31}/\pi$ \\
& & & & & & & & & (mod 1) & (mod 1) \\ \hline
\multirow{5}{*}{NO}   & $U^{VII}_{1,1}$ & $0.7307$ & $0.3390$ & $12.607$ & $0.343$ & $0.02235$ & $0.563$ & $1.700^{*}$ & $0.598^{*}$ & $0.253^{*}$\\ \cline{2-11} & $U^{VII}_{2,1}$ & $0.8112$ & $0.2043$ & $9.952$ & \cellcolor{red!30}$0.322$ & $0.02227$ & $0.461$ & $0$ & $0$ & $0$\\ \cline{2-11} & $U^{VII}_{2,2}$ & $0.0275$ & $0.6463$ & $1.460$ & \cellcolor{red!30}$0.314$ & $0.02222$ & $0.555$ & $1$ & $0$ & $0$\\ \cline{2-11} & $U^{VII}_{2,3}$ & $0.6396$ & $0.2418$ & $8.575$ & \cellcolor{red!30}$0.306$ & $0.02212$ & $0.547$ & $0$ & $0$ & $0$\\ \cline{2-11} & $U^{VII}_{2,4}$ & $0.6430$ & $0.2390$ & $1.479$ & \cellcolor{red!30}$0.314$ & $0.02217$ & $0.459$ & $1$ & $0$ & $0$\\ \cline{2-11}
  \hline \hline
\multirow{2}{*}{IO} & $U^{VII}_{1,1}$ & $0.7302$ & $0.3399$ & $11.276$ & $0.342$ & $0.02251$ & $0.565$ & $1.697^{*}$ & $0.596^{*}$ & $0.252^{*}$\\ \cline{2-11} & $U^{VII}_{1,2}$ & $0.2682$ & $0.6653$ & $19.115$ & $0.345$ & $0.02261$ & $0.445$ & $1.278^{*}$ & $0.399^{*}$ & $0.744^{*}$\\ \cline{2-11}\hline \hline
\end{tabular}}
\caption{\label{tab:best-fit-twopara_A5_Sigma168} The best-fit values of the free parameter and the corresponding lepton mixing observables for two-parameter mixing patterns from the $A_{5}\rtimes H_{CP}$ and $\Sigma(168)\rtimes H_{CP}$ for both normal and inverted neutrino mass orderings. Here $\chi^{2}_{\text{min}}$ the global minimum of the $\chi^{2}$ function occurs at $\left(\theta_{\ell},\theta_{\nu}\right)=\left(\theta^{\text{bf}}_{\ell},\theta^{\text{bf}}_{\nu}\right)$. The remaining conventions follow those used in table~\ref{tab:best-fit-onepara}.}
\end{table}

There are three distinct two-parameter mixing patterns that can be derived from the group $\Delta(6n^{2})\rtimes H_{CP}$, as discussed in section~\ref{sec:two_pars_from_D6n2}. We perform a full $\chi^{2}$ analysis of these three cases for $\Delta(6n^{2})$ with $n\leq 4$. For the mixing pattern $U^{VIII}$, the best-fit values of the two real input parameters and the corresponding predictions for the lepton mixing observables are summarized in table~\ref{tab:twopara_bf_CaseVIII_Delta6nsq}, while the resulting allowed regions of these observables are presented in table~\ref{tab:twopara_3sigma_CaseI_Delta6nsq}.
Taking into account permutations and inequivalent choices of $(\varphi_{1},\varphi_{2})$, we find 10 and 4 phenomenologically viable realizations for the NO and IO mass spectra, respectively. For NO, the viable patterns include $U^{VIII}_{i}$ with $i=4,5,6,7,8,9$. In particular, $U^{VIII}_{4,5,7,8}$ arise from $\Delta(54)$ with $(\varphi_{1},\varphi_{2})=(\frac{\pi}{3},0)$, while $U^{VIII}_{6}$ and $U^{VIII}_{9}$ arise from $\Delta(96)$  with $(\varphi_{1},\varphi_{2})=(\frac{\pi}{4},0)$, $(\frac{\pi}{4},\frac{\pi}{4})$, and $(\frac{\pi}{4},\frac{\pi}{2})$. We note that for $\varphi_{2}=0$, the mixing matrix in Eq.~\eqref{eq:PMNS_caseVIII} becomes real, implying CP-conserving Dirac and Majorana phases. For IO, only $U^{VIII}_{6}$ and $U^{VIII}_{9}$ remain viable, both obtained from $\Delta(96)$ with $(\varphi_{1},\varphi_{2})=(\frac{\pi}{4},\frac{\pi}{4})$ or $(\frac{\pi}{4},\frac{\pi}{2})$. Thus the minimal $\Delta(6n^{2})$ group that can generate viable mixing pattern $U^{VIII}$ is $\Delta(54)$ with CP-conserving phases, and $\Delta(96)$  with CP-violating phases.
Since the mixing pattern $U^{VIII}$ can also originate from $D_{n}\rtimes H_{CP}$ and the group order of $D_{n}$, namely $2n$, is smaller than that of $\Delta(6n^{2})$, we further scan the $D_{n}$ series up to $n\leq 8$. Beyond the viable solutions already found for $\Delta(6n^{2})$ with $n=3,4$ and $\varphi_{2}=0,\frac{\pi}{2}$ that can also be obtained from $D_{n}$, we identify an additional 6 viable cases for NO and 1 for IO, as listed in table~\ref{tab:twopara_bf_CaseVIII_Delta6nsq}. The extra viable NO patterns include $U^{VIII}_{1}$ with $(\varphi_{1},\varphi_{2})=(\frac{\pi}{5},0)$ or $(\frac{\pi}{5},\frac{\pi}{2})$, $U^{VIII}_{4}$ and $U^{VIII}_{7}$ with $(\varphi_{1},\varphi_{2})=(\frac{2\pi}{5},0)$, and $U^{VIII}_{5}$ and $U^{VIII}_{8}$ with $(\varphi_{1},\varphi_{2})=(\frac{2\pi}{7},0)$. For IO, the additional viable case is $U^{VIII}_{1}$ with $(\varphi_{1},\varphi_{2})=(\frac{\pi}{5},\frac{\pi}{2})$. For the case of $D_{n}\rtimes H_{CP}$, the smallest $D_{n}$ group capable of producing a viable $U^{VIII}$ mixing pattern is $D_{3}$, which predicts CP-conserving phases, whereas $D_{4}$ gives rise to nontrivial CP-violating phases, with $\delta_{CP}$ having values close to $3\pi/2$.

\begin{table}[t!]
\centering
\scalebox{0.8}{\begin{tabular}{|c|c|c|c|c|c|c|c|c|c|c|c|c|}
\hline \hline
\multicolumn{13}{|c|}{$\Delta(6n^{2})\rtimes H_{CP}$} \\ \hline
\multirow{2}{*}{Order} & \multirow{2}{*}{Case} & \multirow{2}{*}{$\varphi_1$} & \multirow{2}{*}{$\varphi_2$} & \multirow{2}{*}{$\theta^{\text{bf}}_{\ell}/\pi$} & \multirow{2}{*}{$\theta^{\text{bf}}_{\nu}/\pi$} & \multirow{2}{*}{$\chi^2_{\text{min}}$}   & \multirow{2}{*}{$\sin^2\theta_{12}$} & \multirow{2}{*}{$\sin^2\theta_{13}$} & \multirow{2}{*}{$\sin^2\theta_{23}$}  & \multirow{2}{*}{$\delta_{CP}/\pi$} & $\alpha_{21}/\pi$ & $\alpha_{31}/\pi $   \\
&&&&&&&&&&& (mod 1) & (mod 1)\\ \hline
\multirow{10}{*}{NO} & \multirow{1}{*}{$U^{VIII}_{4}$ }  & $\frac{\pi }{3}$ & $0$ & $0.3893$ & $0.2176$ & $9.952$ & \cellcolor{red!30}$0.322$ & $0.02227$ & $0.461$ & $0$ & $0$ & $0$\\ \cline{2-13}& \multirow{1}{*}{$U^{VIII}_{5}$ }  & $\frac{\pi }{3}$ & $0$  & $0.2178$ & $0.1801$ & $8.575$ & \cellcolor{red!30}$0.306$ & $0.02212$ & $0.547$ & $0$ & $0$ & $0$\\ \cline{2-13}& \multirow{3}{*}{$U^{VIII}_{6}$ }  & $\frac{\pi }{4}$ & $0$  & $0.0676$ & $0.3609$ & $3.040$ & \cellcolor{red!30}$0.308$ & $0.02223$ & $0.511$ & $1$ & $0$ & $0$\\ \cline{3-13}& & $\frac{\pi }{4}$ & $\frac{\pi }{4}$  & $0.0676$ & $0.3489$ & $2.569$ & \cellcolor{red!30}$0.308$ & $0.02223$ & $0.511$ & $1.214$ & $0.853$ & $0.891$\\ \cline{3-13}& & $\frac{\pi }{4}$ & $\frac{\pi }{2}$  & $0.0676$ & $0.6838$ & $4.680$ & \cellcolor{red!30}$0.307$ & $0.02223$ & $0.511$ & $1.458^{*}$ & $0.791^{*}$ & $0.854^{*}$\\ \cline{2-13}& \multirow{1}{*}{$U^{VIII}_{7}$ }  & $\frac{\pi }{3}$ & $0$  & $0.6056$ & $0.7756$ & $1.460$ & \cellcolor{red!30}$0.314$ & $0.02222$ & $0.555$ & $1$ & $0$ & $0$\\ \cline{2-13}& \multirow{1}{*}{$U^{VIII}_{8}$ }  & $\frac{\pi }{3}$ & $0$  & $0.7789$ & $0.8171$ & $1.479$ & \cellcolor{red!30}$0.314$ & $0.02217$ & $0.459$ & $1$ & $0$ & $0$\\ \cline{2-13}& \multirow{3}{*}{$U^{VIII}_{9}$ }  & $\frac{\pi }{4}$ & $0$  & $0.9326$ & $0.2648$ & $1.660$ & \cellcolor{red!30}$0.308$ & $0.02206$ & $0.489$ & $1$ & $0$ & $0$\\ \cline{3-13}& & $\frac{\pi }{4}$ & $\frac{\pi }{4}$  & $0.9326$ & $0.2798$ & $1.564$ & \cellcolor{red!30}$0.308$ & $0.02206$ & $0.489$ & $1.271$ & $0.146$ & $0.094$\\ \cline{3-13}& & $\frac{\pi }{4}$ & $\frac{\pi }{2}$  & $0.9327$ & $0.3161$ & $3.751$ & \cellcolor{red!30}$0.307$ & $0.02205$ & $0.489$ & $1.541$ & $0.209$ & $0.146$\\ \cline{2-13}
  \hline \hline
\multirow{4}{*}{IO}& \multirow{2}{*}{$U^{VIII}_{6}$ }  & $\frac{\pi }{4}$ & $\frac{\pi }{4}$  & $0.9324$ & $0.2798$ & $6.045$ & \cellcolor{red!30}$0.307$ & $0.02224$ & $0.511$ & $1.729^{*}$ & $0.853^{*}$ & $0.905^{*}$\\ \cline{3-13}& & $\frac{\pi }{4}$ & $\frac{\pi }{2}$  & $0.9324$ & $0.6840$ & $3.186$ & \cellcolor{red!30}$0.308$ & $0.02223$ & $0.511$ & $1.458$ & $0.791$ & $0.854$\\ \cline{2-13}& \multirow{2}{*}{$U^{VIII}_{9}$ }  & $\frac{\pi }{4}$ & $\frac{\pi }{4}$  & $0.9324$ & $0.2799$ & $5.740$ & \cellcolor{red!30}$0.307$ & $0.02221$ & $0.489$ & $1.271$ & $0.147$ & $0.095$\\ \cline{3-13}& & $\frac{\pi }{4}$ & $\frac{\pi }{2}$  & $0.9324$ & $0.3160$ & $2.602$ & \cellcolor{red!30}$0.308$ & $0.02221$ & $0.489$ & $1.542$ & $0.209$ & $0.146$\\ \cline{2-13}
\hline \hline
\multicolumn{13}{|c|}{$D_{n}\rtimes H_{CP}$} \\ \hline
\multirow{2}{*}{Order} & \multirow{2}{*}{Case} & \multirow{2}{*}{$\varphi_1$} & \multirow{2}{*}{$\varphi_2$} & \multirow{2}{*}{$\theta^{\text{bf}}_{\ell}/\pi$} & \multirow{2}{*}{$\theta^{\text{bf}}_{\nu}/\pi$} & \multirow{2}{*}{$\chi^2_{\text{min}}$}   & \multirow{2}{*}{$\sin^2\theta_{12}$} & \multirow{2}{*}{$\sin^2\theta_{13}$} & \multirow{2}{*}{$\sin^2\theta_{23}$}  & \multirow{2}{*}{$\delta_{CP}/\pi$} & $\alpha_{21}/\pi$ & $\alpha_{31}/\pi $   \\
&&&&&&&&&&& (mod 1) & (mod 1)\\ \hline
\multirow{6}{*}{NO} & \multirow{2}{*}{$U^{VIII}_{1}$ }  & $\frac{\pi }{5}$ & $0$ & $0.1929$ & $0.4186$ & $4.258$ & \cellcolor{red!30}$0.331$ & $0.02210$ & $0.471$ & $1$ & $0$ & $0$\\ \cline{3-13}& & $\frac{\pi }{5}$ & $\frac{\pi }{2}$  & $0.7398$ & $0.4186$ & $6.204$ & \cellcolor{red!30}$0.331$ & $0.02210$ & $0.471$ & $1.508^{*}$ & $0^{*}$ & $0.017^{*}$\\ \cline{2-13}& \multirow{1}{*}{$U^{VIII}_{4}$ }  & $\frac{2\pi }{5}$ & $0$ & $0.6645$ & $0.7561$ & $1.287$ & \cellcolor{red!30}$0.301$ & $0.02217$ & $0.480$ & $1$ & $0$ & $0$\\ \cline{2-13}& \multirow{1}{*}{$U^{VIII}_{5}$ }  & $\frac{2\pi }{7}$ & $0$ & $0.2467$ & $0.1176$ & $1.271$ & \cellcolor{red!30}$0.306$ & $0.02219$ & $0.544$ & $1$ & $0$ & $0$\\ \cline{2-13}& \multirow{1}{*}{$U^{VIII}_{7}$ }  & $\frac{2\pi }{5}$ & $0$ & $0.6702$ & $0.7636$ & $10.328$ & \cellcolor{red!30}$0.315$ & $0.02209$ & $0.498$ & $0$ & $0$ & $0$\\ \cline{2-13}& \multirow{1}{*}{$U^{VIII}_{8}$ }  & $\frac{2\pi }{7}$ & $0$ & $0.2499$ & $0.1203$ & $8.719$ & \cellcolor{red!30}$0.312$ & $0.02212$ & $0.460$ & $0$ & $0$ & $0$\\ \cline{2-13}
  \hline \hline
\multirow{1}{*}{IO}& \multirow{1}{*}{$U^{VIII}_{1}$ }  & $\frac{\pi }{5}$ & $\frac{\pi }{2}$ & $0.7676$ & $0.4183$ & $3.721$ & \cellcolor{red!30}$0.331$ & $0.02226$ & $0.550$ & $1.486^{*}$ & $0^{*}$ & $0.971^{*}$\\ \cline{2-13}\cline{2-13}
\hline \hline
\end{tabular}}
\caption{\label{tab:twopara_bf_CaseVIII_Delta6nsq}
The best-fit values of the free parameter and the corresponding lepton mixing observables for two-parameter mixing patterns
$U^{VIII}$ originate from $\Delta(6n^{2})\rtimes H_{CP}$ with $n\leq 4$ and from $D_{n}\rtimes H_{CP}$ with $n\leq 8$. The discrete parameters $(\varphi_{1},\varphi_{2})$ specify the residual symmetries associated with flavour groups $\Delta(6n^{2})$ and $D_{n}$.
Here $\chi^{2}_{\text{min}}$ the global minimum of the $\chi^{2}$ function occurs at $\left(\theta_{\ell},\theta_{\nu}\right)=\left(\theta^{\text{bf}}_{\ell},\theta^{\text{bf}}_{\nu}\right)$. The remaining conventions follow those used in table~\ref{tab:best-fit-onepara}.}
\end{table}

\begin{figure}[htbp]
    \centering
    \includegraphics[scale=0.75]{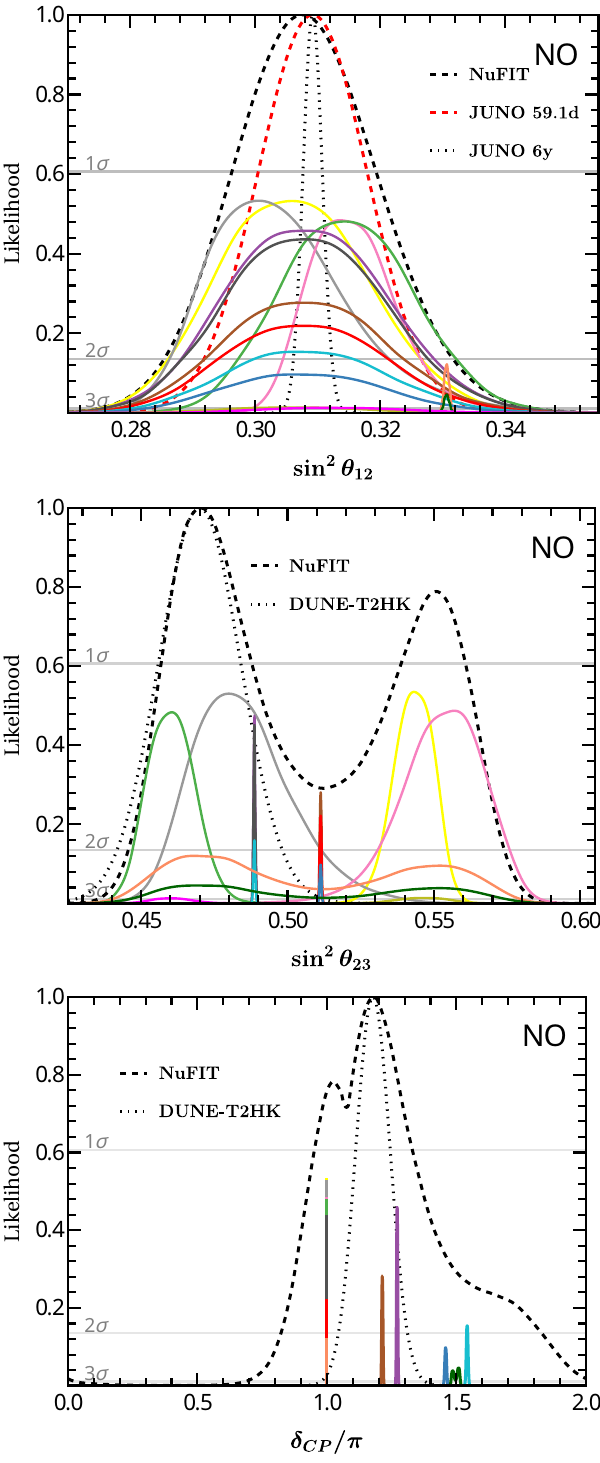}
    \includegraphics[scale=0.75]{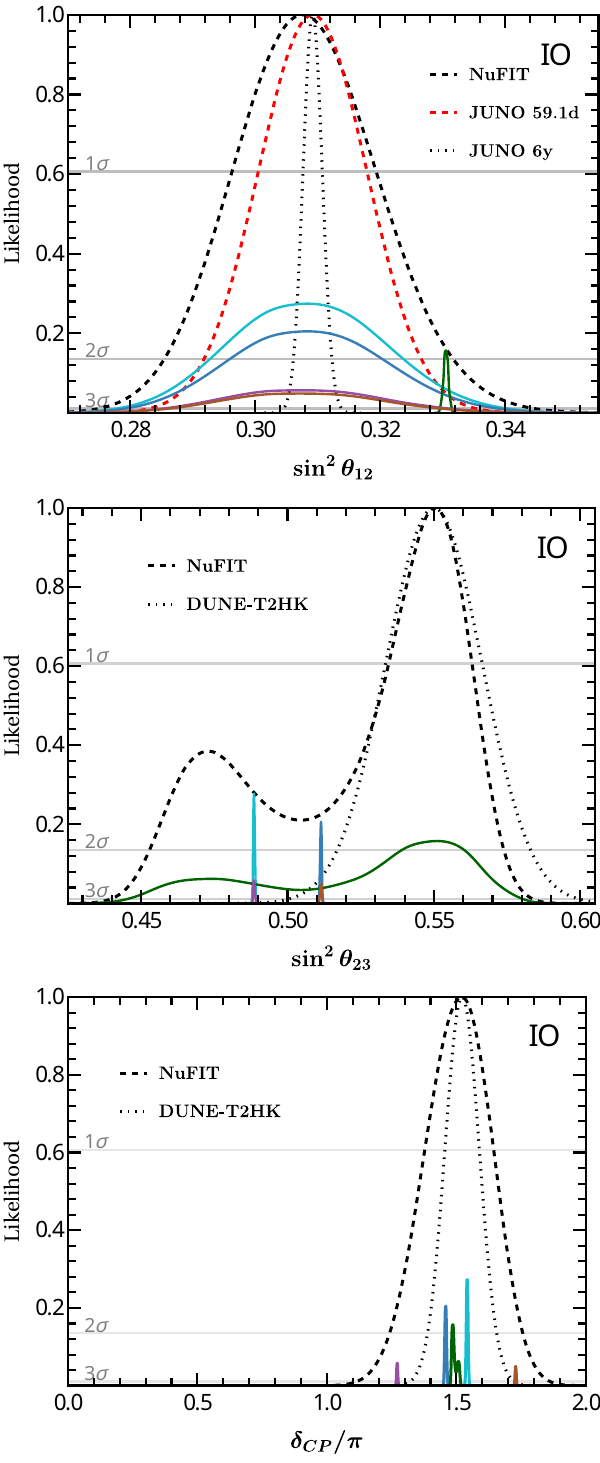}
    \includegraphics[scale=0.75]{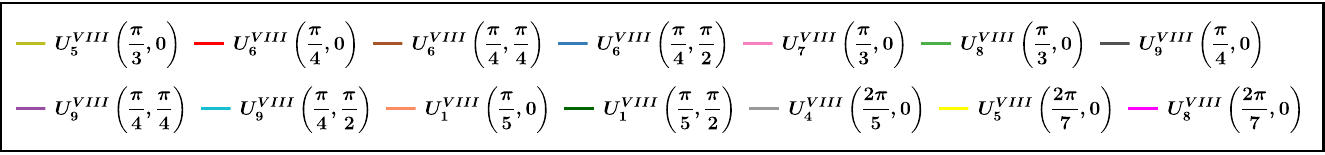}
    \caption{The predictions of likelihood profiles for $\sin^{2}\theta_{12}$, $\sin^{2}\theta_{23}$ and $\delta_{CP}$ obtained using the current global data on the neutrino mixing parameters for two-parameter mixing patterns $U^{VIII}$ that originate from $\Delta(6n^{2})\rtimes H_{CP}$ and $D_{n}\rtimes H_{CP}$. The remaining conventions follow those used in figure~\ref{fig:one_para_likelihood_NO_IO}.}
    \label{fig:two_para_VIII}
\end{figure}

We show in figure~\ref{fig:two_para_VIII} the likelihood profiles of the lepton observables for all viable $U^{VIII}$ mixing patterns with $\chi^{2}_{\text{min}}\leq 9$. Regarding $\sin^{2}\theta_{12}$, we find that the pattern $U^{VIII}_{1}$ exhibits a very narrow allowed region. This behaviour originates from the correlation between $\sin^{2}\theta_{12}$ and $\sin^{2}\theta_{13}$ in Eq.~\eqref{eq:correlation_VIII}. For $\varphi_{1}=\frac{\pi}{5}$, the value of $\sin^{2}\theta_{12}$ is tightly restricted around $0.331$. Consequently, the pattern $U^{VIII}_{1}$ is disfavoured by the current JUNO 59.1-day data at the $2\sigma$ level and may be excluded by JUNO  given its projected six-year precision. In contrast, the predictions for $\sin^{2}\theta_{12}$ in the other viable patterns remain compatible with both the present and future prospective JUNO precision data. 

For $U^{VIII}_{6}$ and $U^{VIII}_{9}$, the value of $\sin^{2}\theta_{23}$ is also determined by $\sin^{2}\theta_{13}$ via Eq.~\eqref{eq:correlation_VIII}, leading to very narrow likelihood profiles. We suppose in the discussion that the current \texttt{NuFIT} best fit values of $\sin^{2}\theta_{23}$ would not change significantly. We observe that $U^{VIII}_{4}$, $U^{VIII}_{8}$ and $U^{VIII}_{9}$ predict $\sin^{2}\theta_{23}<0.5$, which lies within the assumed future $3\sigma$ sensitivity of DUNE and T2HK for NO. In contrast, the patterns $U^{VIII}_{7}$, $U^{VIII}_{5}$ and $U^{VIII}_{6}$ which are obtained from $U^{VIII}_{4}$, $U^{VIII}_{8}$ and $U^{VIII}_{9}$ by exchanging the second and third rows—predict $\sin^{2}\theta_{23}>0.5$, and could therefore be incompatible with considered future DUNE and T2HK high precision data in the NO case. For IO, the predicted $\sin^{2}\theta_{23}$ of $U^{VIII}_{9}$ can be tested by the future DUNE and T2HK data while that of $U^{VIII}_{6}$ remains within the projected $3\sigma$ ranges. 

The Dirac CPV phase $\delta_{CP}$ is strongly constrained in the $U^{VIII}$ patterns. For NO, only $U^{VIII}_{6}$ and $U^{VIII}_{9}$ with $(\varphi_{1},\varphi_{2})=(\frac{\pi}{4},\frac{\pi}{4})$  might survive the future high precision tests of their predictions for $\delta_{CP}$. For IO, the patterns that could be compatible with the future DUNE and T2HK high precision data on $\delta_{CP}$ are $U^{VIII}_{6}$ and $U^{VIII}_{9}$ with $(\varphi_{1},\varphi_{2})=(\frac{\pi}{4},\frac{\pi}{2})$, as well as $U^{VIII}_{1}$ with $(\varphi_{1},\varphi_{2})=(\frac{\pi}{5},\frac{\pi}{2})$. The preceding discussion should be considered as indicative only since at present it is impossible to foresee the results of the experimental determination of the Dirac CPV phase $\delta_{CP}$.

\begin{table}[!t]
\centering
\scalebox{0.8}{\begin{tabular}{|c|c|c|c|c|c|c|c|c|c|c|c|c|}
\hline \hline
\multicolumn{13}{|c|}{$\Delta(6n^{2})\rtimes H_{CP}$ (Case IX)} \\ \hline
\multirow{2}{*}{Order} & \multirow{2}{*}{Case} & \multirow{2}{*}{$\varphi_1$} & \multirow{2}{*}{$\varphi_2$} & \multirow{2}{*}{$\theta^{\text{bf}}_{\ell}/\pi$} & \multirow{2}{*}{$\theta^{\text{bf}}_{\nu}/\pi$} & \multirow{2}{*}{$\chi^2_{\text{min}}$}   & \multirow{2}{*}{$\sin^2\theta_{12}$} & \multirow{2}{*}{$\sin^2\theta_{13}$} & \multirow{2}{*}{$\sin^2\theta_{23}$}  & \multirow{2}{*}{$\delta_{CP}/\pi$} & $\alpha_{21}/\pi$ & $\alpha_{31}/\pi $   \\
&&&&&&&&&&& (mod 1) & (mod 1)\\
\hline
\multirow{16}{*}{NO} & \multirow{3}{*}{$U^{IX}_{1}$ }  & $0$ & $0$ & $0.0852$ & $0.0217$ & $9.952$ & \cellcolor{red!30}$0.322$ & $0.02227$ & $0.461$ & $0$ & $0$ & $0$\\ \cline{3-13}& & $0$ & $\frac{\pi }{4}$  & $0.0833$ & $0.0308$ & $8.442$ & \cellcolor{red!30}$0.325$ & $0.02225$ & $0.460$ & $1.922$ & $0.014$ & $0.507$\\ \cline{3-13}  & & $\frac{\pi }{4}$ & $\frac{\pi }{2}$  & $0.1666$ & $0.1461$ & $8.981$ & \cellcolor{red!30}$0.331$ & $0.02221$ & $0.461$ & $1.876^{*}$ & $0.546^{*}$ & $0.746^{*}$\\ \cline{2-13}& \multirow{5}{*}{$U^{IX}_{2}$ }  & $0$ & $0$  & $0.4782$ & $0.6240$ & $8.576$ & \cellcolor{red!30}$0.306$ & $0.02212$ & $0.547$ & $0$ & $0$ & $0$\\ \cline{3-13}& & $0$ & $\frac{\pi }{4}$  & $0.9131$ & $0.9724$ & $5.573$ & \cellcolor{red!30}$0.305$ & $0.02210$ & $0.553$ & $1.891$ & $0.991$ & $0.491$\\ \cline{3-13}  & & $\frac{\pi }{4}$ & $0$  & $0.4675$ & $0.6258$ & $5.365$ & \cellcolor{red!30}$0.310$ & $0.02213$ & $0.551$ & $1.882^{*}$ & $0.926^{*}$ & $0.954^{*}$\\ \cline{3-13} & & $\frac{\pi }{4}$ & $\frac{\pi }{4}$  & $0.8892$ & $0.9629$ & $6.028$ & \cellcolor{red!30}$0.318$ & $0.02221$ & $0.567$ & $1.756^{*}$ & $0.767^{*}$ & $0.358^{*}$\\ \cline{3-13} & & $\frac{\pi }{2}$ & $\frac{\pi }{4}$  & $0.6558$ & $0.6776$ & $4.784$ & \cellcolor{red!30}$0.312$ & $0.02214$ & $0.557$ & $1.827$ & $0.121$ & $0.395$\\ \cline{2-13}& \multirow{3}{*}{$U^{IX}_{3}$ }  & $0$ & $0$ & $0.3015$ & $0.5797$ & $1.460$ & \cellcolor{red!30}$0.314$ & $0.02222$ & $0.555$ & $1$ & $0$ & $0$\\ \cline{3-13}& & $0$ & $\frac{\pi }{4}$  & $0.0873$ & $0.0407$ & $2.062$ & \cellcolor{red!30}$0.319$ & $0.02224$ & $0.557$ & $1.101^{*}$ & $0.980^{*}$ & $0.489^{*}$\\ \cline{3-13}  & & $\frac{\pi }{4}$ & $\frac{\pi }{2}$  & $0.1816$ & $0.1613$ & $4.269$ & \cellcolor{red!30}$0.331$ & $0.02216$ & $0.549$ & $1.154$ & $0.503$ & $0.287$\\ \cline{2-13}& \multirow{5}{*}{$U^{IX}_{4}$ }  & $0$ & $0$  & $0.4746$ & $0.6210$ & $1.477$ & \cellcolor{red!30}$0.315$ & $0.02219$ & $0.460$ & $1$ & $0$ & $0$\\ \cline{3-13}& & $0$ & $\frac{\pi }{4}$  & $0.9182$ & $0.9808$ & $2.014$ & \cellcolor{red!30}$0.317$ & $0.02217$ & $0.459$ & $1.075^{*}$ & $0.006^{*}$ & $0.506^{*}$\\ \cline{3-13}  & & $\frac{\pi }{4}$ & $0$  & $0.4638$ & $0.6242$ & $2.617$ & \cellcolor{red!30}$0.317$ & $0.02226$ & $0.453$ & $1.131$ & $0.082$ & $0.050$\\ \cline{3-13} & & $\frac{\pi }{4}$ & $\frac{\pi }{4}$  & $0.8962$ & $0.9688$ & $9.628$ & \cellcolor{red!30}$0.330$ & $0.02228$ & $0.441$ & $1.247$ & $0.217$ & $0.630$\\ \cline{3-13} & & $\frac{\pi }{2}$ & $\frac{\pi }{4}$  & $0.6469$ & $0.6762$ & $4.438$ & \cellcolor{red!30}$0.306$ & $0.02221$ & $0.445$ & $1.133^{*}$ & $0.852^{*}$ & $0.588^{*}$\\ \cline{2-13}
  \hline \hline
\multirow{11}{*}{IO} & \multirow{1}{*}{$U^{IX}_{1}$ }  & $\frac{\pi }{4}$ & $\frac{\pi }{2}$  & $0.1746$ & $0.1537$ & $15.915$ & \cellcolor{red!30}$0.330$ & $0.02245$ & $0.456$ & $1.859^{*}$ & $0.520^{*}$ & $0.729^{*}$\\ \cline{2-13}& \multirow{4}{*}{$U^{IX}_{2}$ }  & $0$ & $\frac{\pi }{4}$  & $0.9112$ & $0.9694$ & $10.878$ & \cellcolor{red!30}$0.301$ & $0.02222$ & $0.558$ & $1.879$ & $0.990$ & $0.490$\\ \cline{3-13}  & & $\frac{\pi }{4}$ & $0$  & $0.4645$ & $0.6245$ & $10.208$ & \cellcolor{red!30}$0.315$ & $0.02227$ & $0.548$ & $1.872^{*}$ & $0.920^{*}$ & $0.951^{*}$\\ \cline{3-13} & & $\frac{\pi }{4}$ & $\frac{\pi }{4}$  & $0.8897$ & $0.9635$ & $6.488$ & \cellcolor{red!30}$0.319$ & $0.02233$ & $0.566$ & $1.756^{*}$ & $0.769^{*}$ & $0.359^{*}$\\ \cline{3-13} & & $\frac{\pi }{2}$ & $\frac{\pi }{4}$  & $0.6676$ & $0.6802$ & $6.548$ & \cellcolor{red!30}$0.320$ & $0.02231$ & $0.561$ & $1.775$ & $0.085$ & $0.373$\\ \cline{2-13}& \multirow{2}{*}{$U^{IX}_{3}$ }  & $0$ & $\frac{\pi }{4}$  & $0.0880$ & $0.0419$ & $10.942$ & \cellcolor{red!30}$0.318$ & $0.02240$ & $0.559$ & $1.104^{*}$ & $0.980^{*}$ & $0.488^{*}$\\ \cline{3-13}  & & $\frac{\pi }{4}$ & $\frac{\pi }{2}$  & $0.1938$ & $0.1722$ & $10.368$ & \cellcolor{red!30}$0.332$ & $0.02244$ & $0.556$ & $1.185$ & $0.541$ & $0.313$\\ \cline{2-13}& \multirow{4}{*}{$U^{IX}_{4}$ } & $0$ & $\frac{\pi }{4}$  & $0.9160$ & $0.9778$ & $14.432$ & \cellcolor{red!30}$0.312$ & $0.02234$ & $0.454$ & $1.086^{*}$ & $0.007^{*}$ & $0.507^{*}$\\ \cline{3-13}  & & $\frac{\pi }{4}$ & $0$  & $0.4616$ & $0.6233$ & $12.699$ & \cellcolor{red!30}$0.321$ & $0.02234$ & $0.455$ & $1.139$ & $0.086$ & $0.053$\\ \cline{3-13} & & $\frac{\pi }{4}$ & $\frac{\pi }{4}$  & $0.8968$ & $0.9695$ & $15.468$ & \cellcolor{red!30}$0.331$ & $0.02241$ & $0.442$ & $1.247$ & $0.215$ & $0.629$\\ \cline{3-13} & & $\frac{\pi }{2}$ & $\frac{\pi }{4}$  & $0.6572$ & $0.6775$ & $13.819$ & \cellcolor{red!30}$0.313$ & $0.02241$ & $0.443$ & $1.181^{*}$ & $0.884^{*}$ & $0.607^{*}$\\ \cline{2-13}
\hline \hline
\multicolumn{13}{|c|}{$\Delta(6n^{2})\rtimes H_{CP}$ (Case X)} \\ \hline
\multirow{2}{*}{Order} & \multirow{2}{*}{Case} & \multirow{2}{*}{$\varphi_1$} & \multirow{2}{*}{$\varphi_2$} & \multirow{2}{*}{$\theta^{\text{bf}}_{\ell}/\pi$} & \multirow{2}{*}{$\theta^{\text{bf}}_{\nu}/\pi$} & \multirow{2}{*}{$\chi^2_{\text{min}}$}   & \multirow{2}{*}{$\sin^2\theta_{12}$} & \multirow{2}{*}{$\sin^2\theta_{13}$} & \multirow{2}{*}{$\sin^2\theta_{23}$}  & \multirow{2}{*}{$\delta_{CP}/\pi$} & $\alpha_{21}/\pi$ & $\alpha_{31}/\pi $   \\
&&&&&&&&&&& (mod 1) & (mod 1)\\
\hline
\multirow{4}{*}{NO} & \multirow{2}{*}{$U^{X}_{1}$ }  & $0$ & $0$  & $0.0676$ & $0.6391$ & $3.040$ & \cellcolor{red!30}$0.308$ & $0.02223$ & $0.511$ & $1$ & $0$ & $0$\\ \cline{3-13}
& & $\frac{\pi }{4}$ & $0$  & $0.0676$ & $0.6511$ & $2.569$ & \cellcolor{red!30}$0.308$ & $0.02223$ & $0.511$ & $1.214$ & $0.853$ & $0.891$\\ \cline{2-13}& \multirow{2}{*}{$U^{X}_{2}$ }  & $0$ & $0$  & $0.9326$ & $0.7352$ & $1.660$ & \cellcolor{red!30}$0.308$ & $0.02206$ & $0.489$ & $1$ & $0$ & $0$\\ \cline{3-13}
& & $\frac{\pi }{4}$ & $0$  & $0.9326$ & $0.7202$ & $1.564$ & \cellcolor{red!30}$0.308$ & $0.02206$ & $0.489$ & $1.271$ & $0.146$ & $0.094$\\ \cline{3-13}
  \hline \hline
\multirow{2}{*}{IO} & $U^{X}_{1}$  & $\frac{\pi }{4}$ & $0$  & $0.9324$ & $0.7202$ & $6.045$ & \cellcolor{red!30}$0.307$ & $0.02224$ & $0.511$ & $1.729^{*}$ & $0.853^{*}$ & $0.905^{*}$\\ \cline{2-13}
& $U^{X}_{2}$ & $\frac{\pi }{4}$ & $0$  & $0.9324$ & $0.7201$ & $5.740$ & \cellcolor{red!30}$0.307$ & $0.02221$ & $0.489$ & $1.271$ & $0.147$ & $0.095$\\ \cline{3-13}
\hline \hline
\end{tabular}}
\caption{\label{tab:twopara_bf_CaseIX_X_Delta6nsq}
The best-fit values of the free parameter and the corresponding lepton mixing observables for two-parameter mixing patterns
$U^{IX}$ and $U^{X}$ originate from $\Delta(6n^{2})\rtimes H_{CP}$ with $n\leq 4$. The discrete parameters $(\varphi_{1},\varphi_{2})$ specify the residual symmetries associated with flavour groups $\Delta(6n^{2})$.
Here $\chi^{2}_{\text{min}}$ the global minimum of the $\chi^{2}$ function occurs at $\left(\theta_{\ell},\theta_{\nu}\right)=\left(\theta^{\text{bf}}_{\ell},\theta^{\text{bf}}_{\nu}\right)$. The remaining conventions follow those used in table~\ref{tab:best-fit-onepara}.}
\end{table}

The second two-parameter mixing pattern obtained from $\Delta(6n^{2})\rtimes H_{CP}$ is denoted by $U^{IX}$. As shown in Eq.~\eqref{eq:permu_caseIX}, four distinct realizations arise from different permutations of rows and columns. The $\chi^{2}$-minimization results for $U^{IX}_{1,2,3,4}$ are summarized in table~\ref{tab:twopara_bf_CaseIX_X_Delta6nsq}. There are in total 16 and 11 viable cases for NO and IO respectively. The smallest $\Delta(6n^{2})$ group that can reproduce the experimental data is $S_{4}$ $(n=2)$. In this case, all four mixing patterns $U^{IX}_{1,2,3,4}$ accommodate the \texttt{NuFIT} results for NO with $(\varphi_{1},\varphi_{2})=(0,0)$, and the CPV phases are CP-conserving because the PMNS matrix is real,  with $\delta_{CP} = 0$ or $\pi$. 

The next group capable of explaining the data is $\Delta(96)$ $(n=4)$. For this group, the patterns $U^{IX}_{1}$ and $U^{IX}_{3}$ with $(\varphi_{1},\varphi_{2})=(\frac{\pi}{4},\frac{\pi}{2})$ are viable for both NO and IO. When $(\varphi_{1},\varphi_{2})=(0,\frac{\pi}{4})$, the pattern $U^{IX}_{3}$ is viable for both NO and IO, whereas $U^{IX}_{1}$ is viable only for NO. For $U^{IX}_{2}$ and $U^{IX}_{4}$, agreement with the \texttt{NuFIT} data can be achieved for both mass orderings when $(\varphi_{1},\varphi_{2})$ takes one of the values $(0,\frac{\pi}{4})$, $(\frac{\pi}{4},0)$, $(\frac{\pi}{4},\frac{\pi}{4})$, or $(\frac{\pi}{4},\frac{\pi}{2})$. The corresponding allowed regions of the lepton mixing observables are summarized in table~\ref{tab:twopara_3sigma_CaseII_Delta6nsq}. 

\begin{figure}[htbp]
    \centering
    \includegraphics[scale=0.75]{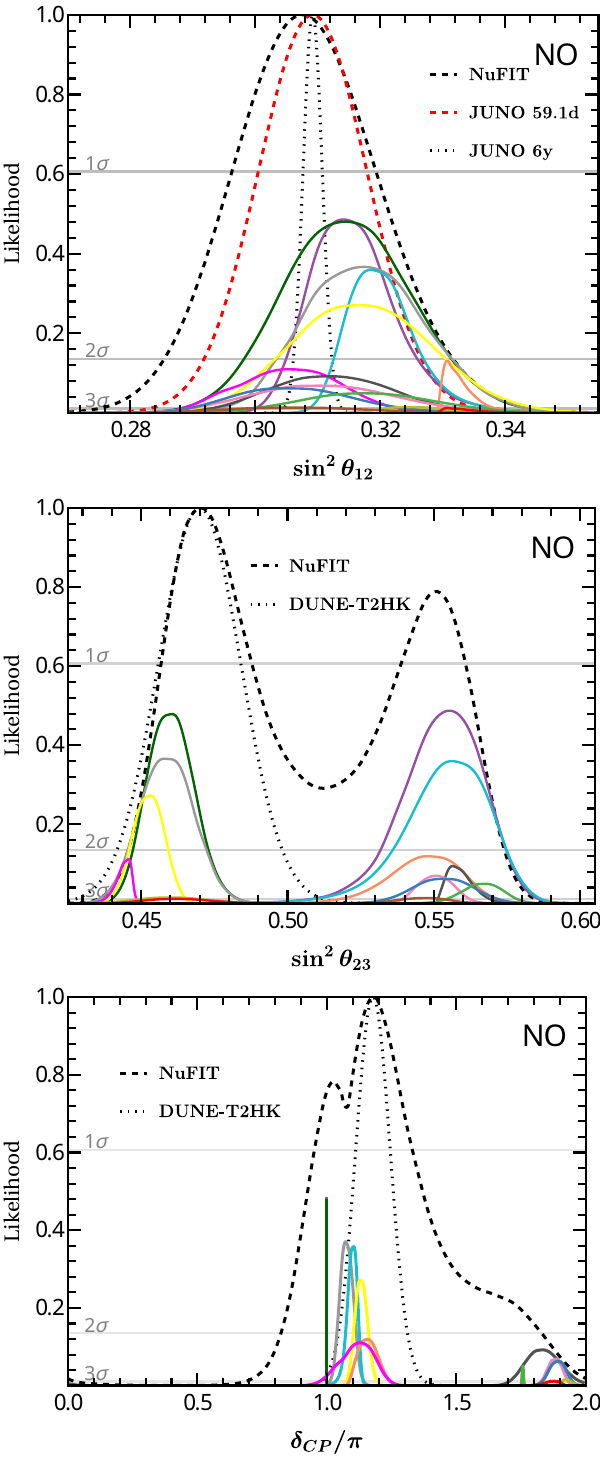}
    \includegraphics[scale=0.75]{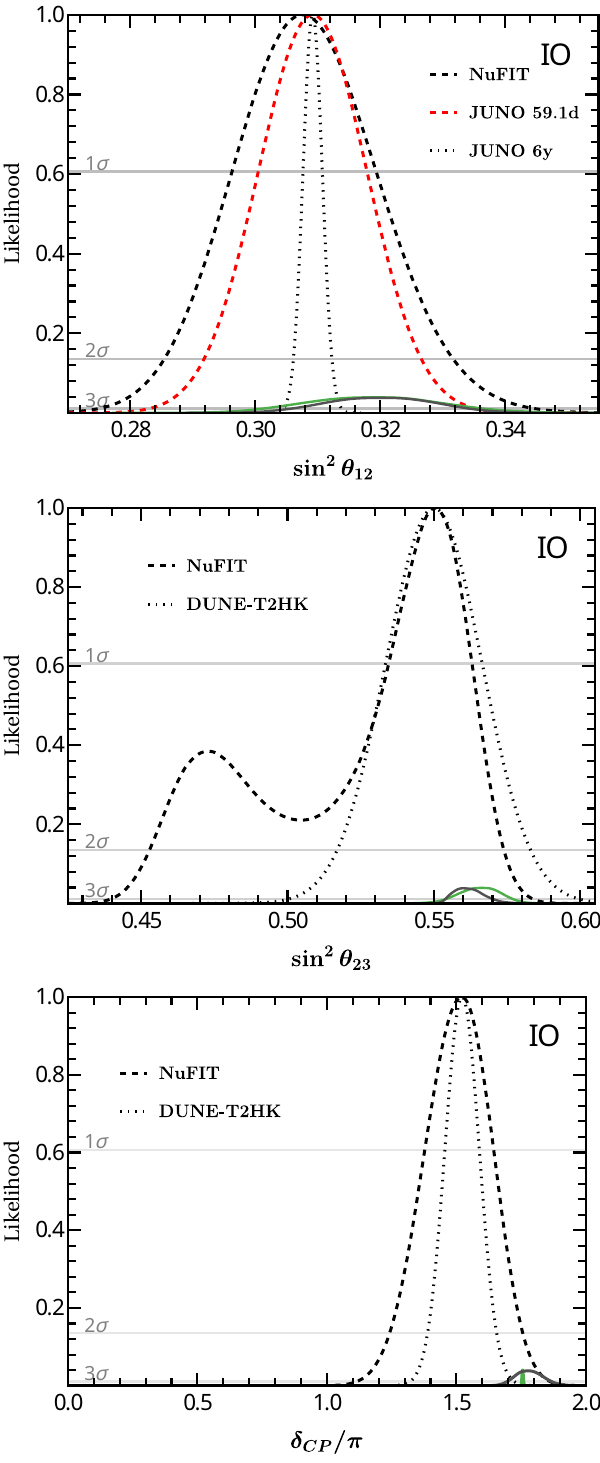}
    \includegraphics[scale=0.75]{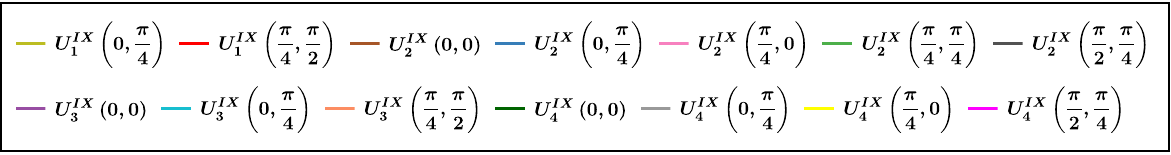}
    \caption{The predictions of likelihood profiles for $\sin^{2}\theta_{12}$, $\sin^{2}\theta_{23}$ and $\delta_{CP}$ obtained using the current global data on the neutrino mixing parameters for two-parameter mixing patterns $U^{IX}$ that originate from $\Delta(6n^{2})\rtimes H_{CP}$. The remaining conventions follow those used in figure~\ref{fig:one_para_likelihood_NO_IO}.}
    \label{fig:two_para_VIX}
\end{figure}

In figure~\ref{fig:two_para_VIX}, we show the predicted likelihood profile for case $U^{IX}$ with $\chi^{2}_{\text{min}} \leq 9$. We find that all selected viable patterns are compatible with the current JUNO 59.1-day data for both NO and IO. The cases $U^{IX}_{1}$ and $U^{IX}_{3}$ with $(\varphi_{1}, \varphi_{2}) = \left( \frac{\pi}{4}, \frac{\pi}{2} \right)$ will be most likely excluded for NO by the projected six-year JUNO precision data. From the plots of $\sin^{2}\theta_{23}$, we can conclude that the patterns $U^{IX}_{1}$ and $U^{IX}_{4}$ predict $\sin^{2}\theta_{23} < 0.5$, while $U^{IX}_{2}$ and $U^{IX}_{3}$ predict $\sin^{2}\theta_{23} > 0.5$. Both sets of patterns can be tested for both the NO and IO scenarios by the future DUNE and T2HK prospective results on $\sin^{2}\theta_{23}$ and  $\delta_{CP}$.

\begin{figure}[htbp]
    \centering
    \includegraphics[scale=0.75]{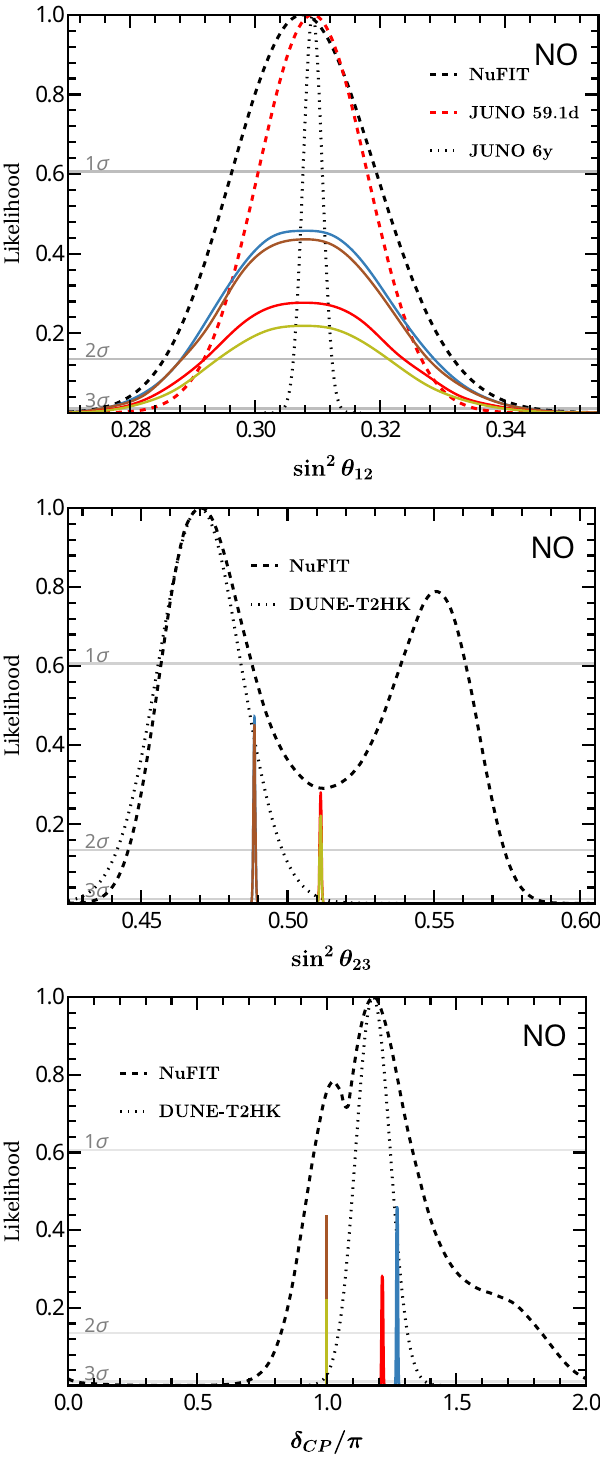}
    \includegraphics[scale=0.75]{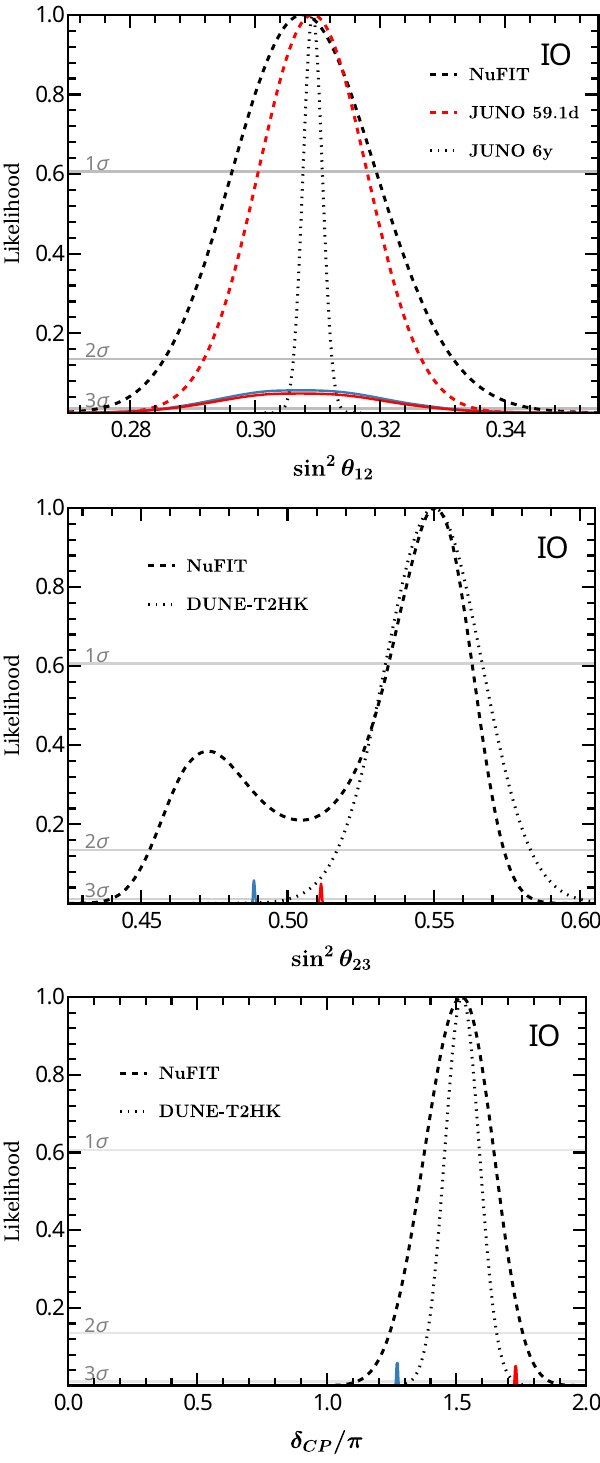}
    \includegraphics[scale=0.75]{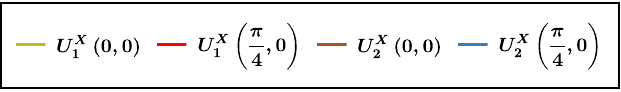}
    \caption{ The predictions of likelihood profiles for $\sin^{2}\theta_{12}$, $\sin^{2}\theta_{23}$ and $\delta_{CP}$ obtained using the current global data on the neutrino mixing parameters for two-parameter mixing patterns $U^{X}$ that originate from $\Delta(6n^{2})\rtimes H_{CP}$. The remaining conventions follow those used in figure~\ref{fig:one_para_likelihood_NO_IO}.}
    \label{fig:two_para_X}
\end{figure}

The numerical analysis for the pattern $U^{X}$ is summarized in table~\ref{tab:twopara_bf_CaseIX_X_Delta6nsq}. Unlike $U^{IX}$, only 4 NO and 2 IO realizations are viable. As in the previous case, the smallest group capable of accommodating the data is $S_{4}\cong \Delta(6\cdot 2^2)$, where the realizations $U^{X}_{1,2}$ with $(\varphi_{1},\varphi_{2})=(0,0)$ imply CP conservation due to the reality of the PMNS matrix, with $\delta_{CP} = \pi$, which for NO is not excluded by the current
data. Moving to the next viable group, $\Delta(96)=\Delta(6\cdot 4^2)$,  the predictions of both the $U^{X}_{1}$ and $U^{X}_{2}$ patterns are compatible with the current data for $(\varphi_{1},\varphi_{2})=(\frac{\pi}{4},0)$. Figure~\ref{fig:two_para_X} illustrates the corresponding likelihood profiles. Owing to the sum rule in Eq.~\eqref{eq:sum_rules_X}, the atmospheric angle in $U^{X}_{1}$ and $U^{X}_{2}$ is tightly correlated with $\sin^{2}\theta_{13}$, yielding sharply peaked distributions for $\sin^{2}\theta_{23}$. The prospective sensitivities  T2HK and
DUNE would  be sufficient to possibly exclude the pattern $U^{X}_{1}$ for NO and the pattern $U^{X}_{2}$ for IO. For the Dirac phase, only the configuration $U^{X}_{1}$ with $(\varphi_{1},\varphi_{2})=(\frac{\pi}{4},0)$ remain potentially compatible with the considered prospective future long baseline data in the NO scenario, while the remaining options in both mass orderings are expected to be disfavored.

Based on the numerical analysis above, we conclude that almost all two-parameter mixing patterns remain compatible with the current oscillation data, even when the JUNO 59.1-day measurement of $\sin^{2}\theta_{12}$ is taken into account. The discrimination between these patterns will rely on future long baseline measurements of $\sin^{2}\theta_{23}$ and $\delta_{CP}$. For the mixing patterns originating from $A_{5}\rtimes H_{CP}$, the mixing matrix $U^{VI}_{6,1}$, featuring a fixed 
(21) entry of $\frac{1}{2\phi_{g}}$, gives the best fit for NO data, while predicting CP-conserving phases.  The pattern $U^{VI}_{1}$, with a fixed (11) entry of $\frac{\phi_{g}}{2}$, provides a good fit to the experimental data for both NO and IO spectra and predicts non-trivial CP-violating Dirac and Majorana phases. For the patterns $U^{VII}$ arising from $\Sigma(168)\rtimes H_{CP}$, the fixed entry is $\frac{1}{2}$, and the minimal $\chi^{2}_{\rm min}$ among different permutations is obtained for $U^{VII}_{2,2}$. Regarding the three distinct patterns $U^{VIII}$, $U^{IX}$, and $U^{X}$, with fixed entries $\cos\varphi_{1}$, $\frac{1}{2}$ and $\frac{1}{\sqrt{2}}$, respectively, which originate from $\Delta(6n^{2})\rtimes H_{CP}$, the minimal viable groups for the case of CP-conserving phases are $\Delta(54)$, $S_{4}$ and $S_{4}$, respectively. To accommodate CP-violating $\delta_{CP}$, the group $\Delta(96)$ is required, allowing all three patterns to fit the experimental data. Furthermore, the $D_{3}$ and $D_{4}$ groups, which are the symmetry groups respectively of the equilateral triangle and the square, can also be employed to generate CP-conserved and CP-violating viable realizations of $U^{VIII}$, respectively. It is notable that the group order is drastically reduced with respect to $\Delta(54)$ and $\Delta(96)$. We notice that the flavor group $D_{5}$ can generate the viable mixing matrix $U^{VIII}_{1}$ with the fixed $(11)$ entry $\cos(\pi/5)$, leading to $\sin^{2}\theta_{12}$ tightly constrained around $0.331$ as dictated by Eq.~\eqref{eq:correlation_VI}. This region is consistent with the current JUNO data, but lies near the upper bound and will be subject to further scrutiny with improved JUNO precision.

\section{\label{sec:conclusion}Summary and conclusion}

We have carried out a comprehensive assessment of the phenomenological viability of the lepton flavour models based on the non-Abelian discrete groups $A_{5}$, $\Sigma(168)$ and the series $\Delta(6n^{2})$ combined with gCP symmetry. The full symmetry groups $A_{5}\rtimes H_{CP}$, $\Sigma(168)\rtimes H_{CP}$, and $\Delta(6n^{2})\rtimes H_{CP}$ are broken to non-trivial residual subgroups $\mathcal{G}_\ell = G_{\ell}\rtimes H^{\ell}_{CP}$ and $\mathcal{G}_\nu = G_{\nu}\times H^{\nu}_{CP}$ in the charged lepton and neutrino sectors, respectively. The PMNS matrix is completely determined by these residual symmetries, no assumptions about the underlying dynamical origin of symmetry breaking are required. We note that if the residual flavour symmetry $G_{\ell}$ distinguishes the three families,  the inclusion of the gCP symmetry $H^{\ell}_{CP}$ would not add any information on the lepton mixing~\cite{Li:2014eia,Ding:2013bpa}. It is remarkable that we can reach sharp predictions for both Dirac CP violation phase $\delta_{CP}$ and Majorana CP phases $\alpha_{21}$, $\alpha_{31}$ due to the presence of gCP symmetry. 

We classify all viable symmetry–breaking schemes into two categories:
\begin{itemize}
    \item[(i)] $\mathcal{G}_{\ell}=Z_{n}$ ($n\geq 3$) or $K_{4}$ and $\mathcal{G}_{\nu}=Z_{2}\times H^{\nu}_{CP}$;
    \item[(ii)] $\mathcal{G}_{\ell}=Z_{2}\times H^{\ell}_{CP}$ and $\mathcal{G}_{\nu}=Z_{2}\times H^{\nu}_{CP}$.
\end{itemize}
Assuming that the three lepton doublets transform as a faithful triplet of the flavour group, we systematically analyze all possible residual symmetries and the resulting predictions for the lepton mixing matrix. The indicated two classes of breaking patterns lead to qualitatively different parametric structures of the PMNS matrix. Under pattern~(i), the PMNS matrix depends on a single real parameter $\theta\in[0,\pi)$ and one of its columns is completely fixed. Under pattern~(ii), the PMNS matrix depends on two real parameters $(\theta_{\ell},\theta_{\nu})$ in $[0,\pi)$ and one matrix element is fixed. A third possibility, with $\mathcal{G}_{\ell}=Z_{2}$ and $\mathcal{G}_{\nu}=K_{4}\times H^{\nu}_{CP}$, also yields a two–parameter mixing matrix but with one row fixed. This option is incompatible with the \texttt{NuFIT} data and is therefore discarded; see Appendix~\ref{app:Z2-K4}.

First of all, we performed a statistical test of the predicted correlations, in the form of sum rules, between the mixing angles for all patterns obtained in approaches~(i) and~(ii), using the latest global neutrino oscillation data from \texttt{NuFIT}-v6.0~\cite{Esteban:2024eli}. Our analysis selects the mixing patterns whose predictions satisfy the following requirement: the three lepton mixing angles $(\theta_{13},\theta_{12},\theta_{23})$ and the Dirac CPV phase $\delta_{CP}$ must all lie within their experimentally allowed $3\sigma$ ranges~\cite{Esteban:2024eli}. In approach~(i), all mixing patterns that remain consistent with the measured mixing angles and the $\delta_{CP}$ at the $3\sigma$ level~\cite{Esteban:2024eli} can be classified into 5 distinct cases, up to permutations of rows and columns. The corresponding mixing matrices are
\begin{eqnarray}
\nonumber&&U^{I}=\frac{1}{\sqrt{3}}
\begin{pmatrix}
\sqrt{2} \sin \varphi _1 ~&~ e^{i \varphi _2} ~&~ \sqrt{2}\cos\varphi_1 \\
\sqrt{2}\cos\left(\varphi _1-\frac{\pi }{6}\right) ~&~ -e^{i\varphi_2} ~&~ -\sqrt{2} \sin \left(\varphi_1-\frac{\pi}{6}\right) \\
\sqrt{2}\cos\left(\varphi_1+\frac{\pi}{6}\right)   ~&~  e^{i\varphi_2} ~&~ -\sqrt{2}\sin\left(\varphi_1+\frac{\pi}{6}\right)
\end{pmatrix}R_{23}(\theta)Q_{\nu}\,,\\
\nonumber&&U^{II}=\frac{1}{\sqrt{3}}
 \left(
\begin{array}{ccc}
 e^{i \varphi _1} & 1 & e^{i \varphi _2} \\
 \omega  e^{i \varphi _1} & 1 & \omega ^2 e^{i \varphi _2} \\
 \omega ^2 e^{i \varphi _1} & 1 & \omega  e^{i \varphi _2} \\
\end{array}
\right)R_{13}(\theta)Q_{\nu}\,, \\
\nonumber&&U^{III}=\left(
\begin{array}{ccc}
-i\sqrt{\frac{\phi_g}{\sqrt{5}}} & \sqrt{\frac{1}{\sqrt{5}\phi_g}} & 0 \\
i\sqrt{\frac{1}{2\sqrt{5}\phi_g}} & \sqrt{\frac{\phi_g}{2\sqrt{5}}} & -\frac{1}{\sqrt{2}} \\
i\sqrt{\frac{1}{2\sqrt{5}\phi_g}} & \sqrt{\frac{\phi_g}{2\sqrt{5}}} & \frac{1}{\sqrt{2}}
\end{array}
\right)R_{13}(\theta)Q_{\nu}\,,\\
\nonumber&&U^{IV}=\frac{1}{2}\left(
\begin{array}{ccc}
 \phi_g  & 1 & \phi_g -1 \\
 \phi_g -1 & -\phi_g  & 1 \\
 1 & 1-\phi_g  & -\phi_g
\end{array}
\right)R_{23}(\theta)Q_{\nu}\,,\\
&&U^{V}=\frac{1}{2\sqrt{3}}\left(
\begin{array}{ccc}
\sqrt{3}-1 &~ 2e^{-i \varphi } &~ -(\sqrt{3}+1)e^{\frac{3\pi i}{4}} \\
-\sqrt{3}-1 &~ 2e^{-i \varphi } &~ (\sqrt{3}-1)e^{\frac{3\pi i}{4}} \\
2 &~ 2e^{-i \varphi } &~ 2e^{\frac{3\pi i}{4}}
\end{array}
\right)R_{13}(\theta)Q_{\nu}\,.
\end{eqnarray}
Notably, $U^{I}$ and $U^{II}$ arise from the breaking of $\Delta(6n^{2})$ flavour symmetries combined with the generalized CP (gCP) symmetry. Furthermore, $U^{II}$ can also be realized by breaking $\Delta(3n^{2})\rtimes H_{CP}$~\cite{Ding:2015rwa} if nontrivial singlets are absent from the model. The mixing matrices $U^{III}$ and $U^{IV}$ are minimally realized using the alternating group $A_5$ in combination with the gCP symmetry, while $U^{V}$ is reproduced by the flavour symmetry group $\Sigma(168) \rtimes H_{CP}$.
In the breaking approach (ii), our statistical analysis identifies 11 viable breaking patterns. Among them, six arise from the breaking of $A_{5}\rtimes H_{CP}$, yielding the PMNS matrices $U^{VI}_{i}$ with $i=1,\dots,6$ in Eq.~\eqref{eq:PMNS_caseVI}. Two additional patterns originate from $\Sigma(168)\rtimes H_{CP}$, corresponding to $U^{VII}_{1}$ and $U^{VII}_{2}$ in Eq.~\eqref{eq:PMNS_caseVII}. The remaining three patterns stem from $\Delta(6n^{2})\rtimes H_{CP}$ and lead to the mixing matrices $U^{VIII}$, $U^{IX}$, and $U^{X}$, shown in Eq.~\eqref{eq:PMNS_caseVIII}, Eq.~\eqref{eq:UPMNS_caseIX}, and Eq.~\eqref{eq:PMNS_X}, respectively. Furthermore, when the left-handed lepton doublets transform as a direct sum of a singlet and a doublet, the mixing matrix $U^{VIII}$ in Eq.~\eqref{eq:PMNS_caseVIII} can also be realized through the breaking of the dihedral group $D_{n}$ combined with the gCP symmetry. We list the 
fixed columns for the one-parameter mixing patterns $U^{I}$, $U^{II}$, $U^{III}$, $U^{IV}$ and $U^{V}$, and the fixed entries for the two-parameter mixing patterns $U^{VI}$, $U^{VII}$, $U^{VIII}$, $U^{IX}$ and $U^{X}$ in table~\ref{tab:fixed_column_entry}.

\begin{table}[t!]
\centering
\renewcommand{\arraystretch}{1.1}
\scalebox{1.0}{\begin{tabular}{|c|c|c|c|} \hline \hline
Case & Fixed column & Case & Fixed entry \\ \hline
$U^{I}$ & $\sqrt{\frac{2}{3}}\left(\sin\varphi_{1},\cos(\varphi_{1}-\frac{\pi}{6}),\cos(\varphi_{1}+\frac{\pi}{6})\right)^{T}$ & $U^{VI}$ & $\frac{\phi_{g}}{2}$, $\frac{1}{2}$, $\frac{1}{2\phi_{g}}$ \\ \hline
$U^{II}$ & $\frac{1}{\sqrt{3}}\left(1,1,1\right)^{T}$ & $U^{VII}$ & $\frac{1}{2}$ \\ \hline
$U^{III}$ & $\left(\sqrt{\frac{1}{\sqrt{5}\phi_{g}}},\sqrt{\frac{\phi_{g}}{2\sqrt{5}}},\sqrt{\frac{\phi_{g}}{2\sqrt{5}}}\right)^{T}$ & $U^{VIII}$ & $\cos\varphi_{1}$ \\ \hline
$U^{IV}$ & $\frac{1}{2}\left(\phi_{g},\phi_{g}-1,1\right)^{T}$ & $U^{IX}$ & $\frac{1}{2}$ \\ \hline
$U^{V}$ & $\frac{1}{\sqrt{3}}\left(1,1,1\right)^{T}$ & $U^{X}$ & $\frac{1}{\sqrt{2}}$ \\
\hline \hline
\end{tabular}}
\caption{\label{tab:fixed_column_entry} The fixed columns for the one-parameter mixing patterns $U^{I}$, $U^{II}$, $U^{III}$, $U^{IV}$ and $U^{V}$, and the fixed entries for the two-parameter mixing patterns $U^{VI}$, $U^{VII}$, $U^{VIII}$, $U^{IX}$ and $U^{X}$. Here $\phi_{g} = (\sqrt{5}+1)/2$ is the golden ratio, and the parameter $\varphi_{1}$ in $U^{I}$ and $U^{VIII}$ takes the forms given in Eq.~\eqref{eq:phi12_I} and Eq.~\eqref{eq:phi12_VIII}, respectively.}
\end{table}

In order to further quantitatively assess how well a mixing determined by a given breaking pattern can describe the current experiment data (NuFIT and JUNO 59.1-day results), we perform a parameter scan and compute the minimum $\chi^{2}$ (defined in Eq.~\eqref{eq:chisq_def}) value for the derived one- and two-parameter mixing patterns. In the cases of $\Delta(6n^{2}) \rtimes H_{CP}$ and $D_{n}\rtimes H_{CP}$, this analysis is restricted
to groups respectively with $n\leq4$ and  $n\leq 8$. The best-fit values of the input parameter and the lepton mixing parameters for the mixing patterns with one input parameter are shown for both NO and IO cases in table~\ref{tab:best-fit-onepara}. Those for the mixing patterns with two input parameters arising from the groups from $A_{5}\rtimes H_{CP}$, $\Sigma(168)\rtimes H_{CP}$, $\Delta(6n^{2})\rtimes H_{CP}$ and $D_{n}\rtimes H_{CP}$ are shown respectively in tables~\ref{tab:best-fit-twopara_A5_Sigma168}, \ref{tab:twopara_bf_CaseVIII_Delta6nsq}
and~\ref{tab:twopara_bf_CaseIX_X_Delta6nsq}.

We scanned the parameter spaces of each mixing matrix to identify viable ranges consistent with neutrino oscillation data at the $3\sigma$ level. This procedure yielded predictions for the lepton mixing parameters shown in detail in tables~\ref{tab:3sigma-onepara}, \ref{tab:3sigma-twopara_A5_Sigma168}, \ref{tab:twopara_3sigma_CaseI_Delta6nsq} and \ref{tab:twopara_3sigma_CaseII_Delta6nsq}. Furthermore, we can determine the minimal value of $\chi^{2}_{\text{min}}$ for the mixing angles and the Dirac CPV phase for each considered mixing matrix. The results for the mixing patterns with one and two free parameters are shown in figures~\ref{fig:onepara_bfpoints_plots_NO_IO} and \ref{fig:twopara_bfpoints_plots_NO_IO}, respectively. We then assessed the robustness of these predictions under future projected experimental conditions. This assessment assumed that: (1) the current best-fit values for the mixing angles remain unchanged, and (2) the relative $1\sigma$ uncertainties on $\sin^{2}\theta_{12}$, $\sin^{2}\theta_{23}$ and $\delta_{CP}$ are scaled to $0.5\%$, $3\%$ and $12^{\circ}$  of their current best-fit values, respectively. These projected uncertainties, corresponding to 6 years of JUNO~\cite{JUNO:2022mxj} data and 15 years of DUNE~\cite{DUNE:2020ypp} and T2HK~\cite{Hyper-Kamiokande:2018ofw} data, are visualized in figures~\ref{fig:one_para_likelihood_NO_IO}, \ref{fig:two_para_A5_Sigma168}, \ref{fig:two_para_VIII}, \ref{fig:two_para_VIX} and \ref{fig:two_para_X} in which the likelihood profiles for $\sin^{2}\theta_{12}$, $\sin^{2}\theta_{23}$ and $\delta_{CP}$ are displayed  for selected viable patterns with $\chi^{2}_{\min}\leq 9$.

For the one-parameter mixing patterns, all five candidates are capable of accommodating the current \texttt{NuFIT} data and yield sharply constrained predictions for the lepton mixing parameters for both the NO and IO neutrino mass spectra. The only exceptions are $U^{IV}_{1}$ and $U^{IV}_{2}$, whose predictions imply a CP-conserving Dirac phase $\delta_{CP}$; this is compatible with the \texttt{NuFIT} constraints only in the NO case. The smallest $\Delta(6n^{2})$ group that can realize the viable patterns $U^{I}$ and $U^{II}$ is $\Delta(24)\cong S_{4}$ with $n=2$. We note that the minimal group capable of generating $U^{II}$ is $A_{4}$ if the nontrivial singlet representations are absent from the model~\cite{Li:2016nap}. For $n=2$, the choice of the relevant discrete parameters $(\varphi_{1},\varphi_{2})=\left(-\frac{\pi}{2},\,\frac{\pi}{2}\right)$ reduces $U^{I}$ to the well-known TM1 mixing form. The recent JUNO 59.1-day results further sharpen these constraints. At the $3\sigma$ level, the viable parameter space of the one-parameter patterns is significantly reduced. In the NO case, only TM1 mixing ($U^{I}$) and the patterns $U^{IV}_{1}$ ($U^{IV}_{2}$), which feature fixed columns $(\phi_{g},\phi_{g}-1,1)^{T}/2$ ($(\phi_{g},1,\phi_{g}-1)^{T}/2$), remain compatible after incorporating the JUNO likelihood, as shown in figure~\ref{fig:one_para_likelihood_NO_IO}. For the IO spectrum, $U^{I}$ (TM1) is the sole surviving option. Mixing patterns such as $U^{II}$ (TM2), $U^{III}$ (GR2), and $U^{V}$ (TM2) are disfavored once the current JUNO constraint on $\sin^{2}\theta_{12}$ is taken into account. We note that, for the mixing pattern $U^{I}$, higher values of $n$ in $\Delta(6n^{2})$ admit additional viable textures—beyond the TM1 form—that are consistent with current JUNO data. Looking into the future, the full set of one-parameter patterns is expected to come under decisive scrutiny: the projected precision of JUNO on $\sin^{2}\theta_{12}$, together with forthcoming measurements of $\sin^{2}\theta_{23}$ and $\delta_{CP}$ from DUNE and T2HK, will provide a definitive test of these scenarios.

Nearly all two-parameter neutrino mixing patterns remain consistent with current \texttt{NuFIT} data and JUNO 59.1-day measurement of $\sin^{2}\theta_{12}$. Moreover, future high-precision JUNO results are unlikely to exclude the majority of these two-parameter scenarios.
Distinguishing among these patterns will require improved long-baseline measurements of $\sin^{2}\theta_{23}$ and $\delta_{CP}$. For the mixing patterns from $A_5 \rtimes H_{CP}$, those with fixed elements $\frac{\phi_g}{2}$, $\frac{1}{2}$ and $\frac{1}{2\phi_g}$ are viable. The best experimental agreement is furnished by the $U^{VI}_{6,1}$ pattern, which has $\frac{1}{2\phi_g}$ fixed at the $(21)$ entry and gives the best fit for the NO data, while conserving the CP-symmetry. In contrast, $U^{VI}_{1}$, with the fixed element $\frac{\phi_{g}}{2}$ at the $(11)$ entry, 
offers a good fit in both NO and IO cases and predicts nontrivial CP violation effects in neutrino oscillations. For the mixing patterns derived from $\Sigma(168)\rtimes H_{CP}$, the fixed entry is $\frac12$. The smallest $\chi^{2}_{\rm min}$ among permutations occurs for $U^{VII}_{2,2}$, with the fixed $(31)$ entry. Among the three mixing patterns $U^{VIII}$, $U^{IX}$ and $U^{X}$ obtained from $\Delta(6n^{2})\rtimes H_{CP}$---with fixed entries $\cos\varphi_{1}$, $\frac{1}{2}$ and $\frac{1}{\sqrt{2}}$, respectively---the minimal flavour groups realizing these patterns with CP-conserving Dirac and Majorana phases are $\Delta(54)$ for $U^{VIII}$ and $S_{4}$ for both $U^{IX}$ and $U^{X}$. Allowing for CP violation requires extending the symmetry to $\Delta(96)$. Then $U^{VIII}_{9}$ for $\varphi_{1}=\varphi_{2}=\frac{\pi}{4}$ and $U^{X}_{2}$ for $(\varphi_{1},\varphi_{2})=(\frac{\pi}{4},0)$ give rise to $\frac{1}{\sqrt2}$ at $(33)$ entry, yielding the smallest $\chi^{2}_{\rm min}$. It is remarkable that the dihedral groups $D_{3}$ and $D_{4}$ can also give rise to the viable mixing pattern $U^{VIII}$ with CP-conserving and CP-violating phases, respectively.

In summary, we have shown that the discrete flavour symmetries  $A_{5}\rtimes H_{CP}$, $\Sigma(168)\rtimes H_{CP}$, $\Delta(6n^{2})\rtimes H_{CP}$ and  $D_{n}\rtimes H_{CP}$, when broken to suitable residual symmetries in the charged lepton and neutrino sectors, give rise to a well defined set of viable neutrino mixing scenarios characterized by one or two real parameters. These constructions lead to sharp, testable predictions for the mixing angles $\theta_{12}$, $\theta_{23}$ and the Dirac phase $\delta_{CP}$. We find that the smallest flavour group capable of generating phenomenologically viable one-parameter mixing patterns is $S_{4}$ (of order 24). For two-parameter mixing patterns, the minimal realizations arise from the dihedral groups $D_{4}$ (order 8) or $D_{3}$ (order 6), the latter giving rise to  CP-conserving Dirac and Majorana phases. The recent JUNO 59.1-day dataset has reduced the $3\sigma$ allowed region of $\sin^{2}\theta_{12}$ with respect to that of \texttt{NuFIT} by a factor of roughly 1.3, exerting a particularly strong impact on one-parameter scenarios, while the constraints on two-parameter patterns remain comparatively mild. A decisive assessment of the phenomenological viability of the surviving mixing patterns will require the next generation of precision measurements. In particular, the full JUNO dataset, together with the upcoming long-baseline experiments DUNE and T2HK, will be essential for thoroughly testing and potentially ruling out, or providing evidence for, the residual symmetry-based frameworks, originating form the non-Abelian discrete symmetry approach to the lepton mixing problem in particle physics.

\section*{Acknowledgements}

CCL is supported by Natural Science Basic Research Program of Shaanxi (Program No. 2024JC-YBQN-0004), and the National Natural Science Foundation of China under Grant No. 12247103. JNL is supported by the Grant No. NSFC-12505133. GJD is supported by the National Natural Science Foundation of China under Grant Nos.~12375104, 12547106 and Guizhou Provincial Major Scientific and Technological Program XKBF (2025)010. The work of S. T. P. was supported in part by the European Union’s Horizon Europe research and innovation programme under the Marie Sk\l{}odowska-Curie Staff Exchange grant agreement No 101086085 – ASYMMETRY, by the Italian INFN program on Theoretical Astroparticle Physics and by the
World Premier International Research Center Initiative (WPI Initiative, MEXT), Japan.

\section*{Appendix}

\begin{appendix}

\section{\label{sec:group_theory}Group theory of $A_{5}$, $\Sigma(168)$, $\Delta(6n^{2})$ and $D_n$}

\subsection{The group $A_{5}$}

$A_5$ is the group of even permutations of five objects. It can be generated by the two generators $S$ and $T$ which satisfy the multiplication rules~\cite{Ding:2011cm}:
\begin{equation}
  S^{2}=T^{5}=(ST)^{3}=1\,.
\end{equation}
The $A_5$ group has five irreducible representations: one singlet representation $\bm{1}$, two three-dimensional representations $\bm{3}$ and $\bm{3^{\prime}}$, one four-dimensional representation $\bm{4}$ and one five-dimensional representation $\bm{5}$. In the present work, our analysis uses the basis of Refs.~\cite{Ding:2011cm,Li:2015jxa,Novichkov:2018nkm,Ding:2019xna}. The explicit forms of the generators $S$ and $T$ in the five irreducible representations are as follows:
\begin{eqnarray}
\begin{array}{cll}
\bm{1:} &   ~S=1\,,~ &  T=1 \,, ~\\[-14pt] \\[4pt]
\bm{3:} &   ~S=\frac{1}{\sqrt{5}}
\begin{pmatrix}
1 & -\sqrt{2} & -\sqrt{2} \\
-\sqrt{2} & -\phi_{g}  & \phi_{g}-1 \\
-\sqrt{2} & \phi_{g}-1 & -\phi_{g}
\end{pmatrix}\,,~
& T=\text{diag}(1,\omega_{5},\omega^4_{5})\,,~~\\[-14pt] \\[4pt]
\bm{3^{\prime}:} &  ~S=\frac{1}{\sqrt{5}}
\begin{pmatrix}
-1 & \sqrt{2} & \sqrt{2} \\
\sqrt{2} & 1-\phi_{g} & \phi_{g}  \\
\sqrt{2} & \phi_{g}  & 1-\phi_{g}
\end{pmatrix}\,, ~
&T=
\text{diag}(1,\omega^2_{5},\omega^3_{5})\,,~~\\[-14pt] \\[4pt]
\bm{4:} &  ~S=\frac{1}{\sqrt{5}}
\begin{pmatrix}
1 & \phi_{g}-1 & \phi_{g}  & -1 \\
\phi_{g}-1 & -1 & 1 & \phi_{g}  \\
\phi_{g}  & 1 & -1 & \phi_{g}-1 \\
-1 & \phi_{g}  & \phi_{g}-1 & 1
\end{pmatrix}\,,~
&T=
\text{diag}(\omega_{5},\omega^2_{5},\omega^3_{5},\omega^4_{5})\,,~~\\[-14pt] \\[4pt]
\bm{5:} &  ~S=\frac{1}{5}
\begin{pmatrix}
-1 & \sqrt{6} & \sqrt{6} & \sqrt{6} & \sqrt{6} \\
\sqrt{6} & (\phi_{g}-1)^{2} & -2 \phi_{g}  & 2(\phi_{g}-1) & \phi_{g} ^2 \\
\sqrt{6} & -2\phi_{g}  & \phi_{g}^2& (\phi_{g}-1)^{2} & 2(\phi_{g}-1) \\
\sqrt{6} & 2(\phi_{g}-1) & (\phi_{g}-1)^{2} & \phi_{g} ^2 & -2 \phi_{g}  \\
\sqrt{6} & \phi_{g}^2 & 2(\phi_{g}-1) & -2 \phi_{g}  & (\phi_{g}-1)^{2}
\end{pmatrix}\,,~
&T=\text{diag}(1,\omega_{5},\omega^2_{5},\omega^3_{5},\omega^4_{5})\,,~~
\end{array}
\end{eqnarray}
where $\omega_5=e^{2\pi i/5}$ denotes the quintic root of unit and $\phi_{g}=(1+\sqrt{5})/2$ is the golden ratio.
In our working basis, the generator $T$ represented by a diagonal matrix and $S$ by a real symmetric matrix in all irreducible representations. As a result, the consistent gCP transformation matrices $X_{\bm{r}}$ coincide in form with the family group transformation matrices $\rho_{\bm{r}}(g)$ with $g\in A_{5}$.

\subsection{The group $\Sigma(168)$}

The group $\Sigma(168)\cong PSL(2,Z_{7})$ is a non-Abelian finite subgroup of $SU(3)$ of order $168$. $\Sigma(168)$ group can be generated by two generators $S$ and $T$ which satisfy the multiplication rules:
\begin{equation}
S^2=(ST)^{3}= T^7= (ST^{3})^{4}=1.
\end{equation}
The $\Gamma_{7}$ group has six irreducible representations: one singlet representation $\bm{1}$, two three-dimensional representations $\bm{3}$ and $\bm{\bar{3}}$, one six-dimensional representation $\bm{6}$, one seven-dimensional representation $\bm{7}$ and one eight-dimensional representation $\bm{8}$. The explicit forms of the generators $S$ and $T$ in the five irreducible representations are chosen as follows~\cite{Ding:2020msi}:
\begin{eqnarray}
\nonumber &&    \bm{1}:   ~S=1\,, \qquad    T=1 \,, ~\\
\nonumber &&  \bm{3}:    ~S=\frac{2}{\sqrt{7}}
\begin{pmatrix}
 -s_2 &~ -s_1 ~& s_3 \\
 -s_1 &~ s_3 ~& -s_2 \\
 s_3 &~ -s_2 ~& -s_1 \\
\end{pmatrix}\,,~ \qquad
\bm{\bar{3}} :  ~S=\frac{2}{\sqrt{7}}
\begin{pmatrix}
 -s_2 &~ -s_1 ~& s_3 \\
 -s_1 &~ s_3 ~& -s_2 \\
 s_3 &~ -s_2 ~& -s_1 \\
\end{pmatrix}\,,
~~\\
\nonumber && \bm{6}:  ~S=\frac{2\sqrt{2}}{7}
\begin{pmatrix}
 \frac{1-c_{2}}{\sqrt{2}} & \frac{1-c_{1}}{\sqrt{2}} & c_{2}-c_{1} & \frac{1-c_{3}}{\sqrt{2}} & c_{3}-c_{2} & c_{1}-c_{3} \\
 \frac{1-c_{1}}{\sqrt{2}} & \frac{1-c_{3}}{\sqrt{2}} & c_{1}-c_{3} & \frac{1-c_{2}}{\sqrt{2}} & c_{2}-c_{1} & c_{3}-c_{2} \\
 c_{2}-c_{1} & c_{1}-c_{3} & \frac{1-c_{1}}{\sqrt{2}} & c_{3}-c_{2} & \frac{1-c_{2}}{\sqrt{2}} & \frac{1-c_{3}}{\sqrt{2}} \\
 \frac{1-c_{3}}{\sqrt{2}} & \frac{1-c_{2}}{\sqrt{2}} & c_{3}-c_{2} & \frac{1-c_{1}}{\sqrt{2}} & c_{1}-c_{3} & c_{2}-c_{1} \\
 c_{3}-c_{2} & c_{2}-c_{1} & \frac{1-c_{2}}{\sqrt{2}} & c_{1}-c_{3} & \frac{1-c_{3}}{\sqrt{2}} & \frac{1-c_{1}}{\sqrt{2}} \\
 c_{1}-c_{3} & c_{3}-c_{2} & \frac{1-c_{3}}{\sqrt{2}} & c_{2}-c_{1} & \frac{1-c_{1}}{\sqrt{2}} & \frac{1-c_{2}}{\sqrt{2}} \\
\end{pmatrix}\,, ~  \\
\nonumber && \bm{7}:   ~S=\frac{2}{7}
\begin{pmatrix}
 -\frac{1}{2} & \sqrt{2} & \sqrt{2} & \sqrt{2} & \sqrt{2} & \sqrt{2} & \sqrt{2} \\
 \sqrt{2} & \frac{s_{2}+4 s_{3}}{\sqrt{7}} & \frac{s1-4 s_2}{\sqrt{7}} & \frac{2 s_1-2 s_2-4 s_3}{\sqrt{7}} & \frac{-4 s_1-s_3}{\sqrt{7}} & \frac{4 s_1+2 s_2+2 s_3}{\sqrt{7}} & \frac{-2 s_1+4 s_2-2 s_3}{\sqrt{7}} \\
 \sqrt{2} & \frac{s_1-4 s_2}{\sqrt{7}} & \frac{-4 s_1-s_3}{\sqrt{7}} & \frac{-2 s_1+4 s_2-2 s_3}{\sqrt{7}} & \frac{s_2+4 s_3}{\sqrt{7}} & \frac{2 s_1-2 s_2-4 s_3}{\sqrt{7}} & \frac{4 s_1+2 s_2+2 s_3}{\sqrt{7}} \\
 \sqrt{2} & \frac{2 s_1-2 s_2-4 s_3}{\sqrt{7}} & \frac{-2 s_1+4 s_2-2 s_3}{\sqrt{7}} & \frac{s_1-4 s_2}{\sqrt{7}} & \frac{4 s_1+2 s_2+2 s_3}{\sqrt{7}} & \frac{s_2+4 s_3}{\sqrt{7}} & \frac{-4 s_1-s_3}{\sqrt{7}} \\
 \sqrt{2} & \frac{-4 s_1-s_3}{\sqrt{7}} & \frac{s_2+4 s_3}{\sqrt{7}} & \frac{4 s_1+2 s_2+2 s_3}{\sqrt{7}} & \frac{s_1-4 s_2}{\sqrt{7}} & \frac{-2 s_1+4 s_2-2 s_3}{\sqrt{7}} & \frac{2 s_1-2 s_2-4 s_3}{\sqrt{7}} \\
 \sqrt{2} & \frac{4 s_1+2 s_2+2 s_3}{\sqrt{7}} & \frac{2 s_1-2 s_2-4 s_3}{\sqrt{7}} & \frac{s_2+4 s_3}{\sqrt{7}} & \frac{-2 s_1+4 s_2-2 s_3}{\sqrt{7}} & \frac{-4 s_1-s_3}{\sqrt{7}} & \frac{s_1-4 s_2}{\sqrt{7}} \\
 \sqrt{2} & \frac{-2 s_1+4 s_2-2 s_3}{\sqrt{7}} & \frac{4 s_1+2 s_2+2 s_3}{\sqrt{7}} & \frac{-4 s_1-s_3}{\sqrt{7}} & \frac{2 s_1-2 s_2-4 s_3}{\sqrt{7}} & \frac{s_1-4 s_2}{\sqrt{7}} & \frac{s_2+4 s_3}{\sqrt{7}} \\
\end{pmatrix}\,, \\
\nonumber && \bm{8} :  ~S=\frac{2\sqrt{6}}{7}
\begin{pmatrix}
 \frac{2 c_{2}-c_{1}-c_{3}}{2 \sqrt{6}} & \frac{c_{1}-c_{3}}{2 \sqrt{2}} & \frac{c_{1}+c_{2}-2 c_{3}}{2 \sqrt{3}} & \frac{c_{1}-2 c_{2}+c_{3}}{2 \sqrt{3}} & \frac{c_{2}+c_{3}-2 c_{1}}{2 \sqrt{3}} & \frac{2 c_{1}-c_{2}-c_{3}}{2 \sqrt{3}} & \frac{2 c_{2}-c_{1}-c_{3}}{2 \sqrt{3}} & \frac{2 c_{3}-c_{1}-c_{2}}{2 \sqrt{3}} \\
 \frac{c_{1}-c_{3}}{2 \sqrt{2}} & \frac{c_{1}-2 c_{2}+c_{3}}{2 \sqrt{6}} & \frac{c_{1}-c_{2}}{2} & \frac{c_{3}-c_{1}}{2} & \frac{c_{2}-c_{3}}{2} & \frac{c_{3}-c_{2}}{2} & \frac{c_{1}-c_{3}}{2} & \frac{c_{2}-c_{1}}{2} \\
 \frac{c_{1}+c_{2}-2 c_{3}}{2 \sqrt{3}} & \frac{c_{1}-c_{2}}{2} & \frac{c_{2}-c_{1}}{\sqrt{6}} & \frac{c_{1}-c_{3}}{\sqrt{6}} & \frac{1-c_{3}}{\sqrt{6}} & \frac{c_{2}-c_{3}}{\sqrt{6}} & \frac{c_{1}-1}{\sqrt{6}} & \frac{c_{2}-1}{\sqrt{6}} \\
 \frac{c_{1}-2 c_{2}+c_{3}}{2 \sqrt{3}} & \frac{c_{3}-c_{1}}{2} & \frac{c_{1}-c_{3}}{\sqrt{6}} & \frac{c_{3}-c_{2}}{\sqrt{6}} & \frac{1-c_{2}}{\sqrt{6}} & \frac{c_{1}-c_{2}}{\sqrt{6}} & \frac{c_{3}-1}{\sqrt{6}} & \frac{c_{1}-1}{\sqrt{6}} \\
 \frac{c_{2}+c_{3}-2 c_{1}}{2 \sqrt{3}} & \frac{c_{2}-c_{3}}{2} & \frac{1-c_{3}}{\sqrt{6}} & \frac{1-c_{2}}{\sqrt{6}} & \frac{c_{1}-c_{3}}{\sqrt{6}} & \frac{c_{1}-1}{\sqrt{6}} & \frac{c_{1}-c_{2}}{\sqrt{6}} & \frac{c_{2}-c_{3}}{\sqrt{6}} \\
 \frac{2 c_{1}-c_{2}-c_{3}}{2 \sqrt{3}} & \frac{c_{3}-c_{2}}{2} & \frac{c_{2}-c_{3}}{\sqrt{6}} & \frac{c_{1}-c_{2}}{\sqrt{6}} & \frac{c_{1}-1}{\sqrt{6}} & \frac{c_{1}-c_{3}}{\sqrt{6}} & \frac{1-c_{2}}{\sqrt{6}} & \frac{1-c_{3}}{\sqrt{6}} \\
 \frac{2 c_{2}-c_{1}-c_{3}}{2 \sqrt{3}} & \frac{c_{1}-c_{3}}{2} & \frac{c_{1}-1}{\sqrt{6}} & \frac{c_{3}-1}{\sqrt{6}} & \frac{c_{1}-c_{2}}{\sqrt{6}} & \frac{1-c_{2}}{\sqrt{6}} & \frac{c_{3}-c_{2}}{\sqrt{6}} & \frac{c_{1}-c_{3}}{\sqrt{6}} \\
 \frac{2 c_{3}-c_{1}-c_{2}}{2 \sqrt{3}} & \frac{c_{2}-c_{1}}{2} & \frac{c_{2}-1}{\sqrt{6}} & \frac{c_{1}-1}{\sqrt{6}} & \frac{c_{2}-c_{3}}{\sqrt{6}} & \frac{1-c_{3}}{\sqrt{6}} & \frac{c_{1}-c_{3}}{\sqrt{6}} & \frac{c_{2}-c_{1}}{\sqrt{6}} \\
\end{pmatrix}\,, \\
\nonumber && \bm{3}:~T=\text{diag}(\omega_{7},\omega_{7}^2,\omega_{7}^4)\,, \qquad \bm{\bar{3}}:~T=\text{diag}(\omega_{7}^6,\omega_{7}^5,\omega_{7}^3)\,, \\
\nonumber &&\bm{6}:~T=\text{diag}(\omega_{7}^1,\omega_{7}^2,\omega_{7}^3,\omega_{7}^4,\omega_{7}^5,\omega_{7}^6)\,, \qquad \bm{7}:~T=\text{diag}(1,\omega_{7},\omega_{7}^2,\omega_{7}^3,\omega_{7}^4,\omega_{7}^5,\omega_{7}^6)\,, \\
&&
\bm{8}:~T=\text{diag}(1,1,\omega_{7},\omega_{7}^2,\omega_{7}^3,\omega_{7}^4,\omega_{7}^5,\omega_{7}^6)\,,
\end{eqnarray}
where the parameter $\omega_{7}$ is the seventh unit root $\omega_{7}=e^{2\pi i/7}$, $s_{n}=\sin{\frac{2n\pi}{7}}$ and $c_{n}=\cos{\frac{2n\pi}{7}}$ with $n=1,2,3$. One see that $T$ is represented by diagonal matrices in different $\Sigma(168)$ irreducible representations, and the representation matrices of $S$ real and symmetric. As a result,  the gCP transformation compatible with $\Sigma(168)$  shares an identical structure with the family group transformation in our chosen basis.

\subsection{\label{sec:Delta6n2}The series group $\Delta(6n^{2})$}

$\Delta(6n^{2})$ is a non-abelian finite subgroup of $SU(3)$ and it is isomorphic to $(Z_n\times Z_n)\rtimes S_3$ for an arbitrary positive integer $n$. Its use as a flavour symmetry and the corresponding implications for lepton mixing have been extensively studied in the literature~\cite{King:2013vna,King:2014rwa,Hagedorn:2014wha,Ding:2014ora,Li:2017zmk,Li:2017abz,Lu:2018oxc}. In this work we follow the conventions of Ref.~\cite{Ding:2014ora}. The group is generated by four elements $a$, $b$, $c$ and $d$ obeying~\cite{Ding:2014ora,Escobar:2008vc}
\begin{eqnarray}
\nonumber &&\quad a^3=b^2=(ab)^2=1,\quad c^n=d^n=1,\qquad cd=dc\,,\\
 &&aca^{-1}=c^{-1}d^{-1},\quad ada^{-1}=c,\quad bcb^{-1}=d^{-1}, \quad bdb^{-1}=c^{-1}\,,
\end{eqnarray}
The $\Delta(6n^2)$ group has $6n^2$ elements which can be expressed as
\begin{equation}
g=a^{\alpha}b^{\beta}c^{\gamma}d^{\delta}\,,~~ \quad \alpha=0,1,2,\quad \beta=0,1,\quad \gamma,\delta=0,1,\ldots, n-1\,.
\end{equation}

The group $\Delta(6n^2)$ has one-dimensional, two-dimensional, three-dimensional and six-dimensional irreducible representations~\cite{Ding:2014ora,Escobar:2008vc}. In particular, there are $2(n-1)$ three-dimensional irreducible representations, denoted $\bm{3}^{l}_{k}$, whose generators take the form~\cite{Ding:2014ora}
\begin{equation}\label{eq:rep_3d_Delta6n2}
a=\begin{pmatrix}0 & ~1~ &0 \\ 0&~0~&1 \\
   1&~0~&0\end{pmatrix},~~
b=(-1)^{k}\begin{pmatrix} 0 &~0~ &1 \\ 0&~1~&0 \\
   1&~0~&0\end{pmatrix},~~
c=\begin{pmatrix} \eta^{l}&~0~ &0 \\ 0&~\eta^{-l}~&0 \\
   0&~0~&1\end{pmatrix},~~
d=\begin{pmatrix}1 &~0~ &0 \\ 0&~\eta^{l}~&0 \\
   0&~0~&\eta^{-l}\end{pmatrix}\,,
\end{equation}
where $\eta=e^{2\pi i/n}$, $k=0,1$ and $l=1, 2,\ldots, n-1$. We shall restrict ourselves to working with only faithful irreducible representations of $\Delta(6n^2)$. Only triplets with $l$ not dividing $n$ are faithful and unfaithful ones are discarded in our analysis. The two representations $\bm{3}^{l}_0$ and $\bm{3}^{l}_1$ differ only by an overall sign in the generator $b$. The generator $a$ is the same for all $l$ and the identities $\rho_{\bm{3}^{l}_{k}}(c)=[\rho_{\bm{3}^{1}_{k}}(c)]^{l}$, $\rho_{\bm{3}^{l}_{k}}(d)=[\rho_{\bm{3}^{1}_{k}}(d)]^{l}$ are fulfilled. It is clear that each power of the $c$ and $d$ generators will appear in every faithful three-dimensional irreducible representation. Without loss of generality, we embed the three left-handed lepton fields into the faithful representation $\bm{3}^{1}_0$ which is denoted as $\bm{3}$.

It has been demonstrated that CP symmetry can be consistently formulated within $\Delta(6n^2)$ flavour symmetry, provided $n$ is not a multiple of 3. Even when $n$ is divisible by 3, physically well-defined CP transformations may still be realized in a model, as long as no fields transform under the $\Delta(6n^2)$ doublet representations $\bm{2_2}$, $\bm{2_3}$, or $\bm{2_4}$~\cite{Ding:2014ora}.  As has been shown in Ref.~\cite{Ding:2014ora}, the most general CP transformation consistent with the $\Delta(6n^2)$ flavour symmetry is of the same form as the flavour symmetry transformation in the basis of Eq.~\eqref{eq:rep_3d_Delta6n2}, i.e.
\begin{equation}
X_{\bm{r}}=\rho_{\bm{r} }(g), \qquad g\in \Delta(6n^2)\,,
\end{equation}
where $\rho_{\bm{r}}(g)$ denotes the representation matrix of the element $g$ in the irreducible representation $\bm{r}$ of the $\Delta(6n^2)$ group.

\subsection{\label{sec:Dn}The series group $D_{n}$}

The dihedral group $D_{n}$ is the symmetry group of a regular $n$-gon for $n>1$.
A regular $n$-gon admits $n$ rotations and $n$ reflections, so $D_{n}$ has $2n$ elements.
For $n>2$ the group is non-abelian, while $D_{1}\simeq Z_{2}$ and $D_{2}\simeq Z_{2}\times Z_{2}$.
It can be written as the semidirect product $Z_{n}\rtimes Z_{2}$.
A convenient presentation is given in terms of two generators $R$ and $S$ satisfying
\begin{equation}
\label{eq:Dn_multi_rules}
R^{n}=S^{2}=(RS)^2=1\,,
\end{equation}
where $R$ represents a rotation and $S$ a reflection.
Every group element can be written as
\begin{equation}
g=S^{\alpha}R^{\beta}\,,\qquad \alpha=0,1,\ \beta=0,1,\dots,n-1\,.
\end{equation}

The group $D_{n}$ admits only real one- and two-dimensional irreducible representations.
The representation content depends on whether $n$ is odd or even.
For odd $n$, there are two singlets $\mathbf{1}_{1,2}$ and $(n-1)/2$ doublets $\mathbf{2}_{j}$:
\begin{equation}
\mathbf{1}_{1}:~R=S=1\,, \qquad
\mathbf{1}_{2}:~R=1,~S=-1\,,
\end{equation}
\begin{equation}
\mathbf{2}_{j}:~
R=\begin{pmatrix} e^{2\pi i j/n} & 0 \\ 0 & e^{-2\pi i j/n} \end{pmatrix},\qquad
S=\begin{pmatrix} 0 & 1 \\ 1 & 0 \end{pmatrix},
\quad j=1,\dots,\frac{n-1}{2}.
\end{equation}
For even $n$, the group has four singlets $\mathbf{1}_{1,2,3,4}$ and $(n/2-1)$ doublets $\mathbf{2}_{j}$:
\begin{equation}
\begin{aligned}
&\mathbf{1}_{1}:~R=S=1\,, \qquad\qquad \mathbf{1}_{2}:~R=1,~S=-1\,,\\
&\mathbf{1}_{3}:~R=-1,~S=1\,, \qquad\quad \mathbf{1}_{4}:~R=S=-1\,,\\
&\mathbf{2}_{j}:~
R=\begin{pmatrix} e^{2\pi i j/n} & 0 \\ 0 & e^{-2\pi i j/n} \end{pmatrix},\quad
S=\begin{pmatrix} 0 & 1 \\ 1 & 0 \end{pmatrix},
\quad j=1,\dots,\frac{n}{2}-1.
\end{aligned}
\end{equation}
As shown in Ref.~\cite{Lu:2019gqp}, the generalized CP transformations compatible with $D_{n}$ take the form
\begin{equation}
\label{eq:GCP_all}
X_{\mathbf{r}}=\rho_{\mathbf{r}}(g)\,,\qquad g\in D_{n}\,.
\end{equation}
Thus in the adopted basis, all admissible CP transformations coincide with flavour transformations of $D_{n}$.

\section{\label{sec:3sigma-ranges}Allowed ranges of lepton mixing observables for viable one- and two-parameter patterns }

In this section, we summarize the allowed ranges of the lepton mixing angles $\theta_{12}$, $\theta_{13}$, $\theta_{23}$ and the CPV phases $\delta_{CP}$, $\alpha_{21}$, and $\alpha_{31}$ for all viable one- and two-parameter mixing patterns discussed in section~\ref{sec:num_one} and section~\ref{sec:num_two}.
The allowed regions are obtained by scanning the full parameter space and imposing the requirement that the three mixing angles and the Dirac CPV phase $\delta_{CP}$ lie within their experimentally preferred $3\sigma$ intervals from \texttt{NuFIT}~v6.0, as summarized in table~\ref{tab:lepton-data}.  The resulting allowed ranges of the lepton mixing observables for the viable patterns listed in table~\ref{tab:best-fit-onepara}, table~\ref{tab:best-fit-twopara_A5_Sigma168}, table~\ref{tab:twopara_bf_CaseVIII_Delta6nsq}, and table~\ref{tab:twopara_bf_CaseIX_X_Delta6nsq} are presented in table~\ref{tab:3sigma-onepara}, table~\ref{tab:3sigma-twopara_A5_Sigma168}, table~\ref{tab:twopara_3sigma_CaseI_Delta6nsq}, and table~\ref{tab:twopara_3sigma_CaseII_Delta6nsq}, respectively.

\begin{table}[t!]\tabcolsep=0.11cm
\begin{center}
\scalebox{0.8}{\begin{tabular}{|c|c|c|c|c|c|c|c|c|}
\hline\hline
Order & Case &$(\varphi_1,\varphi_2)$ & $\sin^2\theta_{12}$ & $\sin^2\theta_{13}$ & $\sin^2\theta_{23}$ & $\delta_{CP}/\pi$ & $\alpha_{21}/\pi \text{(mod 1)}$ & $\alpha_{31}/\pi \text{(mod 1)}$   \\ \hline
\multirow{8}{*}{NO} & \multirow{1}{*}{$U^{I}$} & $(-\frac{\pi}{2},\frac{\pi}{2})$ & $[0.3171,0.3195]$ & $[0.02030,0.02388]$ & $0.5$ & $1.5$ & $0$ & $0$\\  \cline{2-9}
& \multirow{1}{*}{$U^{II}_{1}$} & $(0,\pi)$ & $[0.3402,0.3415]$ & $[0.02030,0.02388]$ & $0.5$ & $1.5$ & $0$ & $0$\\  \cline{2-9}
& \multirow{1}{*}{$U^{II}_{2}$} & $(0,\pi)$ & $[0.3402,0.3415]$ & $[0.02030,0.02388]$ & $0.5$ & $1.5$ & $0$ & $0$\\  \cline{2-9}
& $U^{III}$ & --- & $[0.2821,0.2832]$ & $[0.02030,0.02388]$ & $0.5$ & $1.5$ & $0$ & $0$\\  \cline{2-9} & $U^{IV}_{1}$ &--- & $[0.3295,0.3319]$ & $[0.02030,0.02388]$ & $[0.515,0.532]$ & $1$ & $0$ & $0$\\  \cline{2-9} & $U^{IV}_{2}$ & ---& $[0.3295,0.3319]$ & $[0.02030,0.02388]$ & $[0.468,0.485]$ & $0$ & $0$ & $0$\\  \cline{2-9} & $U^{V}_{1}$ &--- & $[0.3407,0.3415]$ & $[0.02153,0.02388]$ & $[0.541,0.585]$ & $[1.210,1.382]^{*}$ & $[0.153,0.165]^{*}$ & $[0.544,0.565]^{*}$\\  \cline{2-9} & $U^{V}_{2}$ & ---& $[0.3407,0.3415]$ & $[0.02153,0.02388]$ & $[0.435,0.459]$ & $[1.618,1.710]$ & $[0.835,0.841]$ & $[0.435,0.446]$\\  \cline{2-9}
\hline\hline
\multirow{6}{*}{IO} & \multirow{1}{*}{$U^{I}$} & $(-\frac{\pi}{2},\frac{\pi}{2})$ & $[0.3169,0.3193]$ & $[0.02060,0.02409]$ & $0.5$ & $1.5$ & $0$ & $0$\\ \cline{2-9}
& \multirow{1}{*}{$U^{II}_{1}$} & $(0,\pi)$ & $[0.3403,0.3416]$ & $[0.02060,0.02409]$ & $0.5$ & $1.5$ & $0$ & $0$\\  \cline{2-9}
& \multirow{1}{*}{$U^{II}_{2}$} & $(0,\pi)$ & $[0.3404,0.3415]$ & $[0.02060,0.02409]$ & $0.5$ & $1.5$ & $0$ & $0$\\ \cline{2-9}
& $U^{III}$ &--- & $[0.2822,0.2832]$ & $[0.02060,0.02409]$ & $0.5$ & $1.5$ & $0$ & $0$\\  \cline{2-9}  & $U^{V}_{1}$ &--- & $[0.3407,0.3416]$ & $[0.02153,0.02406]$ & $[0.539,0.584]$ & $[1.214,1.386]^{*}$ & $[0.154,0.165]^{*}$ & $[0.545,0.566]^{*}$\\  \cline{2-9} & $U^{V}_{2}$ &--- & $[0.3407,0.3416]$ & $[0.02153,0.02406]$ & $[0.440,0.461]$ & $[1.614,1.690]$ & $[0.835,0.840]$ & $[0.434,0.443]$\\  \cline{2-9}
\hline \hline
\end{tabular}}
\caption {\label{tab:3sigma-onepara}The allowed regions of lepton mixing observables for viable one-parameter mixing patterns given in table~\ref{tab:best-fit-onepara}.}
\end{center}
\end{table}

\begin{table}[t!]
\centering
\renewcommand{\arraystretch}{1.1}
\scalebox{0.75}{\begin{tabular}{|c|c|c|c|c|c|c|c|}
\hline \hline
\multicolumn{8}{|c|}{$A_{5}\rtimes H_{CP}$} \\ \hline
Order & Case &  $\sin^2\theta_{12}$ & $\sin^2\theta_{13}$ & $\sin^2\theta_{23}$  & $\delta_{CP}/\pi$ & $\alpha_{21}/\pi (\text{mod 1})$ & $\alpha_{31}/\pi (\text{mod 1})$   \\ \hline
\multirow{10}{*}{NO}
& $U^{VI}_{1}$ & $[0.330,0.332]$ & $[0.02052,0.02371]$ & $[0.435,0.585]$ & $[0,0.021]\cup[0.689,2]$ & $0$ & $[0,1]$ \\ \cline{2-8} & $U^{VI}_{2}$ & $[0.330,0.332]$ & $[0.02044,0.02382]$ & $[0.435,0.585]$ & $1$ & $0$ & $0$\\  \cline{2-8} & $U^{VI}_{3,2}$ & $[0.330,0.343]$ & $[0.02030,0.02388]$ & $[0.515,0.585]$ & $[0.703,1.297]$ & $[0,0.134]\cup[0.867,1]$ & $[0,0.083]\cup[0.917,1]$\\  \cline{2-8} & $U^{VI}_{4,1}$ & $[0.305,0.345]$ & $[0.02030,0.02388]$ & $[0.435,0.511]$ & $0$ & $0$ & $0$\\  \cline{2-8} & $U^{VI}_{4,2}$ & $[0.297,0.345]$ & $[0.02030,0.02388]$ & $[0.489,0.585]$ & $1$ & $0$ & $0$\\  \cline{2-8} & $U^{VI}_{4,3}$ & $[0.275,0.345]$ & $[0.02030,0.02388]$ & $[0.513,0.572]$ & $0$ & $0$ & $0$\\  \cline{2-8} & $U^{VI}_{4,4}$ & $[0.277,0.345]$ & $[0.02030,0.02388]$ & $[0.435,0.486]$ & $1$ & $0$ & $0$\\  \cline{2-8} & $U^{VI}_{5}$ & $[0.330,0.341]$ & $[0.02030,0.02388]$ & $[0.515,0.585]$ & $[0.780,1.220]$ & $[0,0.175]\cup[0.826,1]$ & $[0,0.126]\cup[0.874,1]$\\  \cline{2-8} & $U^{VI}_{6,1}$ & $[0.275,0.345]$ & $[0.02030,0.02388]$ & $[0.435,0.549]$ & $1$ & $0$ & $0$\\  \cline{2-8} & $U^{VI}_{6,2}$ & $[0.275,0.345]$ & $[0.02030,0.02388]$ & $[0.452,0.568]$ & $0$ & $0$ & $0$\\  \cline{2-8}
  \hline \hline
\multirow{4}{*}{IO} & $U^{VI}_{1}$ & $[0.329,0.332]$ & $[0.02061,0.02408]$ & $[0.440,0.584]$ & $[1.122,1.860]$ & $0$ & $[0.001,0.999]$ \\  \cline{2-8} & $U^{VI}_{3,1}$ & $[0.333,0.339]$ & $[0.02060,0.02408]$ & $[0.440,0.471]$ & $[1.760,1.861]^{*}$ & $[0.881,0.930]^{*}$ & $[0.927,0.957]^{*}$\\  \cline{2-8} & $U^{VI}_{3,2}$ & $[0.332,0.343]$ & $[0.02060,0.02409]$ & $[0.525,0.584]$ & $[1.117,1.298]^{*}$ & $[0.060,0.135]^{*}$ & $[0.036,0.083]^{*}$\\  \cline{2-8} & $U^{VI}_{5}$ & $[0.333,0.341]$ & $[0.02060,0.02409]$ & $[0.535,0.584]$ & $[1.117,1.220]^{*}$ & $[0.824,0.912]^{*}$ & $[0.873,0.936]^{*}$\\  \cline{2-8}     \hline \hline
\multicolumn{8}{|c|}{$\Sigma(168)\rtimes H_{CP}$} \\ \hline
Order & Case & $\sin^2\theta_{12}$ & $\sin^2\theta_{13}$ & $\sin^2\theta_{23}$  & $\delta_{CP}/\pi$ & $\alpha_{21}/\pi (\text{mod 1})$ & $\alpha_{31}/\pi (\text{mod 1})$   \\ \hline
\multirow{5}{*}{NO} & $U^{VII}_{1,1}$ & $[0.332,0.345]$ & $[0.02052,0.02388]$ & $[0.549,0.585]$ & $[1.627,1.742]^{*}$ & $[0.584,0.602]^{*}$ & $[0.242,0.257]^{*}$\\  \cline{2-8} & $U^{VII}_{2,1}$ & $[0.305,0.345]$ & $[0.02030,0.02387]$ & $[0.435,0.511]$ & $0$ & $0$ & $0$\\  \cline{2-8} & $U^{VII}_{2,2}$ & $[0.297,0.345]$ & $[0.02030,0.02388]$ & $[0.489,0.585]$ & $1$ & $0$ & $0$\\  \cline{2-8} & $U^{VII}_{2,3}$ & $[0.275,0.345]$ & $[0.02030,0.02388]$ & $[0.513,0.572]$ & $0$ & $0$ & $0$\\  \cline{2-8} & $U^{VII}_{2,4}$ & $[0.277,0.345]$ & $[0.02030,0.02388]$ & $[0.435,0.486]$ & $1$ & $0$ & $0$\\  \cline{2-8}
  \hline \hline
\multirow{2}{*}{IO} & $U^{VII}_{1,1}$ & $[0.332,0.345]$ & $[0.02064,0.02409]$ & $[0.547,0.584]$ & $[1.632,1.747]^{*}$ & $[0.584,0.602]^{*}$ & $[0.242,0.257]^{*}$\\  \cline{2-8} & $U^{VII}_{1,2}$ & $[0.339,0.345]$ & $[0.02209,0.02408]$ & $[0.440,0.453]$ & $[1.254,1.291]^{*}$ & $[0.398,0.405]^{*}$ & $[0.743,0.748]^{*}$\\  \cline{2-8}
\hline \hline
\end{tabular}}
\caption{\label{tab:3sigma-twopara_A5_Sigma168}The allowed regions of lepton mixing observables for viable two-parameter mixing patterns given in table~\ref{tab:best-fit-twopara_A5_Sigma168}. }
\end{table}

\begin{table}[t!]
\centering
\scalebox{0.75}{\begin{tabular}{|c|c|c|c|c|c|c|c|c|c|}
\hline \hline
\multicolumn{10}{|c|}{$\Delta(6n^{2})\rtimes H_{CP}$} \\ \hline
Order & Case & $\varphi_1$ & $\varphi_2$ & $\sin^2\theta_{12}$ & $\sin^2\theta_{13}$ & $\sin^2\theta_{23}$  & $\delta_{CP}/\pi$ & $\alpha_{21}/\pi \text{(mod 1)}$ & $\alpha_{31}/\pi \text{(mod 1)}$   \\ \hline
\multirow{10}{*}{NO}& \multirow{1}{*}{$U^{VIII}_{4}$ }  & $\frac{\pi }{3}$ & $0$ & $[0.305,0.345]$ & $[0.02030,0.02387]$ & $[0.435,0.511]$ & $0$ & $0$ & $0$\\  \cline{2-10}& \multirow{1}{*}{$U^{VIII}_{5}$ }  & $\frac{\pi }{3}$ & $0$  & $[0.275,0.345]$ & $[0.02030,0.02388]$ & $[0.513,0.572]$ & $0$ & $0$ & $0$\\  \cline{2-10}& \multirow{3}{*}{$U^{VIII}_{6}$ }  & $\frac{\pi }{4}$ & $0$  & $[0.275,0.345]$ & $[0.02041,0.02370]$ & $[0.510,0.512]$ & $0\cup1$ & $0$ & $0$\\  \cline{3-10}& & $\frac{\pi }{4}$ & $\frac{\pi }{4}$  & $[0.275,0.345]$ & $[0.02041,0.02370]$ & $[0.510,0.512]$ & $[1.206,1.222]$ & $[0.843,0.864]$ & $[0.878,0.904]$\\  \cline{3-10}& & $\frac{\pi }{4}$ & $\frac{\pi }{2}$  & $[0.275,0.345]$ & $[0.02041,0.02370]$ & $[0.510,0.512]$ & $[1.448,1.469]^{*}$ & $[0.776,0.806]^{*}$ & $[0.836,0.872]^{*}$\\  \cline{2-10}& \multirow{1}{*}{$U^{VIII}_{7}$ }  & $\frac{\pi }{3}$ & $0$  & $[0.297,0.345]$ & $[0.02030,0.02388]$ & $[0.489,0.585]$ & $1$ & $0$ & $0$\\  \cline{2-10}& \multirow{1}{*}{$U^{VIII}_{8}$ }  & $\frac{\pi }{3}$ & $0$  & $[0.277,0.345]$ & $[0.02030,0.02388]$ & $[0.435,0.486]$ & $1$ & $0$ & $0$\\  \cline{2-10}& \multirow{3}{*}{$U^{VIII}_{9}$ }  & $\frac{\pi }{4}$ & $0$  & $[0.275,0.345]$ & $[0.02044,0.02373]$ & $[0.488,0.490]$ & $0\cup1$ & $0$ & $0$\\  \cline{3-10}& & $\frac{\pi }{4}$ & $\frac{\pi }{4}$  & $[0.275,0.345]$ & $[0.02044,0.02373]$ & $[0.488,0.490]$ & $[1.265,1.278]$ & $[0.136,0.157]$ & $[0.083,0.106]$\\  \cline{3-10}& & $\frac{\pi }{4}$ & $\frac{\pi }{2}$  & $[0.275,0.345]$ & $[0.02044,0.02373]$ & $[0.488,0.490]$ & $[1.531,1.552]$ & $[0.195,0.224]$ & $[0.128,0.164]$\\  \cline{2-10}
  \hline \hline
\multirow{4}{*}{IO}&  \multirow{2}{*}{$U^{VIII}_{6}$ }  & $\frac{\pi }{4}$ & $\frac{\pi }{4}$  & $[0.275,0.345]$ & $[0.02061,0.02392]$ & $[0.511,0.512]$ & $[1.722,1.735]^{*}$ & $[0.843,0.863]^{*}$ & $[0.893,0.917]^{*}$\\  \cline{3-10}& & $\frac{\pi }{4}$ & $\frac{\pi }{2}$  & $[0.275,0.345]$ & $[0.02061,0.02392]$ & $[0.511,0.512]$ & $[1.448,1.469]$ & $[0.775,0.805]$ & $[0.835,0.871]$\\  \cline{2-10}& \multirow{2}{*}{$U^{VIII}_{9}$ } & $\frac{\pi }{4}$ & $\frac{\pi }{4}$  & $[0.275,0.345]$ & $[0.02079,0.02389]$ & $[0.488,0.489]$ & $[1.265,1.278]$ & $[0.138,0.157]$ & $[0.084,0.106]$\\  \cline{3-10}& & $\frac{\pi }{4}$ & $\frac{\pi }{2}$  & $[0.275,0.345]$ & $[0.02079,0.02389]$ & $[0.488,0.489]$ & $[1.531,1.552]$ & $[0.196,0.225]$ & $[0.130,0.165]$\\  \cline{2-10}
\hline \hline
\multicolumn{10}{|c|}{$D_{n}\rtimes H_{CP}$} \\ \hline
Order & Case & $\varphi_1$ & $\varphi_2$ & $\sin^2\theta_{12}$ & $\sin^2\theta_{13}$ & $\sin^2\theta_{23}$  & $\delta_{CP}/\pi$ & $\alpha_{21}/\pi \text{(mod 1)}$ & $\alpha_{31}/\pi \text{(mod 1)}$   \\ \hline
\multirow{6}{*}{NO}& \multirow{2}{*}{$U^{VIII}_{1}$ }  & $\frac{\pi }{5}$ & $0$ & $[0.330,0.332]$ & $[0.02044,0.02382]$ & $[0.435,0.585]$ & $0\cup1$ & $0$ & $0$\\  \cline{3-10}& & $\frac{\pi }{5}$ & $\frac{\pi }{2}$  & $[0.330,0.332]$ & $[0.02044,0.02383]$ & $[0.435,0.585]$ & $[1.475,1.519]^{*}$ & $0^{*}$ & $[0,0.038]^{*}\cup[0.949,1]^{*}$\\  \cline{2-10}& \multirow{1}{*}{$U^{VIII}_{4}$ }  & $\frac{2\pi }{5}$ & $0$ & $[0.275,0.345]$ & $[0.02031,0.02388]$ & $[0.435,0.548]$ & $1$ & $0$ & $0$\\  \cline{2-10}& \multirow{1}{*}{$U^{VIII}_{5}$ }  & $\frac{2\pi }{7}$ & $0$ & $[0.275,0.345]$ & $[0.02030,0.02388]$ & $[0.516,0.563]$ & $1$ & $0$ & $0$\\  \cline{2-10}& \multirow{1}{*}{$U^{VIII}_{7}$ }  & $\frac{2\pi }{5}$ & $0$ & $[0.275,0.345]$ & $[0.02030,0.02388]$ & $[0.452,0.568]$ & $0$ & $0$ & $0$\\  \cline{2-10}& \multirow{1}{*}{$U^{VIII}_{8}$ }  & $\frac{2\pi }{7}$ & $0$ & $[0.275,0.345]$ & $[0.02030,0.02388]$ & $[0.437,0.484]$ & $0$ & $0$ & $0$\\  \cline{2-10}
  \hline \hline
\multirow{1}{*}{IO}& \multirow{1}{*}{$U^{VIII}_{1}$ }  & $\frac{\pi }{5}$ & $\frac{\pi }{2}$ & $[0.329,0.332]$ & $[0.02075,0.02399]$ & $[0.440,0.584]$ & $[1.475,1.518]^{*}$ & $0^{*}$ & $[0,0.036]^{*}\cup[0.950,1]^{*}$\\  \cline{2-10}
\hline \hline
\end{tabular}}
\caption{\label{tab:twopara_3sigma_CaseI_Delta6nsq}The allowed regions of lepton mixing observables for viable two-parameter mixing patterns given in table~\ref{tab:twopara_bf_CaseVIII_Delta6nsq}.}
\end{table}

\begin{table}[t!]
\centering
\scalebox{0.6}{\begin{tabular}{|c|c|c|c|c|c|c|c|c|c|}
\hline \hline
\multicolumn{10}{|c|}{$\Delta(6n^{2})\rtimes H_{CP}$ (Case IX)} \\ \hline
Order & Case & $\varphi_1$ & $\varphi_2$ & $\sin^2\theta_{12}$ & $\sin^2\theta_{13}$ & $\sin^2\theta_{23}$  & $\delta_{CP}/\pi$ & $\alpha_{21}/\pi \text{(mod 1)}$ & $\alpha_{31}/\pi \text{(mod 1)}$   \\ \hline
\multirow{16}{*}{NO}& \multirow{3}{*}{$U^{IX}_{1}$ }  & $0$ & $0$ & $[0.305,0.345]$ & $[0.02030,0.02387]$ & $[0.435,0.511]$ & $0$ & $0$ & $0$\\  \cline{3-10}& & $0$ & $\frac{\pi }{4}$  & $[0.312,0.345]$ & $[0.02030,0.02388]$ & $[0.435,0.511]$ & $[1.883,1.998]$ & $[0,0.023]$ & $[0.500,0.514]$\\  \cline{3-10}  & & $\frac{\pi }{4}$ & $\frac{\pi }{2}$  & $[0.329,0.345]$ & $[0.02030,0.02388]$ & $[0.435,0.494]$ & $[1.764,1.925]^{*}$ & $[0.408,0.692]^{*}$ & $[0.652,0.836]^{*}$\\  \cline{2-10}& \multirow{5}{*}{$U^{IX}_{2}$ }  & $0$ & $0$  & $[0.275,0.345]$ & $[0.02030,0.02388]$ & $[0.513,0.572]$ & $0$ & $0$ & $0$\\  \cline{3-10}& & $0$ & $\frac{\pi }{4}$  & $[0.275,0.345]$ & $[0.02030,0.02388]$ & $[0.513,0.585]$ & $[1.773,1.998]$ & $[0.982,1]$ & $[0.482,0.500]$\\  \cline{3-10}  & & $\frac{\pi }{4}$ & $0$  & $[0.275,0.345]$ & $[0.02031,0.02388]$ & $[0.532,0.585]$ & $[1.623,1.959]^{*}$ & $[0.786,0.972]^{*}$ & $[0.867,0.982]^{*}$\\  \cline{3-10} & & $\frac{\pi }{4}$ & $\frac{\pi }{4}$  & $[0.287,0.345]$ & $[0.02030,0.02388]$ & $[0.547,0.585]$ & $[1.749,1.766]^{*}$ & $[0.725,0.802]^{*}$ & $[0.326,0.384]^{*}$\\  \cline{3-10} & & $\frac{\pi }{2}$ & $\frac{\pi }{4}$  & $[0.289,0.345]$ & $[0.02030,0.02388]$ & $[0.552,0.582]$ & $[0,0.022]\cup[1.636,2]$ & $[0.001,0.246]\cup[0.983,1]$ & $[0.307,0.473]$\\  \cline{2-10}& \multirow{3}{*}{$U^{IX}_{3}$ }  & $0$ & $0$ & $[0.297,0.345]$ & $[0.02030,0.02388]$ & $[0.489,0.585]$ & $1$ & $0$ & $0$\\  \cline{3-10}& & $0$ & $\frac{\pi }{4}$  & $[0.306,0.345]$ & $[0.02030,0.02388]$ & $[0.489,0.585]$ & $[1.002,1.145]^{*}$ & $[0.970,1]^{*}$ & $[0.479,0.500]^{*}$\\  \cline{3-10}  & & $\frac{\pi }{4}$ & $\frac{\pi }{2}$  & $[0.330,0.345]$ & $[0.02030,0.02388]$ & $[0.506,0.585]$ & $[0.722,0.767]\cup[1.075,1.291]$ & $[0.307,0.646]$ & $[0.164,0.387]$\\  \cline{2-10}& \multirow{5}{*}{$U^{IX}_{4}$ }  & $0$ & $0$  & $[0.277,0.345]$ & $[0.02030,0.02388]$ & $[0.435,0.487]$ & $1$ & $0$ & $0$\\  \cline{3-10}& & $0$ & $\frac{\pi }{4}$  & $[0.290,0.345]$ & $[0.02030,0.02388]$ & $[0.435,0.487]$ & $[1.002,1.152]^{*}$ & $[0,0.012]^{*}$ & $[0.500,0.513]^{*}$\\  \cline{3-10}  & & $\frac{\pi }{4}$ & $0$  & $[0.278,0.345]$ & $[0.02030,0.02388]$ & $[0.435,0.469]$ & $[1.047,1.201]$ & $[0.032,0.115]$ & $[0.021,0.068]$\\  \cline{3-10} & & $\frac{\pi }{4}$ & $\frac{\pi }{4}$  & $[0.318,0.345]$ & $[0.02030,0.02388]$ & $[0.435,0.453]$ & $[1.243,1.251]$ & $[0.197,0.233]$ & $[0.616,0.641]$\\  \cline{3-10} & & $\frac{\pi }{2}$ & $\frac{\pi }{4}$  & $[0.283,0.330]$ & $[0.02030,0.02388]$ & $[0.435,0.448]$ & $[0.902,1.280]^{*}$ & $[0.717,0.954]^{*}$ & $[0.503,0.650]^{*}$\\  \cline{2-10}
  \hline \hline
\multirow{11}{*}{IO}& \multirow{1}{*}{$U^{IX}_{1}$ }  & $\frac{\pi }{4}$ & $\frac{\pi }{2}$  & $[0.329,0.334]$ & $[0.02060,0.02409]$ & $[0.440,0.464]$ & $[1.781,1.861]^{*}$ & $[0.428,0.543]^{*}$ & $[0.666,0.744]^{*}$\\  \cline{2-10}& \multirow{4}{*}{$U^{IX}_{2}$ }  & $0$ & $\frac{\pi }{4}$  & $[0.275,0.303]$ & $[0.02060,0.02409]$ & $[0.561,0.584]$ & $[1.780,1.861]$ & $[0.982,0.989]$ & $[0.482,0.489]$\\  \cline{3-10}  & & $\frac{\pi }{4}$ & $0$  & $[0.318,0.345]$ & $[0.02060,0.02409]$ & $[0.531,0.584]$ & $[1.626,1.861]^{*}$ & $[0.785,0.915]^{*}$ & $[0.866,0.948]^{*}$\\  \cline{3-10} & & $\frac{\pi }{4}$ & $\frac{\pi }{4}$  & $[0.288,0.345]$ & $[0.02060,0.02409]$ & $[0.547,0.584]$ & $[1.749,1.765]^{*}$ & $[0.726,0.802]^{*}$ & $[0.327,0.384]^{*}$\\  \cline{3-10} & & $\frac{\pi }{2}$ & $\frac{\pi }{4}$  & $[0.307,0.345]$ & $[0.02060,0.02409]$ & $[0.552,0.582]$ & $[1.636,1.861]$ & $[0.001,0.148]\cup[0.983,1]$ & $[0.308,0.412]$\\  \cline{2-10}& \multirow{2}{*}{$U^{IX}_{3}$ }  & $0$ & $\frac{\pi }{4}$  & $[0.306,0.345]$ & $[0.02061,0.02409]$ & $[0.566,0.584]$ & $[1.117,1.281]^{*}$ & $[0.504,0.511]^{*}\cup[0.970,0.979]^{*}$ & $[0.480,0.487]^{*}\cup[0.928,0.930]^{*}$\\  \cline{3-10}  & & $\frac{\pi }{4}$ & $\frac{\pi }{2}$  & $[0.329,0.345]$ & $[0.02060,0.02409]$ & $[0.507,0.581]$ & $[1.117,1.292]$ & $[0.315,0.647]$ & $[0.168,0.388]$\\  \cline{2-10}& \multirow{4}{*}{$U^{IX}_{4}$ }  & $0$ & $\frac{\pi }{4}$  & $[0.293,0.309]$ & $[0.02061,0.02406]$ & $[0.440,0.446]$ & $[1.117,1.135]^{*}$ & $[0.009,0.011]^{*}$ & $[0.509,0.511]^{*}$\\  \cline{3-10}  & & $\frac{\pi }{4}$ & $0$  & $[0.308,0.345]$ & $[0.02061,0.02408]$ & $[0.445,0.469]$ & $[1.117,1.199]$ & $[0.072,0.114]$ & $[0.045,0.068]$\\  \cline{3-10} & & $\frac{\pi }{4}$ & $\frac{\pi }{4}$  & $[0.326,0.345]$ & $[0.02062,0.02408]$ & $[0.440,0.453]$ & $[1.245,1.252]$ & $[0.198,0.223]$ & $[0.616,0.634]$\\  \cline{3-10} & & $\frac{\pi }{2}$ & $\frac{\pi }{4}$  & $[0.304,0.323]$ & $[0.02061,0.02409]$ & $[0.440,0.448]$ & $[1.117,1.241]^{*}$ & $[0.838,0.924]^{*}$ & $[0.579,0.632]^{*}$\\  \cline{2-10}
\hline \hline
\multicolumn{10}{|c|}{$\Delta(6n^{2})\rtimes H_{CP}$ (Case X)} \\ \hline
Order & Case & $\varphi_1$ & $\varphi_2$  & $\sin^2\theta_{12}$ & $\sin^2\theta_{13}$ & $\sin^2\theta_{23}$  & $\delta_{CP}/\pi$ & $\alpha_{21}/\pi \text{(mod 1)}$ & $\alpha_{31}/\pi \text{(mod 1)}$   \\ \hline
\multirow{4}{*}{NO}& \multirow{2}{*}{$U^{X}_{1}$ }  & $0$ & $0$  & $[0.275,0.345]$ & $[0.02041,0.02370]$ & $[0.510,0.512]$ & $0\cup1$ & $0$ & $0$\\  \cline{3-10}
& & $\frac{\pi }{4}$ & $0$  & $[0.275,0.345]$ & $[0.02041,0.02370]$ & $[0.510,0.512]$ & $[1.206,1.222]$ & $[0.843,0.864]$ & $[0.878,0.904]$\\  \cline{2-10}& \multirow{2}{*}{$U^{X}_{2}$ }  & $0$ & $0$  & $[0.275,0.345]$ & $[0.02044,0.02373]$ & $[0.488,0.490]$ & $0\cup1$ & $0$ & $0$\\  \cline{3-10}
& & $\frac{\pi }{4}$ & $0$  & $[0.275,0.345]$ & $[0.02044,0.02373]$ & $[0.488,0.490]$ & $[1.265,1.278]$ & $[0.136,0.157]$ & $[0.083,0.106]$\\  \cline{3-10}
  \hline \hline
\multirow{2}{*}{IO}& $U^{X}_{1}$  & $\frac{\pi }{4}$ & $0$  & $[0.275,0.345]$ & $[0.02061,0.02392]$ & $[0.511,0.512]$ & $[1.722,1.735]^{*}$ & $[0.843,0.863]^{*}$ & $[0.893,0.917]^{*}$\\  \cline{2-10}
& $U^{X}_{2}$ & $\frac{\pi }{4}$ & $0$  & $[0.275,0.345]$ & $[0.02079,0.02389]$ & $[0.488,0.489]$ & $[1.265,1.278]$ & $[0.138,0.157]$ & $[0.084,0.106]$\\  \cline{3-10}
\hline \hline
\end{tabular}}
\caption{\label{tab:twopara_3sigma_CaseII_Delta6nsq}The allowed regions of lepton mixing observables for viable two-parameter mixing patterns given in table~\ref{tab:twopara_bf_CaseIX_X_Delta6nsq}.}
\end{table}

\section{\label{app:Z2-K4} The symmetry breaking pattern $\mathcal{G}_{\ell}=Z_2$, $\mathcal{G}_{\nu}=K_{4}\times H^{\nu}_{CP}$ and lepton flavour mixing}

This section examines a flavour symmetry breaking pattern where $G_{f}\rtimes H_{CP}$ is broken to $K_{4}\times H^{\nu}_{CP}$ in the neutrino sector and to $Z_2$ in the charged lepton sector. The resulting lepton mixing matrix would depend on only two real free parameters. Under this configuration, the residual symmetry $K_{4}\times H^{\nu}_{CP}$ of neutrino sector imposes strong constraints on the neutrino diagonalization matrix $U_{\nu}$ as specified by the consistency conditions:
\begin{equation}\label{eq:Unu-master-formula_K4CP}
U_{\nu}^{\dagger}\rho_{\bm{3}}(g_{\nu_i})U_\nu
=\text{diag}(\pm 1,\pm 1,\pm 1)\,,\qquad
U_\nu^{\dagger}X_{\nu\bm{3}}U_{\nu}^{*}=\text{diag}(\pm 1,\pm 1,\pm 1)\,.
\end{equation}
where  $g_{\nu_i}$ denote the generators of the residual $K_{4}$ group and $X_{\nu\bm{3}}$ is the corresponding residual CP transformation. Consequently, in this framework, $U_{\nu}$ is determined uniquely up to column permutations $P_{\nu}$ and a diagonal phase matrix $Q_{\nu}=\text{diag}(\pm i,\pm i,\pm i)$ which can shift the Majorana phases $\alpha_{21}$ and $\alpha_{31}$ by $\pi$.

For the charged leptons, the residual symmetry $Z_{2}$ is labeled as  $Z_{2}^{g_{\ell}}$, with generator $g_{\ell}$ satisfying $g_{\ell}^{2}=1$. Under the assumption that the charged lepton sector preserves this $Z_{2}^{g_{\ell}}$ symmetry, the unitary matrix $U_{\ell}$ s restricted to the form~\cite{Lu:2018oxc}:
\begin{equation}\label{eq:Ul_expression}
U_{\ell}=\Sigma_{\ell}U_{23}^{\dagger}(\theta_{\ell},\delta_{\ell})P_{\ell}^{T} Q_{\ell}^{\dagger}\,,
\end{equation}
where $\Sigma_{\ell}$ diagonalizes the representation matrix $\rho_{\bm{3}}(g_{\ell})$ such that
\begin{equation}
\Sigma_{\ell}^{\dagger}\rho_{\bm{3}}(g_{\ell})\Sigma_{\ell}=\text{diag}(1,-1,-1)\,.
\end{equation}
The matrix $U_{23}(\theta_{\ell},\delta_{\ell})$ is a unitary rotation in the 2–3 sector, and $Q_{\ell}$ is a diagonal phase matrix, i.e. \begin{equation}
U_{23}(\theta_{\ell},\delta_{\ell})=\left( \begin{array}{ccc}
1 & 0 & 0 \\
0 & \cos \theta_{\ell} & \sin \theta_{\ell} \\
0 & -\sin \theta_{\ell} & \cos \theta_{\ell} \end{array} \right)
\left( \begin{array}{ccc}
1 & 0 & 0 \\
0 & e^{i\delta_{\ell}} & 0 \\
0 & 0 & e^{-i\delta_{\ell}} \end{array} \right),~~
Q_{\ell}=\left( \begin{array}{ccc}
e^{-i\gamma_{1}} & 0 & 0 \\
0 & e^{-i\gamma_{2}} & 0 \\
0 & 0 & e^{-i\gamma_{3}}  \end{array} \right)\,,
\end{equation}
in which $\theta_{\ell}$, $\delta_{\ell}$ and $\gamma_{1,2,3}$ are unconstrained real parameters. Hence the residual symmetry $\mathcal{G}_{\ell}=Z_2$, $\mathcal{G}_{\nu}=K_{4}\times H^{\nu}_{CP}$ constrains the lepton mixing matrix to be of the following form,
\begin{equation}\label{eq:PMNS_z2K4cp}
U=Q_{\ell}P_{\ell}U_{23}(\theta_{\ell},\delta_{\ell})\Sigma_{\ell}^{\dagger}
U_{\nu}P_{\nu}Q_{\nu}\,,
\end{equation}
where the diagonal phase matrix $Q_{\ell}$ can be absorbed into charged lepton field redefinitions, and $Q_{\nu}$ only affects possible $\pi$ shifts in the Majorana phases. The resulting mixing matrix in Eq.~\eqref{eq:PMNS_z2K4cp} thus depends on only two free parameters $\theta_{\ell}$ and $\delta_{\ell}$, and one row of the mixing matrix is fixed by residual symmetry. Furthermore, the PMNS matrix in Eq.~\eqref{eq:PMNS_z2K4cp} exhibits the following symmetry
properties:
\begin{eqnarray}
\nonumber &&U(\theta_{\ell}+\pi,\delta_{\ell},\theta_{\nu})=P_{\ell}\text{diag}(1,-1,-1)P_{\ell}^{T}U(\theta_{\ell},\delta_{\ell},\theta_{\nu})\,\\
\nonumber &&U(\pi-\theta_{\ell},\delta_{\ell},\theta_{\nu})=P_{\ell}\text{diag}(1,-i,i)P_{\ell}^{T}U(\theta_{\ell},\delta_{\ell}-\pi/2,\theta_{\nu})\,\\
&&U(\theta_{\ell},\delta_{\ell}+\pi,\theta_{\nu})=P_{\ell}\text{diag}(1,-1,-1)P_{\ell}^{T}U(\theta_{\ell},\delta_{\ell},\theta_{\nu})\,
\end{eqnarray}
Since the diagonal matrices $P_{\ell}\text{diag}(1,-1,-1)P_{\ell}^{T}$ and $P_{\ell}\text{diag}(1,-i, i)P_{\ell}^{T}$ can be absorbed into $Q_{\ell}$, the fundamental domains for the parameters are restricted to $\theta_{\ell}\in[0,\pi/2]$ and $\delta_{\ell}\in[0,\pi)$.

When $A_{5}\rtimes H_{CP}$ is broken down to the residual symmetries $Z_{2}$ and $K_{4}\times H^{\nu}_{CP}$ in the charged lepton and neutrino sectors respectively, there are two independent such kind of residual symmetries and accordingly the absolute values of the fixed row is found to be
\begin{eqnarray}
\nonumber \mathcal{G}_{\ell}=Z^{S}_{2} ,\quad \mathcal{G}_{\nu}=K_{4}^{(ST^2ST^3S,TST^4)}\times H^{\nu}_{CP}\quad  &:&\quad  (\frac{\phi_{g}}{2},\frac{1}{2},\frac{\phi_{g}-1}{2})\,, \\
\mathcal{G}_{\ell}=Z^{TST^{4}}_{2} ,\quad \mathcal{G}_{\nu}=K_{4}^{(ST^2ST^3S,TST^4)}\times H^{\nu}_{CP}  \quad &:& \quad (1,0,0)\,,
\end{eqnarray}
with $\rho_{\bm{r}}(T^2)\in H^{\nu}_{CP}$. The two predicted columns are both outside the $3\sigma$ ranges of the magnitude of the elements of the lepton mixing matrix~\cite{Esteban:2024eli}. 

When the residual symmetries originate from  $\Sigma(168)\rtimes H_{CP}$, the only independent breaking pattern that yields a fixed row with no zero elements is $G_{\ell}=Z^{S}_{2}$ and $G_{\nu}=K_{4}^{(T^2ST^5,ST^3ST^4S)}$, and the corresponding unitary matrices $\Sigma_{\ell}$ and $U_{\nu}$ are
\begin{equation}
\Sigma_{\ell}=\frac{1}{2\sqrt{2}}\left(
\begin{array}{ccc}
 \frac{1}{s_2} & \frac{2 s_2-1}{s_2 \sqrt{4 c_2
   (c_2+s_3)+1}} & \frac{-2 s_2-1}{s_2 \sqrt{4
   c_2 (c_2-s_3)+1}} \\
 -\frac{1}{s_3} & \frac{2 c_2+2 s_3+1}{s_2 \sqrt{4
   c_2 (c_2+s_3)+1}} & \frac{2 c_2-2
   s_3+1}{s_2 \sqrt{4 c_2 (c_2-s_3)+1}} \\
 \frac{1}{s_1} & \frac{-2 c_3}{s_2 \sqrt{4 c_2
   (c_2+s_3)+1}} & \frac{-2 c_3}{s_2 \sqrt{4
   c_2 (c_2-s_3)+1}} \\
\end{array}
\right), \quad U_{\nu}=\frac{1}{2 \sqrt{2}}\left(
\begin{array}{ccc}
 \frac{(-1)^{4/7}}{s_{2}} & \frac{(-1)^{5/7}}{s_{3}} & \frac{1}{s_{1}} \\
 \frac{(-1)^{1/7}}{s_{3}} & -\frac{(-1)^{3/7}}{s_{1}} & \frac{1}{s_{2}} \\
 \frac{(-1)^{2/7}}{s_{1}} & \frac{(-1)^{6/7}}{s_{2}} & -\frac{1}{s_{3}} \\
\end{array}
\right)\,,
\end{equation}
where the parameters $s_{n}=\sin{\frac{2n\pi}{7}}$ and $c_{n}=\cos{\frac{2n\pi}{7}}$ with $n=1,2,3$. Then it is easy to check that the breaking pattern leads to a fixed row $(1,1,\sqrt{2})/2$ in the PMNS matrix. The full lepton mixing matrix can then be constructed using Eq.~\eqref{eq:PMNS_z2K4cp} with $P_{\ell}=P_{231}$ and $P_{\nu}=P_{132}$. The three lepton mixing angles depend on two free parameters $\theta_{l}$ and $\delta_{l}$ and we can straightforwardly extract them as follows
\begin{eqnarray}
\nonumber && \sin^2\theta_{13}=\frac{1}{4} \left[1+\sin 2 \theta_{\ell}  \sin \left(2 \delta_{\ell} -\frac{\pi}{4}\right)\right]\,, \quad \sin^2\theta_{23}=\frac{1-\sin 2 \theta_{\ell}  \sin \left(2 \delta_{\ell} -\frac{\pi}{4}\right) }{3-\sin 2 \theta_{\ell}  \sin \left(2 \delta_{\ell} -\frac{\pi}{4}\right)}\,, \\
&&\sin^{2}\theta_{12}=\frac{3-\sqrt{7} \cos 2 \theta_{\ell} +\sqrt{2} \cos 2 \delta_{\ell}  \sin 2 \theta_{\ell} }{6-2\sin 2 \theta_{\ell}  \sin \left(2 \delta_{\ell} -\frac{\pi}{4}\right)}\,.
\end{eqnarray}
It is easy to check that the atmospheric and reactor mixing angles have the correlation
\begin{equation}
\cos^{2}\theta_{13}\cos^{2}\theta_{23}=\frac{1}{2}\,.
\end{equation}
Using the $3\sigma$ range of the reactor mixing angle $0.02030\leq\sin^{2}\theta_{13}\leq0.02388$~\cite{Esteban:2024eli}, we have $\sin^{2}\theta_{23}\in[0.4878,0.4896]$.  Comprehensively scanning over the values of $\theta_{\ell}$ and $\delta_{\ell}$, while requiring $\theta_{13}$ in its experimentally favored $3\sigma$ region, we identify
\begin{equation}\label{eq:theta12_caseiii}
    0.3455\leq\sin^2\theta_{12}\leq 0.6545\,,
\end{equation}
which lies above the $3\sigma$ upper limit of $\sin^2\theta_{12}$ by JUNO~\cite{JUNO:2025gmd}. Hence this mixing pattern is ruled out.

When the residual symmetries originate from  $\Delta(6n^{2})\rtimes H_{CP}$, the only independent breaking pattern that gives a fixed row without zero elements is $G_{\ell}= Z^{bc^{x}d^{x}}_{2}$, $G_{\nu}=K^{(c^{n/2},abc^{y})}_{4}$ and $X_{\nu\bm{r}}=\rho_{\bm{r}}(c^{\delta}d^{2y+2\gamma})$. The corresponding lepton mixing matrices are given by
\begin{eqnarray}
\nonumber &&U^{XI(a)}=\frac{1}{2}R_{13}(\theta_{\ell})\text{diag}(e^{i\delta_{\ell}},1,e^{-i\delta_{\ell}})\left(
\begin{array}{ccc}
 \sqrt{2} e^{i \varphi _1} & -\sqrt{2} e^{i \varphi _1} & 0 \\
 1 & 1 & -\sqrt{2} e^{i \varphi _2} \\
 1 & 1 & \sqrt{2} e^{i \varphi _2}
\end{array}
\right)Q^{\dagger}_{\nu},\\
\nonumber &&=\text{diag}(e^{i\delta_{\ell}},1,e^{i\delta_{\ell}})R_{13}(\theta_{\ell})\left(
\begin{array}{ccc}
 \sqrt{2} e^{i (\varphi _1+2\delta_{\ell})} & -\sqrt{2} e^{i (\varphi _1+2\delta_{\ell})} & 0 \\
 1 & 1 & -\sqrt{2} e^{i \varphi _2} \\
 1 & 1 & \sqrt{2} e^{i \varphi _2}
\end{array}
\right)Q^{\dagger}_{\nu}, \\
\label{eq:PMNS_Case_XI} &&U^{XI(b)}=P_{132}U^{VIII(a)}_{PMNS}\,,
\end{eqnarray}
where the parameters $\varphi_1$ and $\varphi_2$ are determined as
\begin{equation}
\Delta(6n^2)~~:~~  \varphi_{1}=\frac{3\delta+2x+2y}{n}\pi,\qquad
\varphi_{2}=-\frac{3\gamma+2x+2y}{n}\pi\,,
\end{equation}
These two parameters $\varphi_1$ and $\varphi_2$ are interdependent of each other, and they can take the discrete values
\begin{equation}
\Delta(6n^2)~~:~~  \varphi_1, \varphi_2~(\textrm{mod}~2\pi)=0, \frac{1}{n}\pi, \frac{2}{n}\pi,\ldots,\frac{2n-1}{n}\pi\,,
\end{equation}
Obvious $U^{XI(b)}_{PMNS}$ is obtained from $U^{XI(a)}_{PMNS}$ by exchanging the second and third rows. In this case, the row fixed by residual symmetry is $(1,1,-\sqrt{2}e^{i\varphi_2})/2$, and it could be the second or the third row of the PMNS mixing matrix. The sum rule among the reactor and solar mixing angles is given by~\cite{Yao:2016zev}
\begin{equation}
\sin^2\theta_{12}=\frac{1}{2}\pm\tan\theta_{13}\sqrt{1-\tan^2\theta_{13}}\,\cos(\varphi_1+2\delta_{\ell})\,,
\end{equation}
where the ``+'' and ``$-$'' signs are valid $0<\theta_{\ell}<\pi/2$ and $\pi/2<\theta_{\ell}<\pi$ respectively. Freely varying the phase $\delta_{\ell}$ and limiting $\theta_{13}$ in its $3\sigma$ region~\cite{Esteban:2024eli}, we find that the solar angle can vary in the range $0.3455\leq\sin^2\theta_{12}\leq 0.6545$ which is identical with that of Eq.~\eqref{eq:theta12_caseiii} and ruled out by JUNO.

In summary, the breaking pattern with $\mathcal{G}_{\ell}=Z_{2}$ and $\mathcal{G}_{\nu}=K_{4}\times H^{\nu}_{CP}$ is not viable, since the experimental data of solar mixing angle $\theta_{12}$ can not be accommodated.

\end{appendix}


\providecommand{\href}[2]{#2}\begingroup\raggedright\endgroup

\end{document}